\documentclass[apj]{emulateapj}

\newcommand{\etal}{{et al.~}}
\newcommand{\msunh}{\>h^{-1}\rm M_\odot}
\newcommand{\msunhh}{\>h^{-2}\rm M_\odot}
\newcommand{\Msun}{\>{\rm M_\odot}}

\newcommand{\mpch}{\>h^{-1}{\rm {Mpc}}}
\newcommand{\kpch}{\>h^{-1}{\rm {kpc}}}

\newcommand{\calP}{{\cal P}}

\newcommand{\rmh}{{\rm h}}
\newcommand{\rmg}{{\rm g}}
\newcommand{\rmd}{{\rm d}}
\newcommand{\rmpp}{{\rm p}}
\newcommand{\rmc}{{\rm c}}
\newcommand{\rms}{{\rm s}}
\newcommand{\rmm}{{\rm m}}
\newcommand{\rmb}{{\rm b}}
\newcommand{\rme}{{\rm e}}

\def\gtsima{$\; \buildrel > \over \sim \;$}
\def\ltsima{$\; \buildrel < \over \sim \;$}
\def\gta{\lower.7ex\hbox{\gtsima}}
\def\lta{\lower.7ex\hbox{\ltsima}}

\shorttitle{The Galaxy - Dark Matter Connection}
\shortauthors{Yang et al.}

\begin{document}
            

\title{Evolution of the Galaxy - Dark Matter Connection and \\
       the Assembly of Galaxies in Dark Matter Halos}

\author{Xiaohu Yang\altaffilmark{1}, H.J. Mo \altaffilmark{2}, Frank C. van
  den Bosch\altaffilmark{3}, Youcai Zhang\altaffilmark{1}, Jiaxin
  Han\altaffilmark{1}}
\altaffiltext{1}{Key Laboratory for Research in Galaxies and Cosmology,
Shanghai Astronomical Observatory; Nandan Road 80, Shanghai 200030, China;
E-mail: xhyang@shao.ac.cn}
\altaffiltext{2}{Department of Astronomy, University of Massachusetts, Amherst
  MA 01003-9305}
\altaffiltext{3} {Astronomy Department, Yale University, P.O. Box 208101, New
  Haven, CT 06520-8101}


\begin{abstract}  
  We present a new model  to describe the galaxy-dark matter connection across
  cosmic time,  which unlike the popular subhalo  abundance matching technique
  is self-consistent in  that it takes account of the  facts that (i) subhalos
  are  accreted at  different  times,  and (ii)  the  properties of  satellite
  galaxies may  evolve after accretion.  Using observations  of galaxy stellar
  mass functions out  to $z \sim 4$, the conditional  stellar mass function at
  $z\sim 0.1$  obtained from SDSS  galaxy group catalogues, and  the two-point
  correlation  function (2PCF)  of galaxies  at $z  \sim 0.1$  as  function of
  stellar mass,  we constrain  the relation between  galaxies and  dark matter
  halos over the  entire cosmic history from $z \sim 4$  to the present.  This
  relation is then used to  predict the median assembly histories of different
  stellar  mass  components  within   dark  matter  halos  (central  galaxies,
  satellite galaxies, and halo stars).  We also make predictions for the 2PCFs
  of high-$z$ galaxies as function of  stellar mass. Our main findings are the
  following: (i) Our  model reasonably fits all data  within the observational
  uncertainties,  indicating that  the $\Lambda$CDM  concordance  cosmology is
  consistent  with a  wide variety  of  data regarding  the galaxy  population
  across cosmic time.   (ii) At low-$z$, the stellar  mass of central galaxies
  increases with halo  mass as $M^{0.3}$ and $M^{\ga 4.0}$  at the massive and
  low-mass ends, respectively.  The  ratio $M_{\ast,c}/M$ reveals a maximum of
  $\sim 0.03$  at a halo  mass $M \sim  10^{11.8}\msunh$, much lower  than the
  universal baryon fraction ($\sim 0.17$).  At higher redshifts the maximum in
  $M_{\ast,c}/M$ remains close to $\sim 0.03$, but shifts to higher halo mass.
  (iii) The  inferred time-scale for  the disruption of satellite  galaxies is
  about the same as the dynamical  friction time scale of their subhalos. (iv)
  The  stellar  mass  assembly  history  of  central  galaxies  is  completely
  decoupled  from   the  assembly  history   of  its  host  halo;   the  ratio
  $M_{\ast,c}/M$  initially increases rapidly  with time  until the  halo mass
  reaches $\sim 10^{12} \msunh$, at which point $M_{\ast,c}/M \sim 0.03$. Once
  $M  \gta 10^{12}  \msunh$,  there  is little  growth in  $M_{\ast,c}$,
  causing the ratio  $M_{\ast,c}/M$ to decline. In Milky-Way  sized halos more
  than half  of the central stellar mass  is assembled at $z\la  0.5$.  (v) In
  low mass  halos, the accretion  of satellite galaxies contributes  little to
  the formation of their central galaxies, indicating that most of their stars
  must have  formed {\it  in situ}.  In  massive halos  more than half  of the
  stellar mass of the  central galaxy has to be formed {\it  in situ}, and the
  accretion of satellites  can only become significant at  $z\la 2$.  (vi) The
  total mass in  halo stars is more  than twice that of the  central galaxy in
  massive halos, but  less than 10 percent of $M_{\ast,c}$  in Milky Way sized
  halos.   (vii)  The  2PCFs  of  galaxies on  small  scales  holds  important
  information regarding  the evolution of satellite galaxies,  and at high-$z$
  is predicted to be much steeper than at low-$z$, especially for more massive
  galaxies.   We discuss various  implications of  our findings  regarding the
  formation and evolution of galaxies in a $\Lambda$CDM cosmology.
\end{abstract}




\section{Introduction}
\label{sec:intro}

In  recent years,  much  effort has  been  made to  establish the  statistical
connection between  galaxies and dark  matter halos, as parameterized  via the
conditional luminosity function (CLF) (Yang, Mo  \& van den Bosch 2003) or the
halo  occupation  distribution (HOD)  (Jing  et  al.  1998; Peacock  \&  Smith
2000).  This  galaxy-dark  matter   connection  describes  how  galaxies  with
different  properties occupy  halos of  different mass,  and  yields important
insight into how  galaxies form and evolve in dark  matter halos. In practice,
the various  methods that have been  used to constrain  the galaxy-dark matter
connection (galaxy clustering, galaxy-galaxy lensing, galaxy group catalogues,
abundance  matching,  satellite  kinematics)   use  the  fact  that  the  halo
properties,  such as  mass function,  mass profile,  and clustering,  are well
understood in the current $\Lambda$CDM model of structure formation (e.g., Mo,
van den Bosch \& White 2010).

At low-redshift, large  redshift surveys, such as the  two-degree Field Galaxy
Redshift Survey (2dFGRS; Colless \etal  2001) and the Sloan Digital Sky Survey
(SDSS; York  \etal 2000), have  provided accurate estimates of  the luminosity
and stellar  mass functions  of galaxies (e.g.,  Norberg \etal  2002b; Blanton
\etal 2003; Li \& White 2009), of their two-point correlation functions (2PCF)
as function  of various galaxy  properties (e.g., Norberg \etal  2002a; Zehavi
\etal 2005, 2011; Wang \etal  2007), their satellite kinematics (e.g., van den
Bosch \etal  2004; More \etal  2009, 2011), and  even of their  excess surface
densities, a measure for the  tangential shear caused by gravitational lensing
due  to their  mass distributions  (e.g.,  Mandelbaum \etal  2005). All  these
results have  been used  to infer how  galaxies with different  properties are
distributed in  halos of different masses  (e.g.  Jing \etal  1998; Peacock \&
Smith 2000; Yang \etal 2003; van den Bosch \etal 2003a, 2007; Zheng \etal 2005;
Tinker \etal 2005;  Mandelbaum \etal 2006; Brown \etal  2008; More \etal 2009,
2011; Cacciato  \etal 2009;  Neistein et al.  2011a,b; Avila-Reese  \& Firmani
2011). In  addition, these large galaxy  redshift surveys can also  be used to
identify  galaxy systems  (groups), defined  as  those galaxies  that share  a
common dark matter  host halo (Yang et al.  2005a;  2007). Such group catalogs
can be  used to examine the halo  - galaxy connection even  more directly than
the methods  mentioned above  (e.g., Eke \etal  2004; Yang \etal  2005b, 2008,
2009b).   These  analyses  have  revealed  a number  of  important  properties
regarding  the relation  between galaxies  and  their dark  matter halos.  The
stellar  mass  to   halo  mass  ratio  has  a  minimum   for  halos  of  $\sim
10^{12}\msunh$,  and increases  rapidly  towards both  lower  and higher  halo
masses (Yang  et al.   2003, 2005b, 2008,  2009b; van  den Bosch et  al. 2003a;
Tinker \etal 2005; Leauthaud \etal 2012), suggesting that galaxy formation is
most efficient in halos with a present-day mass of $\sim 10^{12}\msunh$, i.e.,
these  halos have  the highest  integrated star  formation  efficiencies. Yet,
their  total stellar  masses  are only  a few  percent  of that  of the  halo,
indicating that the overall star formation efficiency is very low.

Ideally, we would like to carry  out similar analyses at various redshifts, so
as to  investigate how  star formation proceeds  and how galaxies  assemble as
their host halos grow within the  cosmic density field.  Such analyses are now
becoming  possible  at  intermediate   redshift,  $z\sim  1$,  where  reliable
luminosity  and stellar  mass functions  of galaxies  have been  obtained from
various redshift  surveys, such  as the DEEP2  survey (Davis \etal  2003), the
COMBO-17 survey  (Wolf \etal  2004), VVDS (Le  Fevre \etal 2005),  and zCOSMOS
(Lilly \etal 2007).  In addition, these surveys have been  used to measure the
2PCFs of galaxies  as function of their luminosity,  stellar mass and/or color
(e.g., Daddi \etal 2003; Coil \etal 2006; Phleps \etal 2006; Pollo \etal 2006;
McCracken \etal 2008; Meneux \etal 2008, 2009; Foucaud \etal 2010; de la Torre
\etal 2010). These observations have  prompted a series of investigations into
the galaxy-dark matter connection and its evolution between $z \sim 1$ and the
present (e.g.,  Bullock \etal  2002; Moustakas \&  Somerville 2002;  Yan \etal
2003; Zheng 2004; Lee \etal 2006; Hamana \etal 2006; Cooray 2005, 2006; Cooray
\& Ouchi  2006; Conroy \etal 2005,  2007; White \etal 2007;  Zheng \etal 2007;
Conroy \& Wechsler 2009; Wang \& Jing 2010; Wetzel \& White 2010; Wang \& Jing
2010; Leauthaud \etal 2011; Wake \etal 2011).

With the advent of deep, multi  wave-band surveys, it has even become possible
to estimate the luminosity/stellar mass functions of galaxies out to $z\sim 8$
(e.g.  Drory  \etal  2005;  Fontana  \etal 2006;  Perez-Gonzalez  \etal  2008;
Marchesini \etal  2009; Stark  \etal 2009; Bouwens  \etal 2011).  However, the
data  samples are still  small (and  hence subject  to cosmic  variance), with
large discrepancies among different  measurements (see Marchesini \etal 2009).
Furthermore, since reliable clustering measurements are in general unavailable
for these  high redshift galaxy samples, it  is not possible to  carry out the
same  HOD/CLF analyses  for these  high-$z$ galaxies  as for  galaxies  at low
$z$. Nevertheless, attempts  have been made to establish  the relation between
galaxies and their  dark matter halos out to high $z$  using a technique known
as abundance matching, in which galaxies of a given luminosity or stellar mass
are  linked to  dark  matter halos  of  given mass  by  matching the  observed
abundance of the galaxies in question  to the halo abundance obtained from the
halo mass function (typically also accounting for subhalos). This approach was
first used by Mo  \& Fukugita (1996) and Mo, Mao \&  White (1999) to model the
number density and clustering of Lyman-break galaxies.  More recently, several
studies  have used  abundance  matching techniques  to  probe the  galaxy-dark
matter  connection out  to $z  \sim  5$ (e.g.,  Vale \&  Ostriker 2004,  2006;
Conroy,  Wechsler \&  Kravtsov 2006;  Shankar \etal  2006; Conroy  \& Wechsler
2009; Moster \etal 2010; Guo \etal 2010; Behroozi, Conroy \& Weschler 2010).

An  important aspect of  abundance matching  is the  treatment of  dark matter
subhalos.  It is  usually assumed that a central galaxy  resides at the center
of each  halo, and orbiting around  it are satellite  galaxies associated with
the subhalos of  the host halo.  These subhalos  were distinctive (host) halos
themselves  before  they were  accreted  into  their  hosts, namely  satellite
galaxies  were themselves  central galaxies  before their  host halo  became a
subhalo.   The  modeling of  the  total  galaxy  population can  therefore  be
separated into two parts: (i) the formation of central galaxies in dark matter
halos  at  different  redshifts,  and  (ii) the  accretion  and  evolution  of
satellite  galaxies in  their  host  halos. With  the  use of  high-resolution
numerical simulations  and extended  Press-Schechter theory (Bond  \etal 1991;
Lacey \&  Cole 1993), the properties  of the subhalo population  have now been
determined with  great accuracy  (e.g.  Gao \etal  2004; De Lucia  \etal 2004;
Tormen \etal 2004; van den Bosch \etal 2005a; Weller \etal 2005; Diemand \etal
2007; Giocoli  \etal 2008; Li \&  Mo 2009).  If satellite  galaxies are indeed
associated with subhalos, these properties  should be relevant to the modeling
of the galaxy population. In particular, since subhalos were host halos before
accretion, the subhalo  property that seems most relevant is  its mass {\it at
  the time  of accretion}.  Consequently,  many abundance matching  studies to
date  have linked  galaxies of  property $\calP$  (i.e., galaxy  luminosity or
stellar mass) to dark matter (sub)-halos of mass $M$ using %
\begin{eqnarray}\label{N_matching}
\int_{\calP}^{\infty} n_\rmg(\calP,z) \,\rmd\calP & = &
\int_M^{\infty} n_\rmh(M,z) \, \rmd M \nonumber\\
& & +\int_M^{\infty} n_{\rm sub}(m_a,z) \, \rmd m_a\,.
\end{eqnarray}
Here $n_\rmg(\calP,z)$ is the comoving  number density of galaxies of property
$\calP$ at redshift  $z$, $n_\rmh(M,z)$ is the halo  mass function at redshift
$z$,  and $n_{\rm  sub}(m_a,z)$ is  the comoving  number density  of subhaloes
identified at  redshift $z$ which  at accretion have  a mass $m_a$.  Hence, in
such  abundance matching  the property  of a  satellite galaxy  is  assumed to
depend only on its halo mass at accretion.

Although subhalo  abundance matching yields galaxy  correlation functions that
are in remarkably  good agreement with observations (e.g.,  Conroy \etal 2006;
Guo  \etal 2010; Wang  \& Jing  2010), it  implies a  particular path  for the
evolution of satellite  galaxies. Indeed, assuming that the  stellar masses of
satellite galaxies depend only on  their halo mass at accretion implies either
that the relation between $\calP$ and halo mass $M$ is independent of the time
when the subhalo is accreted, and the evolution after accretion is independent
of the host halo into which the  subhalo has been accreted and how long ago it
has  been accreted,  or  that the  effects  of different  accretion times  and
subsequent evolutions in different hosts  conspire to give a stellar mass that
depends  only on  the mass  of the  subhalo at  accretion.  Such  evolution of
satellites is  neither physically motivated  nor the only possible  path which
satellite  galaxies could  take.   In fact,  applying  the abundance  matching
technique as described above to data  at different $z$ results in a $\calP(M)$
relation  that changes  with  redshift  (e.g., Conroy  \etal  2006; Conroy  \&
Wechsler 2009;  Moster \etal 2010;  Wang \& Jing  2010).  If one  accepts that
$\calP$ for central galaxies depends on  both $M$ and $z$, then the properties
of satellite galaxies  do not just depend on the host  halo mass at accretion,
$m_a$, but also on the accretion redshift, $z_a$. Furthermore, after accretion
a satellite galaxy may lose or gain  stellar mass and even be disrupted due to
tidal stripping  and disruption, such  that $\calP$ of satellite  galaxies may
depend on other properties in addition  to both $m_a$ and $z_a$.  So far these
effects have not been modeled in any detail in abundance-matching studies.

In  this  paper, we  develop  a new,  self-consistent  approach  to model  the
relation between galaxies and dark matter halos over cosmic time. Our approach
is based on the model of Yang \etal (2011), which gives subhalo abundance as a
function of both mass at accretion  and accretion time.  This allows us to use
a galaxy-halo  relation that depends on  subhalo accretion time  and to follow
the dynamical evolution after  their accretion.  Furthermore, since satellites
observed in halos (galaxy systems) at the present time are `fossil records' of
both  the  formation  of  central  galaxies at  high-$z$  and  the  subsequent
dynamical  evolution,  the  observed  abundance and  clustering  of  satellite
galaxies  at the  present time  can be  used to  constrain the  galaxy  - halo
relations  at different  redshifts.  In  the present  work, we  use  our model
together with  various observational  data to obtain  the stellar mass  - halo
mass relation as a function of redshift  from $z \sim 4$ to the present.  This
paper is organized as  follows.  In Section~\ref{sec:halomodel} we outline the
halo model ingredients to be used.  In Section~\ref{sec:model} we describe how
we model the stellar mass function, the conditional stellar mass function, and
the two point correlation  function of galaxies. The observational constraints
used  for our  analysis  are presented  in Section~\ref{sec:observation}.  The
results are presented in Section~\ref{sec:res}, while Section~\ref{sec:evolve}
describes  the  implications  regarding  the  assembly  of  galaxies  and  the
evolution of  the galaxy dark matter  connection. We summarize  our results in
Section~\ref{sec:summary}, which also includes a detailed discussion regarding
various implications for galaxy formation and evolution.

Throughout this  paper we  will mostly focus  on a  $\Lambda$CDM `concordance'
cosmology whose  parameters are consistent with the  seventh-year data release
of the WMAP  mission (Komatsu \etal 2011; hereafter  WMAP7). However, in order
to  investigate the  possible cosmology  dependence  of our  results, we  also
present results for the following set of $\Lambda$CDM cosmologies: WMAP1, with
parameters consistent  with the  first-year data release  of the  WMAP mission
(Spergel \etal  2003); WMAP3, with  parameters given in Spergel  \etal (2007);
WMAP5, with parameters given in  Dunkley \etal (2009), and finally a cosmology
(`Millenium') with parameters that are  identical to those adopted in Springel
\etal (2005)  for the Millennium  Simulation.  Table~\ref{tab:cosmology} lists
the parameters for all these cosmologies.
\begin{deluxetable}{lccccl}
  \tabletypesize{\scriptsize} \tablecaption{$\Lambda$CDM
    cosmological models used in this paper.} \tablewidth{0pt}
  \tablehead{Name & $\Omega_\rmm$ & $\Omega_{\Lambda}$ & $n_s$ & $h$ & $\sigma_8$
  } \startdata
WMAP1 & 0.30 & 0.70 & 1.00 & 0.70 & 0.90  \\
WMAP3 & 0.238 & 0.762 & 0.951 & 0.73 & 0.75 \\ 
WMAP5 & 0.258 & 0.742 & 0.963 & 0.719 & 0.796 \\
WMAP7 & 0.275 & 0.725 & 0.968 & 0.702 & 0.816 \\
Millennium & 0.25 & 0.75 & 1.00 & 0.73 & 0.90
\enddata
\label{tab:cosmology}
\end{deluxetable}

\section{Dark matter halos and subhalos}
\label{sec:halomodel}

The main goal of this paper is  to obtain a self-consistent model for the link
between galaxies and  dark matter halos accross cosmic  time.  In this section
we describe our  model ingredients for halos and subhalos  that are needed for
our investigation.

\subsection {Dark Matter Halo Mass Function}
\label{sec:halomf}

\begin{figure*}
\plotone{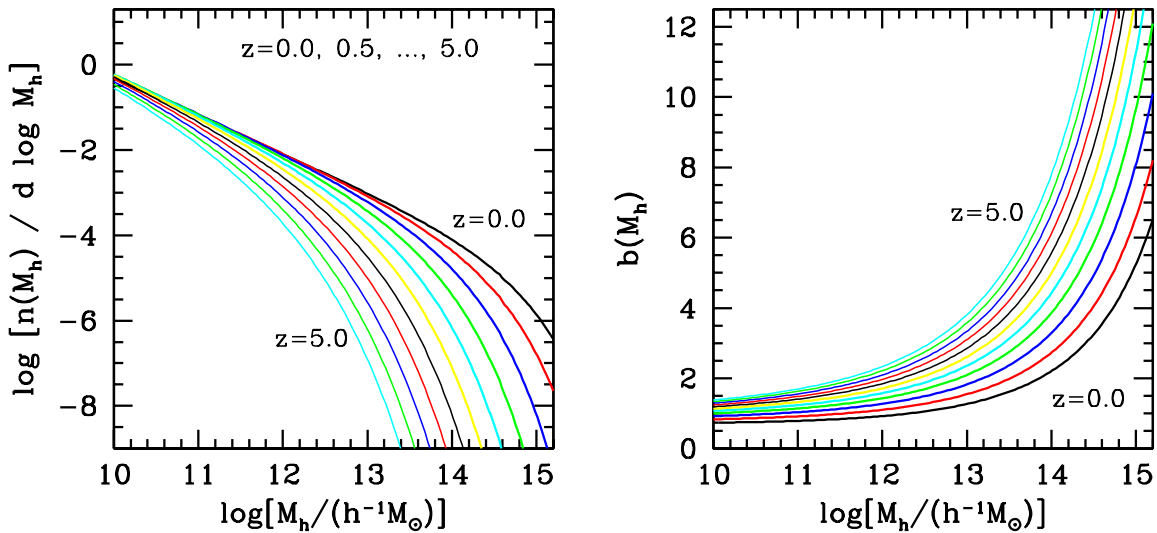}
\caption{Left panel: the  halo mass functions predicted using  the Sheth Mo \&
  Tormen  (2001) model  for the  WMAP7  cosmology at  different redshifts,  as
  indicated.   Right panel:  the bias  parameter of  dark matter  halos, again
  predicted using the Sheth, Mo \& Tormen (2001) model for the WMAP7 cosmology
  at the same redshifts as those indicated in the left panel.}
\label{fig:haloMF}
\end{figure*}

The mass function  of dark matter halos, $n_\rmh(M,z)\,\rmd  M$, describes the
number  density  of  dark matter  halos  of  mass  $M$  at redshift  $z$.  The
Press-Schechter  formalism  (Press \&  Schechter  1974)  yields an  analytical
estimate for $n_\rmh(M,z)$, and we use  the revised form given in Sheth, Mo \&
Tormen (2001) based on the ellipsoidal collapse model: 
\begin{equation}\label{halomf} 
n_\rmh(M,z) \, \rmd M = {{\overline \rho} \over M^2} \nu f(\nu) \, 
\left| {\rmd {\rm ln} \sigma \over \rmd{\rm ln} M}\right| \, \rmd M\,,
\end{equation}
where $\bar{\rho}$ is the mean matter density of the Universe at $z$,
$\sigma(M)$ is the mass variance, and $\nu = \delta_\rmc(z)/\sigma(M)$ with
$\delta_\rmc(z)$ the critical over-density required for collapse at redshift
$z$.  The function $f(\nu)$ is given by
\begin{equation}\label{fnuST} 
\nu \, f(\nu) = 2 A \,\left(1 + {1\over
\nu'{^{2q}}}\right)\ \left({\nu'{^2}\over 2\pi}\right)^{1/2}
\exp\left(-{\nu'{^2} \over 2}\right)\,
\end{equation}
with  $\nu'=\sqrt{a}\,\nu$,  $a=0.707$,  $q=0.3$  and $A\approx  0.322$.   The
resulting  mass function  has been  shown to  be in  excellent  agreement with
numerical simulations, as long as halo masses are defined as the masses inside
a sphere  with an average over-density  of about $180$ that  of the background
mass density (Sheth \& Tormen  1999; Jenkins \etal 2001).  As an illustration,
the left-hand panel of Fig.~\ref{fig:haloMF}  shows the halo mass functions at
several redshifts for the WMAP7 cosmology, predicted using Eq.~(\ref{halomf}).

\subsection {Halo Density Profile}
\label{sec:haloprof}

Throughout  this paper  we assume  that dark  matter halos  are  spherical and
follow an NFW density profile (e.g., Navarro, Frenk \& White 1997): 
\begin{equation}
\label{NFW} 
\rho(r) = \frac{{\overline \delta} \, {\overline \rho}}
{(r/r_\rms)(1+r/r_\rms)^{2}},
\end{equation}
where  $r_\rms$   is  a  characteristic   radius,  and  $\bar{\delta}$   is  a
dimensionless  amplitude  which  can  be   expressed  in  terms  of  the  halo
concentration parameter $c \equiv r_{\rm 180}/r_\rms$ as
\begin{equation}
\label{overdensity} \bar{\delta} = {180 \over 3} \, {c^{3} \over
{\rm ln}(1+c) - c/(1+c)}\,.
\end{equation}
Numerical simulations have shown that  $c$ is correlated with halo mass (e.g.,
NFW; Eke \etal  2001; Jing \& Suto 2000; Bullock \etal  2001; Zhao \etal 2003;
2009;   Macci\`o   \etal   2007).   Throughout   this   paper   we   use   the
concentration-mass relation  of Zhao \etal (2009), properly  corrected for our
definition of halo mass.

\subsection {Halo Bias}
\label{sec:halobias}

Dark matter halos are biased tracers of the dark matter mass distribution. The
amplitude of this bias depends on halo mass and is expressed via the halo bias
function, $b_\rmh(M)$ (e.g., Mo \& White 1996, 2002). This allows one to write
the two point correlation function between  halos of masses $M_1$ and $M_2$ on
large scales as 
\begin{equation} \label{xihalo} 
\xi_{\rm hh}(r|M_1,M_2) = b_\rmh(M_1) \, b_\rmh(M_2) \, \xi_{\rm dm}(r)\,,
\end{equation}
where  $\xi_{\rm  dm}(r)$  is   the  two-point  correlation  function  of  the
(non-linear) dark matter mass distribution. Throughout this paper we adopt the
halo bias function of (Sheth, Mo \& Tormen 2001), which is given by
\begin{eqnarray}\label{bm} 
b_\rmh(M) & = & 1 + {1\over\sqrt{a}\delta_{\rm c}} \
\Bigl[ \sqrt{a}\,(a\nu^2) + \sqrt{a}\,b\,(a\nu^2)^{1-c} - \nonumber \\
& & {(a\nu^2)^c\over (a\nu^2)^c + b\,(1-c)(1-c/2)}\Bigr],
\end{eqnarray}
with $a=0.707$, $b=0.5$, $c=0.6$ and $\nu = \delta_{\rm c}/
\sigma(M)$.  Note that our halo mass function~(\ref{fnuST}) and halo
bias function (\ref{bm}) obey the normalization condition
\begin{equation}
\int b_\rmh(\nu) \, f(\nu) \, \rmd\nu = 1
\end{equation}
which  expresses that  the  distribution  of dark  matter  is, by  definition,
unbiased with respect  to itself. As an illustration,  the right-hand panel of
Fig.~\ref{fig:haloMF}  shows  the halo  bias  factor,  $b_\rmh(M)$ at  several
redshifts for the WMAP7 cosmology, predicted using Eq.~(\ref{bm}).
 
\subsection {The Subhalo Population}
\label{ssec:subhalo}

An important ingredient of our  model describing the link between galaxies and
dark  matter halos is  a statistical  description for  the population  of dark
matter subhalos. In particular, we  need to know the distribution of accretion
masses, $m_a$, and accretion redshifts, $z_a$, for the population of subhalos,
as a function  of the mass of its host  halo.  Let $n_{\rm sub}(m_a,z_a|M,z)\,
\rmd m_a\,  \rmd\ln(1+z_a)$ denote the  number of subhalos  in a host  halo of
mass $M$ at redshift $z$, as  a function of their accretion masses, $m_a$, and
accretion redshifts,  $z_a$.  We  will refer to  $n_{\rm sub}$ as  the subhalo
mass function  (SHMF). In  the past, the  subhalo mass function  has typically
been  studied using $N$-body  simulations or  Monte-Carlo realizations  of the
extended Press-Schechter formalism (e.g.,  Sheth \& Lemson 1999; Somerville \&
Kolatt 1999; Cole  \etal 2000; van den Bosch \etal 2005a; Giocoli \etal 2008a;
Cole \etal  2008; Parkinson  \etal 2008; Fakhouri  \& Ma 2008;  Fakhouri \etal
2010). Recently, however,  Yang \etal (2011; hereafter Y11)  presented a fully
analytical model for $n_{\rm sub}(m_a,z_a|M,z)$ which is the one we will adopt
here. Following their notation we have that 
\begin{equation}\label{eq:shmf0}
n_{{\rm sub},0}(m_a,z_a|M,z) = 
m_a^{-1} \, {\cal N}_a(s_a, \delta_a\vert S,\delta) \,,
\end{equation}
where $s_a = \sigma^2(m_a)$, $S = \sigma^2(M)$, $\delta_a = \delta_\rmc(z_a)$,
$\delta =  \delta_\rmc(z)$, and the subscript  `0' on the  SHMF indicates that
this function  ignores higher order  subhalos (subhalos of subhalos,  etc; see
below).   Y11  considered   three  different   models  for   ${\cal  N}_a(s_a,
\delta_a\vert  S,\delta)$, and  we adopt  their Model  III which  is  the most
accurate. For  host halos  of given $(M,  z)$, this  model uses the  mean halo
assembly  history given  by the  fitting formula  of Zhao  \etal (2009)  and a
log-normal model to describe the  scatter among different halos (see Section~3
of Y11 for details). Note that  this SHMF only includes subhalos that are {\it
  directly}  accreted  onto  the main  branch  of  the  host halo  (hence  the
subscript `0').   However, since the  SHMF given by  Eq.~(\ref{eq:shmf0}) also
applies to  subhalos before  their accretion (when  they themselves  were host
halos),  we  can  in  principle  calculate the  SHMF  including  higher  level
sub-halos, i.e.,  sub-halos within  sub-halos, etc. For  example, the  SHMF of
sub$^i$-subhalos ($i$th level subhalos) can be written as
\begin{eqnarray}\label{eq:subsubi} 
n_{{\rm sub},i}(m_a, z_a\vert M, z) =\int\int n_{{\rm sub},i-1}(m_a, z_a\vert
m_{i-1}, z_{i-1}) &&\nonumber \\ 
\times n_{{\rm sub},0}(m_{i-1}, z_{i-1}\vert M, z) ~{\rm d} m_{i-1}\,{\rm d}\ln
(1+z_{i-1})\,.  ~~~~~~&& 
\end{eqnarray}
Thus the total subhalo population can be described using 
\begin{equation}\label{totshmf}
n_{\rm sub}(m_a, z_a|M, z) = \sum_{i=0}^{N_{\rm max}} 
n_{{\rm sub},i}(m_a, z_a|M, z)\,.
\end{equation}
In practice, we find that the  sub$^i$-subhalos for $i \geq 2$ contribute less
that 5\%  to $n_{\rm sub}(m_a, z_a|M, z)$  in the mass range  of interest here
($\ga 10^{10}\msunh$), and we therefore  adopt $N_{\rm max}=1$ in what follows
(see also Yang, Mo \& van den Bosch 2009a).

\section {Modeling Mass Functions and Correlation Functions}
\label{sec:model}

\subsection{The Stellar Mass Functions}

In the conditional luminosity (stellar mass) function model developed
by Yang \etal (2003), the stellar mass function of galaxies at a given
redshift $z$ can be written as
\begin{equation}
\label{eq:SMF}
\Phi(M_{\ast},z) = \int_0^{\infty} \Phi(M_{\ast}|M,z) \,
n_{\rm h}(M,z) \, {\rm d}M\,,
\end{equation}
where $\Phi(M_{\ast}|M,z)$ is the conditional stellar mass function
(CSMF), which gives the mean number of galaxies with stellar masses in
the range $M_{\ast}\pm {\rm d}M_{\ast}/2$ hosted by halos of mass $M$
at $z$. In general, we split the CSMF in two parts,
\begin{equation}\label{eq:CLF_fit}
\Phi(M_{\ast}|M,z) = \Phi_\rmc(M_{\ast}|M,z) + \Phi_\rms(m_{\ast}|M,z)\,.
\end{equation}
where $\Phi_\rmc(M_{\ast}|M,z)$ and $\Phi_\rms(m_{\ast}|M,z)$ are the
contributions from the central and satellite galaxies, respectively.  Once
$\Phi(M_{\ast}|M,z)$ is known, the average halo occupation number (HON) of
galaxies within a given stellar mass range, $M_{\ast,1} < M_{\ast} <
M_{\ast,2}$ can be written as
\begin{equation}
\label{meanNM}
\langle N|M,z \rangle = \int_{M_{\ast,1}}^{M_{\ast,2}} 
\Phi(M_{\ast} \vert M,z) \,\rmd M_{\ast} \,.
\end{equation}
Using Eq.\,(\ref{eq:CLF_fit}) we write
\begin{equation}
\langle N|M,z \rangle = \langle N_\rmc |M,z\rangle + 
\langle N_\rms |M,z\rangle \,,
\end{equation}
where $N_\rmc$ and $N_\rms$ are the occupation numbers of central and
satellite galaxies, respectively.

\subsubsection{The Conditional Stellar Mass Function of Central Galaxies}

Based on the results of Yang \etal (2009b, hereafter Y09b), we assume
that the CSMF of central galaxies is given by a lognormal distribution:
\begin{equation}\label{eq:phi_c}
\Phi_\rmc(M_{\ast}|M,z) = {1\over {\sqrt{2\pi}\sigma_c}} {\rm exp}
\left[- { {(\log M_{\ast}/M_{\ast, c} )^2 } \over 2\sigma_c^2} \right]\,,
\end{equation}
where $\log M_{\ast, c}$ is  the expectation value of the (10-based) logarithm
of  the   stellar  mass   of  the  central   galaxy  and  $\sigma_c$   is  the
dispersion. For simplicity, throughout the paper, we assume $\sigma_c$ to be 
independent of halo mass.

Numerous studies have attempted to constrain the relation between $M_{\ast,
  c}$ and $M$ and its evolution with redshift (see Section~\ref{sec:intro}).
Following Y09b, we assume that this relation has the form of a broken
power-law:
\begin{equation}\label{eq:Lc_fit}
M_{\ast, c} = M_{\ast, 0} \frac { (M/M_1)^{\alpha +\beta} }{(1+M/M_1)^\beta } \,,
\end{equation}
so that $M_{\ast, c} \propto M^{\alpha+\beta}$ ($M_{\ast, c} \propto
M^{\alpha}$) for $M \ll M_1$ ($M \gg M_1$).  This model contains four free
parameters: an amplitude $M_{\ast, 0}$, a characteristic halo mass, $M_1$, and
two power-law slopes, $\alpha$ and $\beta$. Note that all four parameters may
depend on redshift, which we will parameterize accordingly in
Section~\ref{sec:fitting} below.

\subsubsection{The Conditional Stellar Mass Function of Satellite Galaxies}
\label{sec:CSMF_sat}

The conditional stellar mass function for satellite galaxies can formally be
written as,
\begin{eqnarray}\label{phismodel}
\Phi_{\rm s}(m_{\ast}|M,z)  ~~~~~~~~~~~~~~~~~ ~~~~~~~~~~~~~~~~~~~~~~~~~~~~~~~~~~\nonumber\\
=\int\limits_0^{\infty} \rmd \log m_{\ast}'  \int\limits_0^{\infty} \rmd m_a
\int\limits_{z}^{\infty} {\rmd z_a \over 1+z_a} \int\limits_0^{M} \rmd M_a
\int\limits_0^1 \rmd\eta ~\nonumber\\
P(m_{\ast}| m_{\ast}',  m_a; z_a; M_a; \eta, z)~~~~~~~~~~~~~~~~~~~~~~~~~\nonumber \\
\Phi_\rme(m_{\ast}'|m_a,z_a, z) \, n_{\rm sub}(m_a,z_a|M,z) ~~~~~~~~~~~~~~\nonumber\\
P(M_a,z_a|M,z) \, P(\eta) \,. ~~~~~~~~~~~~~~~~~~~~~~~~~~~~~~~
\end{eqnarray}
Here    $n_{\rm    sub}(m_a,z_a|M,z)$     is    the    SHMF    described    in
Section~\ref{ssec:subhalo}; $\Phi_\rme (m_{\ast}'|m_a,z_a,  z)$ is the CSMF of
galaxies  that were  accreted at  redshift  $z_a$, taking  into account  their
subsequent  mass  evolution  to the  redshift  $z$  in  question due  to  star
formation and  stellar evolution; $P(m_{\ast}|m_{\ast}', z_a;  M_a; m_a; \eta,
z)$ is  the probability that  a satellite galaxy  whose stellar mass  would be
$m_{\ast}'$ in the  absence of stripping or disruption ends  up with a stellar
mass $m_{\ast}$ at redshift $z$ due to such environmwental effects.  This last
probability is written in a form  that depends explicitly on the masses of the
subhalo and host  halo at the accretion epoch  ($m_a$ and $M_a$, respectively)
as well  as on the orbital circularity  of the subhalo, $\eta$  (to be defined
below).   Note  that  the  integrand includes  the  probability  distributions
$P(\eta)$ and $P(M_a,z_a|M,z)$.  The latter describes the probability that the
main progenitor  of a halo  of mass $M$  at redshift $z$  has a mass  $M_a$ at
redshift  $z_a \geq  z$; hence  $P(M_a,z_a|M,z)$ describes  the  mass assembly
history of  the host halo. Note  also that the integration  over $z_a$ implies
that the  CSMF of satellite  galaxies at redshift  $z$ depends on the  CSMF of
centrals at all redshifts $z_a \geq  z$. The orbital circularity is defined as
$\eta \equiv  j/j_\rmc(E)$ ($0  \leq \eta  \leq 1$). Here  $j$ is  the orbital
specific angular momentum and $j_\rmc(E)$  is the specific angular momentum of
a  circular orbit  that has  the same  orbital energy,  $E$, as  the  orbit in
question. Numerical simulations have show that the probability distribution of
$\eta$ of dark matter subhaloes is well approximated by
\begin{equation}\label{eq:P_eta}
P(\eta) \propto \eta^{1.2} (1-\eta)^{1.2}\,,
\end{equation}
independent of redshift or halo mass (e.g., Zentner \etal 2005).

An important quantity that must somehow enter in a description of $P(m_{\ast}|
m_{\ast}', m_a; z_a; M_a; \eta, z)$ is the dynamical friction time
scale, $t_{\rm df}$, defined as the time interval between the accretion of a
subhalo and the epoch at which it is either tidally disrupted or cannibalized
by the central galaxy.  Using $N$-body simulations, Boylan-Kolchin \etal
(2008) have shown that
\begin{equation}\label{DF_time}
{t_{\rm df} \over \tau_{\rm dyn}} = 0.216
{(M_a/m_a)^{1.3} \over \ln (1+M_a/m_a)} \,
\left[{r_\rmc(E)\over r_{\rm vir}(z_a)}\right]\, {\rm e}^{1.9 \eta}\,,
\end{equation}
which is the  functional form we adopt throughout.  Here $r_{\rm vir}(z_a)$ is
the virial radius  of the host halo at $z_a$, and  $\tau_{\rm dyn} \approx 0.1
H^{-1}(z_a)$ is the dynamical time of the halo at $z_a$.  As an approximation,
we assume
\begin{equation}
r_c(E)= r_{\rm vir}(z_a)\,,
\end{equation}
so that $t_{\rm df} = t_{\rm df}(m_a/M_a,z_a,\eta)$, and can thus be
evaluated for the integrand in Eq.~(\ref{phismodel}).

In this paper we assume that a galaxy after becoming a satellite can gain
stellar mass due to star formation and suffer mass loss due to passive
evolution. As a simple model, we assume the overall evolutionary effect can be
modeled as a function of the masses $m_{\ast, z}$ and $m_{\ast, a}$, where
$m_{\ast, z}$ is the expected median stellar mass of central galaxies in halos
of mass $m_a$ at redshift $z$.  Specifically, we write the evolution of the
median stellar mass of a satellite galaxy as
\begin{equation}
{\overline m}_{\ast}' = (1-c) m_{\ast, a} + c m_{\ast, z}\,,
\end{equation}
where $c$ is a parameter which may depend on $z$.  Thus, if $c=0$, then
${\overline m}_{\ast}' = m_{\ast, a}$ so that the stellar mass of a satellite is
equal to its original mass at accretion.  On the other hand if $c=1$, then
${\overline m}_{\ast}' = m_{\ast, z}$ so that the stellar mass of a satellite is
the same as that of central galaxy of the same halo mass at the redshift, $z$,
in question. The $c=1$ case corresponds to the assumption adopted in the
conventional abundance matching represented by equation (\ref{N_matching}).
By treating $c$ as a free parameter, we can examine how the evolution may
deviate from the simplified assumptions. Given this assumption, we write
\begin{equation}\label{eq:phi_c}
\Phi_\rme(m_{\ast}'|m_a,z_a,z) = {1\over {\sqrt{2\pi}\sigma_c'}} {\rm exp}
\left[- { {(\log m_{\ast}'/{\overline m}_{\ast}' )^2 } \over 2\sigma_c'^2} \right]\,.
\end{equation}
As a simple model, we assume $\sigma_c'$ to be the same as 
$\sigma_c$ that specifies the CSMF of central galaxies (see Section 3.3). 
Thus,  in the special case of $c=0$,  $\Phi_\rme(m_{\ast}'|m_a,z_a,z)$
has the same form as the CSMF of central galaxies at the accretion 
redshift $z_a$. In general,  $\Phi_\rme$ is a lognormal distribution 
with a median  ${\overline m}_{\ast}'$ and a dispersion given by $\sigma_c$.
 
For the stripping and disruption of the satellite galaxies, we adopt a simple
form for $P(m_{\ast} | m_{\ast, \rm evo}, z_a; m_a; M_a; \eta, z)$, assuming
that significant tidal stripping occurs only for the dark matter subhalo, but
not for the stellar component that represents the satellite galaxy, at least
not until it is either destroyed by the tidal forces of, or cannibalized by, the
central galaxy. In this case, we have that
\begin{eqnarray}\label{dynevolmod}
P(m_{\ast}| m_{\ast}', z_a; m_a; M_a; \eta, z) =
~~~~~~~~~~~~~~~~~~~~~~~~~&& \nonumber \\ \left\{ \begin{array}{ll}
  \delta^{\rm D}(m_{\ast}-m_{\ast}')~~~~ & \mbox{if $\Delta t < p_t~ t_{\rm df} $} \\ 0 &
  \mbox{otherwise}\,.
\end{array}\right.~~~~&&
\end{eqnarray}
Here
\begin{equation}
\Delta t\equiv t(z) - t (z_a)\,,
\end{equation}
is  the   time  interval  between  $z$   and  $z_a$  (i.e.,   the  time  since
accretion). Thus  we have two  free parameters $c$  and $p_t$ that we  seek to
constrain  using the observed  satellite population.   The parameter  $p_t$ is
intended  to account  for  the tidal  stripping  effect. The  extreme case  of
$p_t=\infty$  corresponds  to   no  tidal  stripping/disruption  of  satellite
galaxies, while the case of $p_t=0$ corresponds to instantaneous disruption of
all    the    satellite    galaxies.     Under    the    above    assumptions,
Eq.~(\ref{phismodel}) reduces to
\begin{eqnarray} \label{phismodeleasy}
\Phi_{\rm s}(m_{\ast}|M,z) =
\int\limits_0^{M} \rmd m_a
\int\limits_{z}^{\infty} {\rmd z_a \over 1+z_a} \int\limits_0^{M} \rmd M_a
\int\limits_0^1 \rmd\eta ~~~~~~~~~~&&\nonumber\\
\Phi_\rme(m_{\ast}|m_a,z_a,z)  \, n_{\rm sub}(m_a,z_a|M,z) ~~~~~~~~~~~~~~~ && \nonumber\\
P(M_a,z_a|M,z) \, P(\eta) \, \Theta(p_t \,t_{\rm df} - \Delta t)\,, ~~~~~~~~~~~~~&&
\end{eqnarray}
with $\Theta(x)$ the Heaviside step function, which is straightforward to
compute numerically.

For the $i$th-order sub-subhalos (note that we only consider $i=1$), we assume
that the  dynamical friction time  is also given by  equation (\ref{DF_time}),
but with  $M_a$ referring  to the mass  of the  main branch of  the $(m_{i-1},
z_{i-1})$ halo at the accretion  time $z_a=z_i$.  This dynamical friction time
has to be compared with $\Delta t\equiv t(z_{i-1}) - t (z_i)$ to decide if the
$(m_i,  z_i)$  halo and  its  central galaxy  were  disrupted  already in  the
$(i-1)$th-order subhalo. Their subsequent  disruption then follows that of the
$(i-1)$th-order (here 0th-order) subhalos.

\subsection{The Two-Point Correlation Function of Galaxies}
\label{sec:2pcf}

In modeling the two-point correlation function (2PCF) of galaxies we
follow the procedures outlined in Wang \etal (2004).  In particular,
we write the 2PCF as
\begin{equation}
\xi(r,z) = \xi_{\rm 1h}(r,z) + \xi_{\rm 2h}(r,z)\,,
\label{2pcf}
\end{equation}
where $\xi_{\rm 1h}$ represents the correlation due to pairs of galaxies
within the same halo (the ``1-halo'' term), and $\xi_{\rm 2h}$ describes the
correlation due to galaxies that occupy different halos (the ``2-halo'' term).

In our  model, we assume  that the radial  number density distribution  of the
satellite galaxies, $n_{\rm s}(r|M,z)$ follows  the same NFW profile as a dark
matter halo  of mass  $M$ at redshift  $z$ out  to the virial  radius, $r_{\rm
  vir}$,  defined  as  the radius  inside  of  which  the average  density  is
$\Delta_{\rm vir}$ times the critical density. To good approximation 
\begin{equation}\label{Deltavir}
\Delta_{\rm vir} = 18\pi^2 + 60 x - 32 x^2
\end{equation}
with $x  = \Omega_\rmm(z) -1$ (Bryan  \& Norman, 1998). Note  that in reality,
the  distribution  of   satellite  galaxies  may  not  follow   an  exact  NFW
profile.  However, as tested  in Yang  et al.  (2004) by  populating satellite
galaxies according  to the real mass  distributions in dark  matter halos, the
change produced in the 2PCFs is quite small, less $20\%$ on small scale in
real space. For the {\it projected} 2PCFs which will be used in this work, the
impact is  even smaller,  less tha $10\%$.   Furthermore, our  {\it projected}
2PCFs are to be estimated on  scales larger than $0.1\mpch$, where the results
are  not  sensitive   to  the  assumption  on  the   concentration  of  galaxy
distribution in dark matter halos (see  Yang et al. 2005).  Thus the foible of
using  NFW profile is  negligible in  comparison to  the uncertainties  in the
observational data  and in  the halo model  itself.  For convenience,  in what
follows   we   use  the   {\it   normalized}   number  density   distribution,
$u_\rms(r|M,z) = n_\rms(r|M,z)/ \langle N_\rms|M,z \rangle$, so that 
\begin{equation}
4 \pi \int_{0}^{r_{\rm vir}} u_\rms(r|M,z) \, r^2 \, \rmd r = 1\,,
\end{equation}
where $r_{\rm vir} = r_{\rm vir}(M,z)$ is the virial radius of the
halo of mass $M$ at redshift $z$.  The 1-halo term of the galaxy
correlation function can then be written as
\begin{eqnarray}\label{xi1h}
\xi_{\rm 1h}(r,z) & = & {2 \over {\overline n}^2_\rmg(z)}
\int_0^{\infty} n_\rmh(M,z) \, \langle N_{\rm pair}|M,z\rangle \nonumber \\
& & f(r|M,z) \, {\rm d}M \,,
\end{eqnarray}
where $\langle N_{\rm pair}|M,z\rangle$ is the mean number of
distinctive galaxy pairs in a halo of mass $M$ at redshift $z$,
$f(r|M,z) 4\pi r^2 \Delta r$ is the fraction of those pairs that have
separations in the range $r\pm \Delta r/2$, and $\overline{n}_{\rm g}(z)$
is the mean number density of galaxies at redshift $z$, given by
\begin{equation}\label{barng} 
{\overline n}_\rmg(z) = \int_{0}^{\infty} n_\rmh(M,z) \, \langle N|M,z\rangle 
\, \rmd M\,.
\end{equation}
For a given $(M,z)$, the mean number of pairs as function of
separation, $\langle N_{\rm pair}\rangle f(r)$, can be divided into
contributions from central-satellite pairs and satellite-satellite
pairs:
\begin{equation}
\label{eq:Npairfr}
\langle N_{\rm pair}\rangle f(r) =
\langle N_{\rm cs}\rangle u_{\rm s}(r) +
\langle N_{\rm ss}\rangle f_{\rm s}(r)\,,
\end{equation}
where $f_{\rm s}(r)$ is the satellite galaxy pair distribution
function within a dark matter halo of mass $M$ at redshift $z$:
\begin{equation}
\label{eq:fs}
f_{\rm s}(r) = 2\pi \int_0^{r_{\rm vir}} u_{\rm s}(s)\, s^2ds \int^\pi_0
u_{\rm s}(|{\bf s}+{\bf r}|) \sin \theta\, d\theta\,,
\end{equation}
with $|{\bf s}+{\bf r}|=(s^2+r^2+2sr\cos{\theta})^{1/2}$.  The number
of central-satellite pairs is
\begin{equation}
\langle N_{\rm cs}|M,z \rangle = \langle N_\rmc|M,z\rangle \,  
\langle N_\rms|M,z\rangle \,.
\end{equation}
For the number of satellite-satellite pairs we assume that $N_\rms$
follows Poisson statistics so that
\begin{equation}
\langle N_{\rm ss}|M,z \rangle = \langle N_\rms|M,z\rangle^2  \,.
\end{equation}

The 2-halo term of the 2PCF for galaxies can be written as,
\begin{equation}
\xi_{\rm 2h}(r,z) = \left[f_{\rm exc}(r,z)\right]^2 \, 
\xi_{\rm dm}(r,z)\,,
\end{equation}
where $\xi_{\rm dm}(r,z)$ is the non-linear 2PCF of dark matter
particles at redshift $z$, which we obtain by Fourier Transforming the
non-linear power spectrum of Smith \etal (2003), and
\begin{eqnarray}
f_{\rm exc}(r,z) & = & {1 \over \overline{n}_{\rm g}(z)} 
\int_{0}^{\infty} n_\rmh(M,z) \, \langle N|M,z\rangle  \nonumber \\
& & b_\rmh(M,z) \, U(r|M,z)\, \rmd M\,.
\end{eqnarray}
Here we have taken into account the halo-halo exclusion effect, in
that the 2-halo galaxy pair can not have an average distance smaller
than $r_{\rm exc}(M,z)$:
\begin{equation}
U(r|M,z) = \left \{ \begin{array}{l} 0 \quad {\rm if} \quad r < r_{\rm exc}(M,z) \\
1 \quad {\rm else} \,.\end{array} \right .
\end{equation}
As shown in Wang \etal (2004), this method of computing the 2PCFs is
accurate at the few percent level as long as $r_{\rm exc}(M,z) = 2
r_{\rm vir}(M,z)$, which is the value we adopt throughout.

Observationally, the direct measurement is not the real-space 2PCF
because of redshift space distortions.  Instead one measures the
projected 2PCF, $w_\rmpp(r_\rmpp)$, which is related to the real
space 2PCF as
\begin{equation}\label{eq:wrp}
w_\rmpp(r_\rmpp) = 2 \int_{0}^{\infty}
\xi\left(\sqrt{r_{\pi}^2 + r_\rmpp^2}\right) \, \rmd r_{\pi}\,,
\end{equation}
where the comoving distance has been decomposed into components
parallel ($r_{\pi}$) and perpendicular ($r_\rmpp$) to the line of
sight.  In practice, the integration is only performed over a finite
range of $r_{\pi}$. Following Wang \etal (2007), we use a maximum
integration limit of $r_{\pi,{\rm max}} = 40\mpch$ (see Appendix A).

\subsection{Redshift Dependence}
\label{sec:zdep}

The formalism  described above allows us  to model the  stellar mass functions
and correlation  function of galaxies once  a set of parameters  is adopted to
specify  the CSMF,  $\Phi_\rmc(M_{\ast}|M,  z)$, and  the dynamical  evolution
function,  $P(m_{\ast}  | m_{\ast,  a},  z_a; m_a;  M_a;  \eta,  z)$.  At  any
particular redshift, our model for  these quantities is fully described by the
following  seven free  parameters:  $M_{\ast, 0}$,  $M_1$, $\alpha$,  $\beta$,
$\sigma_c$, $p_t$ and  $c$.  In order to describe the  evolution of the galaxy
distribution over cosmic time, we need to specify how each of these parameters
changes with redshift.

Throughout this paper we assume the following redshift dependence for
our model parameters:
\begin{eqnarray}
\log[M_{\ast,0} (z)] &=&\log(M_{\ast,0})+ {\gamma_1 z} \nonumber \\
\log[M_1 (z)]        &=&\log(M_1)       + {\gamma_2 z} \nonumber \\
\alpha(z)            &=&\alpha          + {\gamma_3 z} \nonumber \\
\log[\beta(z)] &=&{\rm min} [\log(\beta)+ {\gamma_4 z} + {\gamma_5 z^2} ,2]     \\
\sigma_c(z)&=& {\rm max} [0.173, 0.2 z] \nonumber \\
p_t(z) &=& p_t \nonumber \\
c(z) &=& c \nonumber \,.
\end{eqnarray}
These  functional forms have  been obtained  using trial-and-error  whereby we
tried to minimize  the number of parameters while achieving a  good fit to the
overall data (see Section~\ref{sec:observation} below).  As we have tested and
as  pointed  out  in  Moster  et  al. (2010),  $\sigma_c(z)$  cannot  be  well
constrained by  the SMFs  alone. The form  adopted above for  $\sigma_c(z)$ is
based on the  amount of true scatter  at low redshift obtained by  Y09 and the
uncertainty in the stellar mass measurements at high redshifts (e.g., D05; see
also Conroy et al. 2009).  Note  that the redshift dependence is relevant only
at  high redshift  when  $\sigma_c$ is  dominated  by the  uncertainty in  the
stellar  mass   measurements,  and  so   we  use  $\sigma_c(z)$,   instead  of
$\sigma_c(z_a)$, for $\sigma_c$ in  Eq. \ref{phismodeleasy}. The model is thus
specified  by a total  of 11  free parameters,  four to  describe the  CSMF at
$z=0$, five to describe their evolution with redshift, and two ($c$ and $p_t$)
to describe  the evolution  of satellite galaxies.   Note that we  assume that
both $c$ and $p_t$ are independent of redshift.  As we demonstrate below, this
assumption  is perfectly  compatible with  current data.  However, it  is very
simplified and needs to be re-evaluated once better data becomes available.


\section{Observational Constraints}
\label{sec:observation}

The goal of  this paper is to constrain  the CSMF, $\Phi_\rmc(M_{\ast}|M, z)$,
and the  dynamical evolution  function, $P(m_{\ast} |  m_{\ast, a},  z_a; m_a;
M_a;  \eta,  z)$, and  in  particular  their  evolution with  redshift,  using
observational data. In this section  we summarize the observational data to be
used to constrain our model. All our low redshift data are obtained from the
SDSS DR7 (Abazajian \etal 2009). 

\subsection {Low-Redshift Data}

\subsubsection{The Total Stellar Mass Function}
\label{sec:data_SMF}

\begin{figure}
\plotone{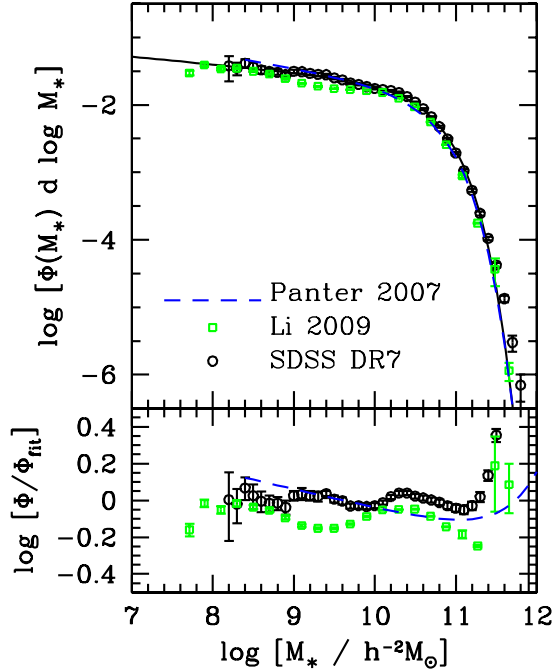}
\caption{Upper  panel: the  stellar mass  function  (SMF) of  galaxies at  low
  redshifts ($z  \sim 0.1$).  Open circles with  errorbars are the  results we
  obtained from the SDSS DR7, and the solid line is the corresponding best-fit
  Schechter form. For comparison, the  dashed line shows  the best-fit of
  the SMF obtained by Panter \etal  (2007), and the squares with errorbars are
  the results  obtained by  Li \&  White (2009). Lower  panel: similar  to the
  upper panel, but here for the ratios between the SMFs and our
  best-fit Schechter form.}
\label{fig:SMF}
\end{figure}

For the low  redshift stellar mass function (SMF) we use  the results that are
obtained using the same method outlined  in Y09b, but here updated to the SDSS
DR7 (Abazajian \etal 2009). Note that in these measurements the galaxy stellar
masses are  estimated using the model of  Bell \etal (2003) in  which a Kroupa
(2001) IMF was adopted (Borch \etal 2006). The resulting SMFs, for galaxies of
different colors and of different group member types, are provided in Appendix
B for reference.  Here  for our purpose, we only use the  SMF of all galaxies,
which is shown  in the upper panel of Fig.~\ref{fig:SMF}  as open circles with
error bars. For comparison, the solid and dashed lines show the best Schechter
fits for  the SMFs of  our measurements and  the one obtained by  Panter \etal
(2007)  from the  SDSS  DR3  (Abazajian \etal  2005),  respectively. The  open
squares correspond to the SMF obtained by Li \& White (2009) from the SDSS DR7
as well but for stellar masses obtained by Blanton \& Roweis (2007). To have a
clearer vision  of the difference between  these measurements, we  show in the
lower panel of  Fig.~\ref{fig:SMF} the ratio of the  SMFs which are normalized
by our  best fit Schechter  form SMF.  There  are $\sim 50\%$ and  $\sim 20\%$
difference at the high and low  mass ends between these three SMFs, which most
likely arise  from differences in (i)  the methods used to  derive the stellar
masses and  (ii) the different  data releases used.   As pointed out in  Li \&
White (2009), srarting from similar  IMF, the systematic difference of stellar
masses obtained by different approaches are at $0.1$dex to $0.3$dex at high to
low mass  ends.  For the  systematic errors, it  is very difficult to  come up
with a realistic covariance model for  the stellar mass function, as it is not
clear the uncertainties  in the assumed IMF and  spectral synthesis model.  In
the simplest assumption, the covariance can be a constant shift in the stellar
mass, where  our results can  be scaled to  other IMFs and  spectral synthesis
models.  Here,  for simplicity, to mimic  such kind of  systematics, we assume
there is at least 20\% error on each data point of our SMF.

\subsubsection{The conditional stellar mass function}
\label{sec:data_CSMF}

In addition to the (global) SMF, we also use the conditional stellar mass
function (CSMF) to constrain our model.  The CSMFs have been obtained by Y09b
from the SDSS DR4 group catalogues constructed by Yang \etal (2007) using the
halo-based group finder of Yang \etal (2005a). Here we update the CSMFs using
the latest SDSS DR7 group catalogues\footnote{see {\it
    http://gax.shao.ac.cn/data/Group.html}}.  The halo masses provided in the
group catalogues are obtained using the halo-abundance matching method (see
Section~4.1 in Yang \etal 2007) where halo mass function for the WMAP7
cosmology is used. The results of the CSMFs are provided in Appendix B for
reference.

\subsubsection{The two-point correlation function}
\label{sec:data_2PCF}

We  have measured  the  projected 2PCFs  for  SDSS DR7  galaxies of  different
stellar masses,  estimated using the model  of Bell \etal  (2003), i.e., these
are the same stellar  mass estimates as used for our SMF  and CSMF. A detailed
description of how we obtained these 2PCFs can be found in Appendix~A. We note
that similar  measurements have  already been carried  out by Li  \etal (2006)
using data from  the SDSS DR2 with stellar masses as  given by Kauffmann \etal
(2003). As we  have tested, our results agree with those  obtained by Li \etal
(2006) at 1-$\sigma$ level. Here we make use of our own measurements with full
covariance matrixes to constrain the CSMFs.

\begin{figure*}
\plotone{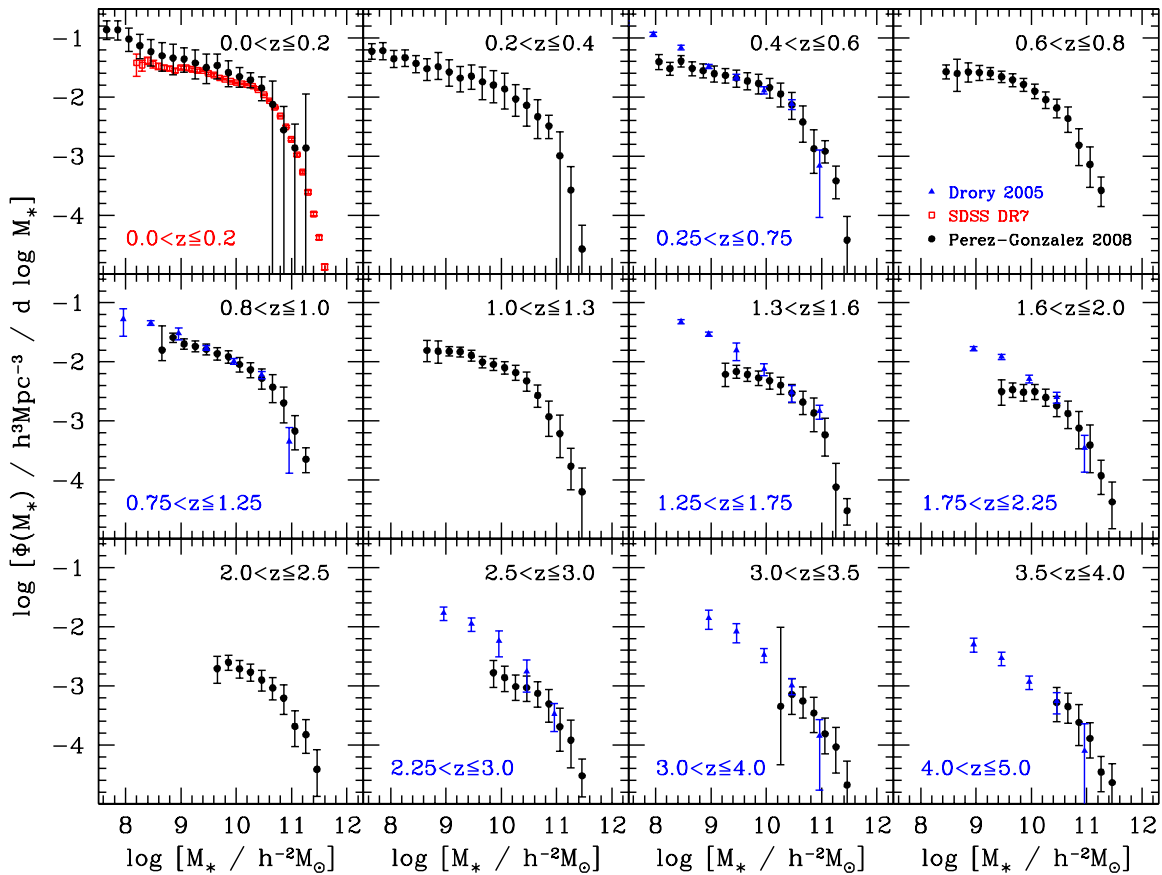}
\caption{The stellar mass functions of  galaxies in different redshift bins as
  indicated in  each panel.   The filled circles  with error-bars  are results
  obtained  from  the  Spitzer  observation  by  Perez-Gonzalez  \etal  (2008,
  PG08). The  open squares with error-bars  are results obtained  in this work
  from  the  SDSS DR7.   The  filled  triangles  with error-bars  are  results
  obtained by Drory  \etal (2005, D05), for which we  have combined their data
  for FORS Deep and the GOODS/CDF-S Fields.  The redshift range labeled in the
  upper right-hand corner of each panel  is for the results of PG08, while the
  range labeled in the lower left-hand corner is for the results obtained from
  SDSS DR7 (first  panel) and D05 (other panels).  Here  the stellar masses of
  PG08 and D05 are  divided by a factor of 1.7 to  take account the systematic
  differences of using different IMFs.}
\label{fig:highz_SMF}
\end{figure*}

\subsection{High-Redshift Data}
\label{sec:data_highz}

In addition to the low-redshift data discussed above, we also use data at high
redshifts.  Numerous  studies  have  measured  SMFs at  high  redshifts  using
different  data sets. Unfortunately,  differences among  the SMFs  obtained by
different authors  are still quite large.  In this paper we  therefore use two
different sets of  data, and use the differences in the  resulting models as a
method to gauge the model  uncertainties arising from the uncertainties in the
data.

The first  data set that  we use are  the SMFs of Perez-Gonzalez  \etal (2008,
hereafter PG08) obtained from Spitzer  data for galaxies in the redshift range
from $0  \lta z \lta  4$. Note  that in their  study, PG08 assumed  a Salpeter
(1955) IMF which  is different from the Kroupa IMF used  in the local universe
for the SDSS stellar masses. As suggested in PG08, the stellar masses based on
the Salpeter  (1955) IMF are systematically  larger by a factor  of $\sim 1.7$
than that based  on the Kroupa IMF.  For consistency,  we therefore devide all
the stellar  masses provided in PG08  by this factor.  The  resulting data are
shown  as   solid  dots  with  errorbars   in  Fig.~\ref{fig:highz_SMF}.   For
comparison, the  open squares in the  upper left-hand panel  correspond to the
local SMF obtained from  the SDSS DR7 in this study.  Note  that the latter is
in  general consistent with  that of  PG08 except  at low  mass end,  which we
believe to be  a result of cosmic variance and/or  the stellar mass estimator.
In what follows we  therefore ignore the SMF of PG08 for the  $0 < z \leq 0.2$
redshift bin.

The  second data set  for high  redshift SMFs  is that  of Drory  \etal (2005;
hereafter D05), which is based on  deep multicolor data in the FORS Deep Field
and the Great Observatories Origins Deep Survey-South/Chandra Deep Field-South
(GOODS/CDF-S)  region.  For  our investigation,  we combine  the  two separate
measurements of these two regions,  taking their average and using the maximum
of their original  error and the difference between the two  as the new error.
For the few data  points for which only FORS Deep Field  data is available, we
update their errors using the percentage  errors of the nearest data points to
partly take account of systematic  errors.  Again, the stellar masses obtained
in D05 assumed  a Salpeter (1955) IMF,  and we divide the stellar  masses by a
factor of  $\sim 1.7$ to correct them  to the Kroupa IMF.   The resulting data
are  shown in  Fig.~\ref{fig:highz_SMF}  as filled  triangles with  errorbars.
Note that  there is large  discrepancy between the  PG08 and the  D05 results,
especially at high  redshift. Since it is difficult to  decide which result is
more reliable,  we will  perform two separate  analyses using either  of these
data sets.  Finally, as  we did for  the low-redshift  data, we also  assume a
minimum of  20\% error for each  data point of  the SMFs to account  for other
possible uncertainties in the data.

Two-point correlation  functions for galaxies  of different stellar  mass bins
are available at  intermediate redshifts (e.g., Meneux \etal  2008 using VVDS;
Meneux  \etal 2009 using  zCOSMOS). However,  as pointed  out in  Meneux \etal
(2009), because of  the small sky coverage of  the surveys, these measurements
suffer significantly  from cosmic  variance. We therefore  decided not  to use
these measurements to constrain our model.  Instead we shall only use them for
comparison with our model predictions.

\subsection{The Fitting Method}
\label{sec:fitting}

%
\begin{figure*}
\plotone{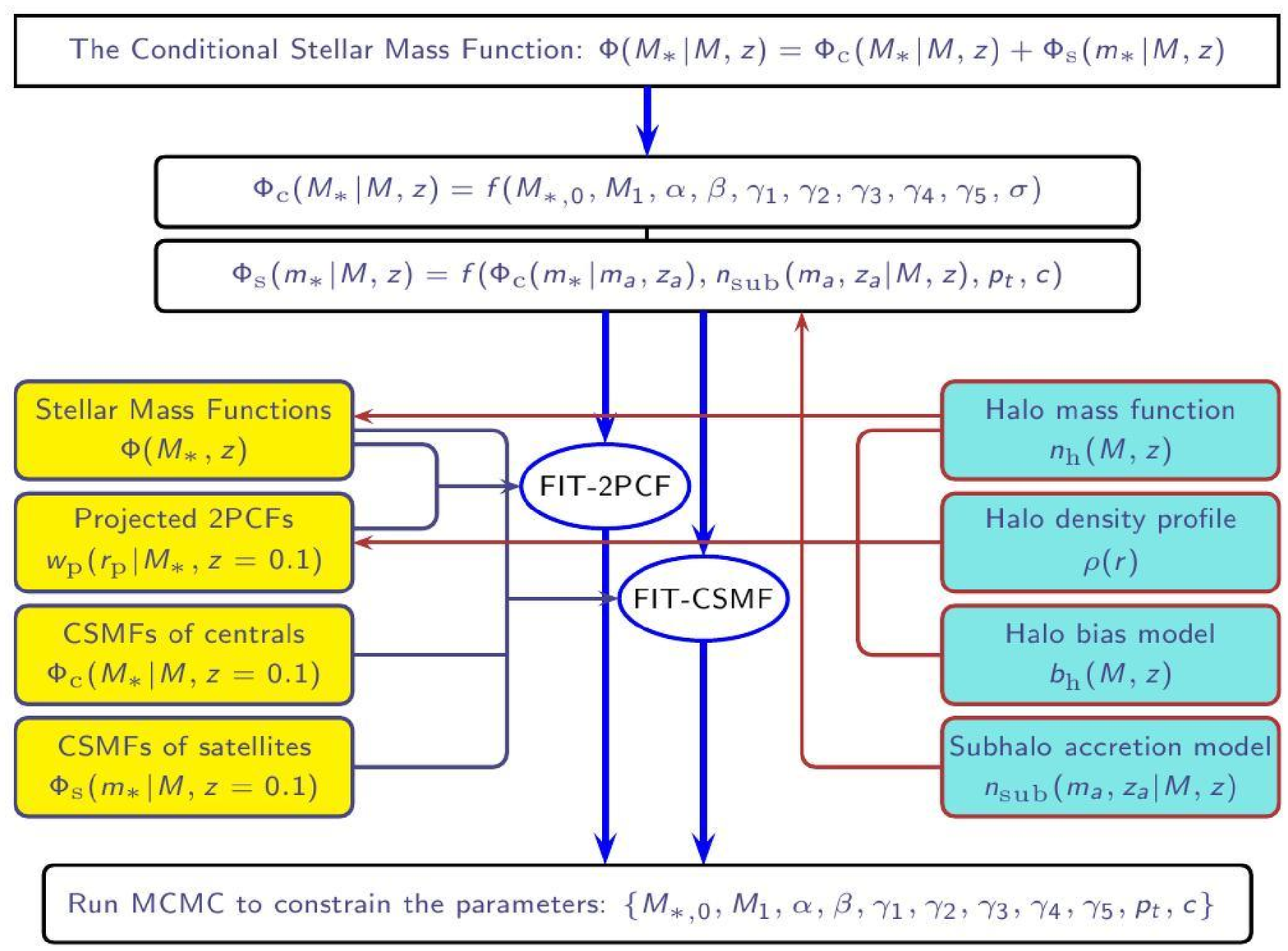}
\caption{A flowchart  showing how we  constrain the paramenters of  CSMFs. The
  filled bars on the left-hand  side list the observational quantities used as
  constraints, while  those on  the right-hand side  list the  ingredients to
  model these quantities.  We use MCMC  to obtain both the best fit parameters
  and their confidence ranges.  The  MCMC analysis is performed separately for
  both FIT-2PCF and FIT-CSMF. }
\label{fig:flowchart}
\end{figure*}
%
%
\begin{figure*}
\center{ \includegraphics[height=17.0cm,width=17.0cm,angle=0]{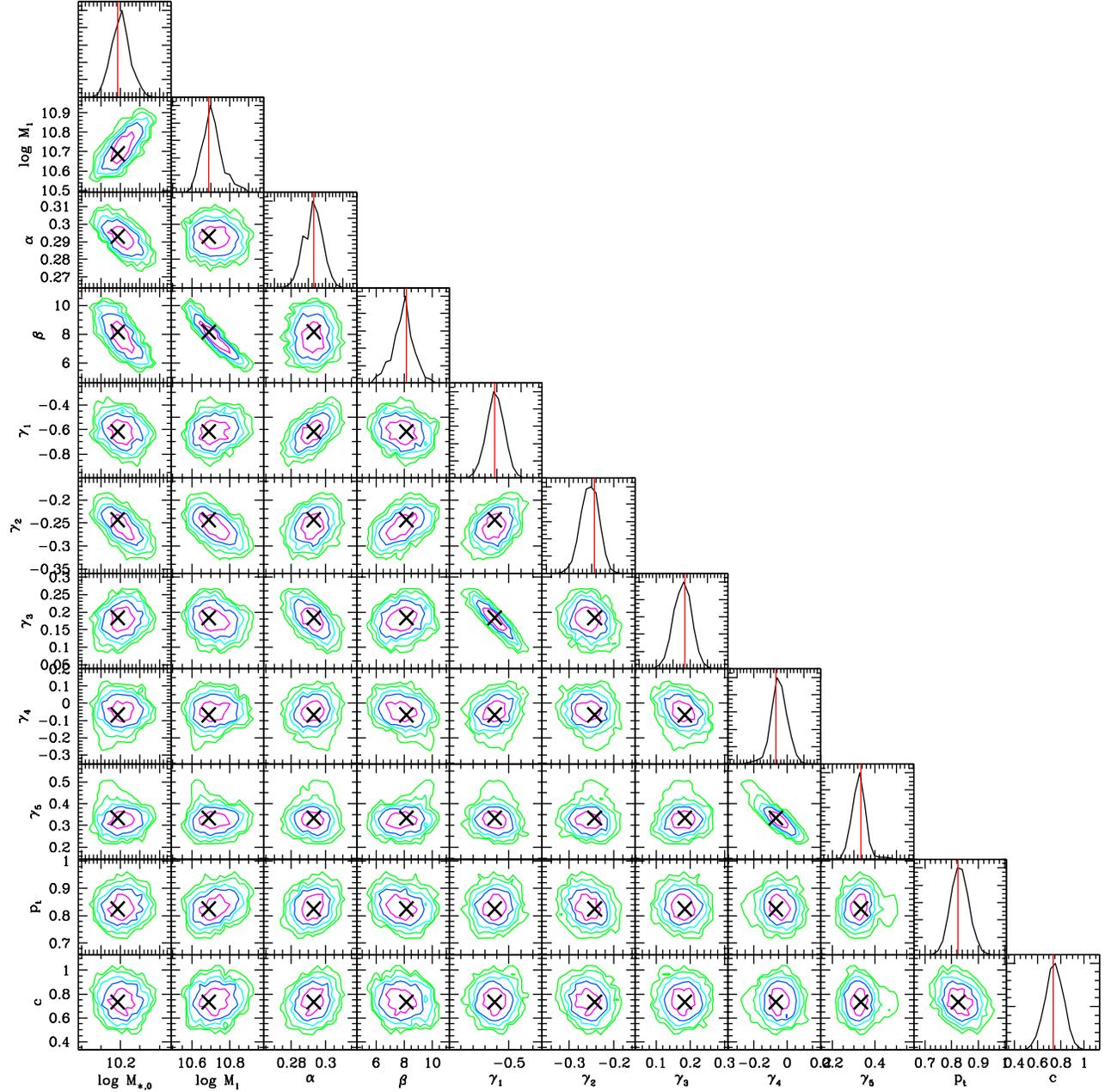}}
\caption{The best fit (cross or  vertical line) and the projected distribution
  of the parameters  in 2-D or 1-D  space. The outer to inner  contours in the
  2-D plane correspond  to the boundaries that enclose  95\%, 68\%, 30\%, 10\%
  and 1\%  models with  the smallest $\chi^2$  values, respectively.   The 1-D
  distributions  are the marginalized  distributions of  individual parameters
  obtained from all $10,000$ models. }
\label{fig:error}
\end{figure*}
%

\begin{deluxetable}{llll}
  \tabletypesize{\scriptsize} \tablecaption{Model Abbreviations}
  \tablewidth{0pt} \tablehead{Model & Data constraint & $c$ & $p_t$ }

\startdata 
FIT0     & SMF            &    - &  0 \\
FIT1     & SMF            &    1 & $\infty$ \\
FIT-2PCF ~~~~ & SMF+2PCF~~~~       &    free~~~~     & free~~~~ \\
FIT-CSMF & SMF+CSMF       &    free     & free \\
\cline{1-4}\\
SMF1  &   SMF=SDSS DR7+PG08 \\
SMF2  &   SMF=SDSS DR7+D05
\enddata

\tablecomments{ SDSS  DR7: the  stellar mass function  at $z=0.1$  we obtained
  from SDSS DR7 using the method of Yang \etal (2009b); PG08: the stellar mass
  functions obtained  by Perez-Gonzalez  \etal (2008) from  Spitzer, excluding
  the data  at $z=0.1$; D05: the  combined stellar mass  functions obtained by
  Drory \etal (2005) using deep multicolor data in the FORS Deep Field and the
  GOODS/CDF-S region.}
\label{tab:acronym}
\end{deluxetable}

To constrain model  parameters using observational data, we  use a Monte-Carlo
Markov  Chain (hereafter  MCMC)  to  explore the  likelihood  function in  the
multi-dimensional parameter space (see Yan, Madgwick \& White 2003 and van den
Bosch  et  al.   2005b for  similar  analyses).   In  order to  check  whether
different  data sets  are  mutually  compatible, we  carry  out two  different
analyses.  The first one, referred to as FIT-2PCF, uses the SMFs together with
the 2PCFs [$w_\rmpp(r_\rmpp)$] of galaxies at low-$z$. The second, referred to
as FIT-CSMF, uses  the SMFs at different redshifts together  with the CSMFs at
low $z$ as constraints.  The corresponding $\chi^2$ are defined as
\begin{equation}\label{eq:chi2W}
\chi_{\rm 2PCF}^2  = \chi^2(\Phi)+ \chi^2(w_\rmpp)\,,
\end{equation}
and
\begin{equation}\label{eq:chi2C}
\chi_{\rm CSMF}^2 =  \chi^2(\Phi)+ \chi^2(\Phi_{\rm CSMF})\,,
\end{equation}
respectively.  Here $\chi^2(\Phi)$, $\chi^2(w_{\rm p})$, $\chi^2(\Phi_{\rm
  CSMF})$ are given by
\begin{equation}\label{eq:chir2_phi}
\chi^2(\Phi)  =  \sum_{i=1}^{N_z}  \sum_{j} 
    \left[ \frac{\Phi_{\rm  mod}(M_{\ast,j},z_i) - 
   \Phi_{\rm obs}(M_{\ast,j},z_i)} {\sigma_{\Phi}(M_{\ast,j},z_i)} \right]^2  \,,
\end{equation}
and
\begin{equation}\label{eq:chir2_wrp}
\chi^2(w_\rmpp) =  \sum_{i=1}^{N_w} \sum_{j,k} 
\Delta w_{\rm p,j} ({\bf C}^{-1})_{j,k}\Delta w_{\rm p,k}\;,
\end{equation}
with
\begin{equation}
\Delta w_{\rm p,j}= w_{\rm p, mod}(r_{\rmpp,j}|M_{\ast,i}) -
   w_{\rm p, obs}(r_{\rmpp,j}|M_{\ast,i})  \,,
\end{equation}
and 
\begin{eqnarray}\label{eq:chir2_phiC}
\chi^2(\Phi_{\rm CSMF})= ~~~~~~~~~~~~~~~~~~~~~~~~~~~~~~~~~~~~~~~~~~~~~~~&&\\  
~~~~{\frac{1}{N_C}}\sum_{i=1}^{N_C} \sum_{j}
 \left[ \frac{\Phi_{\rm  mod}(M_{\ast,j}|M_i)-\Phi_{\rm
      obs}(M_{\ast,j}|M_i)}{\sigma_{\rm CSMF}(M_{\ast,j}|M_{i})}
  \right]^2 &&\nonumber \,.
\end{eqnarray}
In the  above expressions, $\sigma_{\Phi}(M_{\ast,j},  z_i)$, and $\sigma_{\rm
  CSMF} (M_{\ast,j} | M_{i})$  are, respectively, the observational errors for
the  SMF  and  CSMF,  while  $({\bf  C}^{-1})_{j,k}$ is  the  inverse  of  the
covariance matrix of the observed projected 2PCF.  The integers $N_{z}$, $N_w$
and $N_C$  are the  number of  SMFs (for different  redshifts), the  number of
projected 2PCFs  (for different  stellar mass bins),  and the number  of CSMFs
(for  different  halo  mass   bins),  respectively.   The  factor  $1/N_C$  in
Eq. (\ref{eq:chir2_phiC})  is somewhat arbitrary, which is  included to reduce
the weight of  the CSMFs, as both the  CSMFs and the low-$z$ SMF  are based on
the same data and thus not completely independent.

In  addition to  FIT-2PCF and  FIT-CSMF, in  which all  eleven  parameters are
included in the  fitting, and both the  SMFs and 2PCFs (or CSMFs)  are used as
constraints, we also  consider two extreme cases where $p_t$  is assumed to be
either $0$ or $\infty$. The first  case (referred to as FIT0 in the following)
with $p_t=0$ effectively  assumes that all satellite galaxies  are stripped as
soon as they are accreted (i.e., there are no satellite galaxies).  The second
case (referred to as FIT1) assumes $p_t=\infty$ and $c=1$, so that a satellite
is never  disrupted and evolves  in such  a way that,  at the redshift  $z$ in
question, it has the same stellar mass as a central galaxy in a host halo with
mass equal to the halo mass of  the satellite at its accretion time.  In these
two extreme cases we only use  the SMFs at different redshifts as constraints.
These  extreme  cases  are used  to  gauge  the  importance of  the  satellite
population in our modeling (see also Yang \etal 2009a; Wang \& Jing 2010).

As mentioned above,  there are large differences between  the SMFs obtained by
PG08 and D05  at high redshifts, and we therefore will  use them separately to
constrain our model.   More specifically, we consider two  sets of models. The
first uses  the high-$z$ SMFs of  PG08 and is  refered to as SMF1;  the second
uses  the high-$z$  SMFs  of D05  and  referred to  as  SMF2.  For  reference,
Table~\ref{tab:acronym}  lists   all  our  different   models,  including  the
abbreviations that we use to refer to them.

We start our MCMC from an initial guess and allow a `burn-in' of 1,000 steps
for the chain to equilibrate in the likelihood space. At any point in the
chain we generate a new set of model parameters by drawing the shifts in the
eleven free parameters from eleven independent Gaussian distributions. The
Gaussian widths are tuned so that the average accepting rate for the new trial
model is about 0.25.  The probability of accepting the trial model is
\begin{equation}
\label{probaccept}
P_{\rm accept} = \left\{ \begin{array}{ll}
1.0 & \mbox{if $\chi^2_{\rm new} < \chi^2_{\rm old}$} \\
{\rm exp}[-(\chi^2_{\rm new}-\chi^2_{\rm old})/2] & \mbox{if
$\chi^2_{\rm new} \geq \chi^2_{\rm old}$} \end{array} \right.
\end{equation}
with the $\chi^2$ defined in Eq.~(\ref{eq:chi2W}) or in Eq.~(\ref{eq:chi2C}).

We  construct  a  MCMC of  500,000  steps  after  `burn-in', with  an  average
acceptance rate  of $\sim  25$ percent.  Before  the further treatment  of the
MCMC,  we have tested  its convergence  using the  'convergence ratio'  $r$ as
defined in Dunkley  et al.  (2005). In all cases  in consideration, we observe
$r<0.01$ is achieved for each  parameter. In order to suppress the correlation
between neighboring models  in the chain, we thin the chain  by a factor $50$.
This results  in a final MCMC  consisting of $10,000$  independent models that
sample  the posterior  distribution.   As  an illustration,  we  show in  Fig.
\ref{fig:flowchart}  the flowchart  of  our fiting  process.   Based on  these
$10,000$ independent  models, we  obtain the best  parameter set that  has the
smallest $\chi^2$.  The 68\% confidence level (CL) around this set of model is
obtained by simply sorting the models according to their $\chi^2$, and finding
models that have  $\chi^2$ values smaller than that  corresponding to the 68\%
CL.  As an  illustration, the outer to inner  contours in Fig.~\ref{fig:error}
plot the  projected 2-D boundaries in  the parameter space  that enclose 95\%,
68\%, 30\%, 10\% and 1\% of the models with the smallest $\chi^2$ values, with
the best-fit values  indicated by a cross.  The  marginalized 1-D distribution
of each parameter  obtained from all the $10,000$ models is  also shown in the
plot.  As one can see, strong  covariance exists in some parameter pairs, such
as  $\beta$   and  $M_1$,  $\gamma_1$  and  $\gamma_3$,   and  $\gamma_4$  and
$\gamma_5$.  Overall,  most parameters are  well constrained, with  their mean
values much larger than the  corresponding dispersions.  For reference we list
in  Tables~\ref{tab:fitCSMF1} and~\ref{tab:fitCSMF2}  the  best-fit parameters
together with their  68\% CLs.  IDs 1 to 4 correspond  to FIT0, FIT1, FIT-2PCF
and FIT-CSMF, respectively, all  assuming WMAP7 cosmology.  Results are listed
separately for  cases using  SMF1 and SMF2.   For the other  four cosmological
models  listed   in  Table~\ref{tab:cosmology}  we  only   give  the  best-fit
parameters and their 68\% CLs for FIT-CSMF and FIT-2PCF models (IDs 5-12). The
details of all these models are discussed in the following section.

\begin{turnpage}
\begin{deluxetable*}{lcccccccccccccc}
  \tabletypesize{\scriptsize} \tablecaption{The best fit parameters of the
    CSMF based on SMF1. }  \tablewidth{0pt} \tablehead{ID & cosmology & $\log
    M_{\ast,0}$ & $\log M_1$ & $\alpha$ & $\beta$ & $\gamma_1$ & $\gamma_2$ &
    $\gamma_3$ & $\gamma_4$ & $\gamma_5$ & $p_t$ & $c$ & $\chi^2$ &
    FIT-type  \\ \cline{1-15}\\ (1) & (2)
    & (3) & (4) & (5) & (6) & (7) & (8) & (9) & (10) & (11) & (12) & (13) &
    (14) & (15)  } \startdata

1 & WMAP7 & $10.21^{+ 0.09}_{- 0.10}$ & $ 10.73^{+ 0.15}_{- 0.19}$ & $ 0.30^{+
  0.02}_{- 0.02}$ & $ 5.54^{+ 1.95}_{- 1.10}$ & $ -0.67^{+ 0.22}_{- 0.20}$ & $
-0.27^{+ 0.07}_{- 0.06}$ & $ 0.19^{+ 0.07}_{- 0.08}$ & $ 0.14^{+ 0.13}_{-
  0.15}$ & $ 0.25^{+ 0.10}_{- 0.08}$ & 0 & -- & 35.10 & FIT0 \\

2 & & $ 10.12^{+ 0.10}_{- 0.11}$ & $ 10.82^{+ 0.16}_{- 0.16}$ & $ 0.33^{+
  0.02}_{- 0.02}$ & $ 9.40^{+ 2.66}_{- 2.31}$ & $ -0.51^{+ 0.18}_{- 0.26}$ & $
-0.27^{+ 0.06}_{- 0.06}$ & $ 0.15^{+ 0.10}_{- 0.07}$ & $ -0.03^{+ 0.12}_{-
  0.14}$ & $ 0.28^{+ 0.08}_{- 0.08}$ & $\infty$ & 1.00 & 38.91 & FIT1 \\

3 & & $ 10.25^{+ 0.03}_{- 0.05}$ & $ 10.73^{+ 0.06}_{- 0.05}$ & $ 0.28^{+
  0.02}_{- 0.01}$ & $ 8.19^{+ 0.49}_{- 0.59}$ & $ -0.65^{+ 0.15}_{- 0.18}$ & $
-0.26^{+ 0.03}_{- 0.05}$ & $ 0.19^{+ 0.07}_{- 0.05}$ & $ -0.09^{+ 0.09}_{-
  0.13}$ & $ 0.33^{+ 0.09}_{- 0.06}$ & $ 1.31^{+ 0.23}_{- 0.15}$ & $ 0.64^{+
  0.24}_{- 0.58}$ & 113.33 & FIT-2PCF \\

4 & & $ 10.19^{+ 0.08}_{- 0.06}$ & $ 10.69^{+ 0.20}_{- 0.11}$ & $ 0.29^{+
  0.01}_{- 0.02}$ & $ 8.15^{+ 1.91}_{- 2.33}$ & $ -0.62^{+ 0.20}_{- 0.19}$ & $
-0.24^{+ 0.05}_{- 0.07}$ & $ 0.18^{+ 0.07}_{- 0.08}$ & $ -0.07^{+ 0.15}_{-
  0.12}$ & $ 0.33^{+ 0.10}_{- 0.09}$ & $ 0.82^{+ 0.10}_{- 0.08}$ & $ 0.73^{+
  0.23}_{- 0.24}$ & 102.09 & FIT-CSMF \\

\cline{1-15}\\

5 & WMAP1 & $ 10.23^{+ 0.04}_{- 0.07}$ & $ 10.84^{+ 0.02}_{- 0.14}$ & $
0.29^{+ 0.01}_{- 0.01}$ & $ 7.11^{+ 1.64}_{- 0.25}$ & $ -0.62^{+ 0.16}_{-
  0.16}$ & $ -0.23^{+ 0.05}_{- 0.04}$ & $ 0.18^{+ 0.06}_{- 0.06}$ & $ -0.12^{+
  0.10}_{- 0.12}$ & $ 0.35^{+ 0.07}_{- 0.07}$ & $ 1.12^{+ 0.16}_{- 0.29}$ & $
0.47^{+ 0.53}_{- 0.23}$ & 107.07 & FIT-2PCF \\

6 & & $ 10.22^{+ 0.03}_{- 0.07}$ & $ 10.78^{+ 0.03}_{- 0.10}$ & $ 0.28^{+
  0.01}_{- 0.01}$ & $ 7.66^{+ 1.51}_{- 0.11}$ & $ -0.59^{+ 0.14}_{- 0.16}$ & $
-0.22^{+ 0.06}_{- 0.05}$ & $ 0.17^{+ 0.07}_{- 0.05}$ & $ -0.14^{+ 0.14}_{-
  0.08}$ & $ 0.38^{+ 0.06}_{- 0.10}$ & $ 0.95^{+ 0.10}_{- 0.10}$ & $ 0.29^{+
  0.19}_{- 0.15}$ & 103.77 & FIT-CSMF \\

7 & WMAP3 & $ 10.27^{+ 0.06}_{- 0.03}$ & $ 10.72^{+ 0.09}_{- 0.09}$ & $
0.29^{+ 0.01}_{- 0.02}$ & $ 7.39^{+ 1.20}_{- 0.81}$ & $ -0.63^{+ 0.18}_{-
  0.13}$ & $ -0.29^{+ 0.05}_{- 0.04}$ & $ 0.19^{+ 0.05}_{- 0.07}$ & $ 0.05^{+
  0.10}_{- 0.11}$ & $ 0.27^{+ 0.07}_{- 0.06}$ & $ 1.27^{+ 0.44}_{- 0.20}$ & $
0.94^{+ 0.06}_{- 0.38}$ & 151.37 & FIT-2PCF \\

8 & & $ 10.21^{+ 0.04}_{- 0.04}$ & $ 10.64^{+ 0.08}_{- 0.05}$ & $ 0.29^{+
  0.01}_{- 0.01}$ & $ 7.60^{+ 0.59}_{- 0.88}$ & $ -0.62^{+ 0.20}_{- 0.15}$ & $
-0.27^{+ 0.04}_{- 0.04}$ & $ 0.19^{+ 0.04}_{- 0.08}$ & $ 0.10^{+ 0.11}_{-
  0.08}$ & $ 0.25^{+ 0.06}_{- 0.07}$ & $ 0.77^{+ 0.08}_{- 0.06}$ & $ 0.97^{+
  0.03}_{- 0.13}$ & 100.64 & FIT-CSMF \\

9 & WMAP5 & $ 10.22^{+ 0.05}_{- 0.06}$ & $ 10.70^{+ 0.06}_{- 0.09}$ & $
0.29^{+ 0.02}_{- 0.01}$ & $ 8.18^{+ 1.04}_{- 0.68}$ & $ -0.59^{+ 0.14}_{-
  0.18}$ & $ -0.25^{+ 0.04}_{- 0.05}$ & $ 0.18^{+ 0.06}_{- 0.06}$ & $ -0.01^{+
  0.11}_{- 0.10}$ & $ 0.29^{+ 0.07}_{- 0.07}$ & $ 1.16^{+ 0.20}_{- 0.12}$ & $
0.95^{+ 0.05}_{- 0.51}$ & 120.92 & FIT-2PCF \\

10 & & $ 10.22^{+ 0.03}_{- 0.06}$ & $ 10.73^{+ 0.05}_{- 0.07}$ & $ 0.29^{+
  0.02}_{- 0.01}$ & $ 7.05^{+ 0.89}_{- 0.57}$ & $ -0.60^{+ 0.20}_{- 0.16}$ & $
-0.26^{+ 0.05}_{- 0.04}$ & $ 0.18^{+ 0.06}_{- 0.08}$ & $ 0.03^{+ 0.08}_{-
  0.11}$ & $ 0.28^{+ 0.07}_{- 0.06}$ & $ 0.83^{+ 0.08}_{- 0.08}$ & $ 0.60^{+
  0.18}_{- 0.21}$ & 106.20 & FIT-CSMF \\

11 & Millennium & $ 10.27^{+ 0.04}_{- 0.05}$ & $ 10.74^{+ 0.08}_{- 0.10}$ & $
0.27^{+ 0.02}_{- 0.01}$ & $ 7.16^{+ 1.31}_{- 0.80}$ & $ -0.67^{+ 0.22}_{-
  0.11}$ & $ -0.23^{+ 0.06}_{- 0.05}$ & $ 0.20^{+ 0.04}_{- 0.08}$ & $ -0.02^{+
  0.13}_{- 0.10}$ & $ 0.31^{+ 0.06}_{- 0.08}$ & $ 0.94^{+ 0.43}_{- 0.07}$ & $
0.97^{+ 0.03}_{- 0.74}$ & 98.43 & FIT-2PCF \\

12 & & $ 10.22^{+ 0.05}_{- 0.06}$ & $ 10.70^{+ 0.10}_{- 0.08}$ & $ 0.28^{+
  0.02}_{- 0.01}$ & $ 7.25^{+ 0.94}_{- 1.14}$ & $ -0.57^{+ 0.16}_{- 0.18}$ & $
-0.21^{+ 0.04}_{- 0.06}$ & $ 0.16^{+ 0.06}_{- 0.06}$ & $ 0.01^{+ 0.13}_{-
  0.08}$ & $ 0.30^{+ 0.05}_{- 0.07}$ & $ 1.00^{+ 0.10}_{- 0.08}$ & $ 0.35^{+
  0.16}_{- 0.13}$ & 98.28 & FIT-CSMF \\

\enddata

\tablecomments{  Column  (1):  model  ID.   Column  (2):  cosmology  used  for
  fitting. Column  (3-13): the  best fit parameters  for the CSMF.   Note that
  $M_{\ast,0}$ is in units of $h^{-2} M_{\odot}$ and $M_1$ in units of $h^{-1}
  M_{\odot}$, and $\log$ is the 10 based logarithm.  Column (14): the $\chi^2$
  defined in Eq.(\ref{eq:chi2W}) or (\ref{eq:chi2C}). Column (15): the type of
  FIT used to  constrain the models.  Here results are  obtained by using SMF1
  . } \label{tab:fitCSMF1}
\end{deluxetable*}
%

%
\begin{deluxetable*}{lcccccccccccccc}
  \tabletypesize{\scriptsize} \tablecaption{The best fit parameters for the
    CSMFs based on SMF2. } \tablewidth{0pt} \tablehead{ID & cosmology & $\log
    M_{\ast,0}$ & $\log M_1$ & $\alpha$ & $\beta$ & $\gamma_1$ & $\gamma_2$ &
    $\gamma_3$ & $\gamma_4$ & $\gamma_5$ & $p_t$ & $c$ & $\chi^2$ &
    FIT-type  \\ \cline{1-15}\\ (1) & (2)
    & (3) & (4) & (5) & (6) & (7) & (8) & (9) & (10) & (11) & (12) & (13) &
    (14) & (15)  } \startdata

1 & WMAP7 & $ 10.45^{+ 0.07}_{- 0.11}$ & $ 11.17^{+ 0.11}_{- 0.19}$ & $
0.25^{+ 0.03}_{- 0.02}$ & $ 3.23^{+ 0.78}_{- 0.37}$ & $ -0.98^{+ 0.22}_{-
  0.18}$ & $ -0.25^{+ 0.07}_{- 0.05}$ & $ 0.42^{+ 0.09}_{- 0.11}$ & $ -0.07^{+
  0.13}_{- 0.09}$ & $ 0.01^{+ 0.05}_{- 0.01}$ & 0 & -- & 56.03 & FIT0 \\

2 & & $ 10.45^{+ 0.08}_{- 0.11}$ & $ 11.43^{+ 0.15}_{- 0.18}$ & $ 0.27^{+
  0.03}_{- 0.02}$ & $ 3.85^{+ 0.92}_{- 0.65}$ & $ -0.79^{+ 0.23}_{- 0.19}$ & $
-0.24^{+ 0.10}_{- 0.08}$ & $ 0.38^{+ 0.10}_{- 0.12}$ & $ -0.16^{+ 0.09}_{-
  0.09}$ & $ 0.02^{+ 0.04}_{- 0.02}$ & $\infty$ & 1.00 & 52.91 & FIT1 \\

3 & & $ 10.46^{+ 0.08}_{- 0.04}$ & $ 11.21^{+ 0.14}_{- 0.09}$ & $ 0.25^{+
  0.01}_{- 0.03}$ & $ 4.00^{+ 0.54}_{- 0.64}$ & $ -0.93^{+ 0.14}_{- 0.18}$ & $
-0.26^{+ 0.07}_{- 0.03}$ & $ 0.39^{+ 0.11}_{- 0.05}$ & $ -0.15^{+ 0.07}_{-
  0.12}$ & $ 0.04^{+ 0.02}_{- 0.07}$ & $ 1.18^{+ 0.21}_{- 0.23}$ & $ 0.91^{+
  0.09}_{- 0.48}$ & 128.54 & FIT-2PCF \\

4 & & $ 10.36^{+ 0.05}_{- 0.06}$ & $ 11.06^{+ 0.08}_{- 0.15}$ & $ 0.27^{+
  0.01}_{- 0.01}$ & $ 4.34^{+ 0.96}_{- 0.52}$ & $ -0.96^{+ 0.13}_{- 0.19}$ & $
-0.23^{+ 0.05}_{- 0.06}$ & $ 0.41^{+ 0.07}_{- 0.08}$ & $ -0.11^{+ 0.11}_{-
  0.08}$ & $ 0.01^{+ 0.05}_{- 0.07}$ & $ 0.88^{+ 0.07}_{- 0.10}$ & $ 0.98^{+
  0.02}_{- 0.26}$ & 131.21 & FIT-CSMF \\

\cline{1-15}\\

5 & WMAP1 & $ 10.46^{+ 0.06}_{- 0.09}$ & $ 11.30^{+ 0.09}_{- 0.18}$ & $
0.25^{+ 0.02}_{- 0.02}$ & $ 3.69^{+ 0.85}_{- 0.46}$ & $ -0.87^{+ 0.10}_{-
  0.18}$ & $ -0.20^{+ 0.06}_{- 0.06}$ & $ 0.37^{+ 0.09}_{- 0.05}$ & $ -0.18^{+
  0.08}_{- 0.09}$ & $ 0.04^{+ 0.03}_{- 0.05}$ & $ 0.97^{+ 0.23}_{- 0.14}$ & $
0.94^{+ 0.06}_{- 0.50}$ & 123.92 & FIT-2PCF\\

6 & & $ 10.39^{+ 0.04}_{- 0.06}$ & $ 11.17^{+ 0.09}_{- 0.10}$ & $ 0.26^{+
  0.02}_{- 0.01}$ & $ 4.34^{+ 0.62}_{- 0.54}$ & $ -0.93^{+ 0.16}_{- 0.12}$ & $
-0.21^{+ 0.05}_{- 0.09}$ & $ 0.38^{+ 0.06}_{- 0.08}$ & $ -0.17^{+ 0.10}_{-
  0.08}$ & $ 0.03^{+ 0.05}_{- 0.04}$ & $ 1.03^{+ 0.13}_{- 0.09}$ & $ 0.69^{+
  0.18}_{- 0.21}$ & 131.82 & FIT-CSMF \\

7 & WMAP3 & $ 10.49^{+ 0.03}_{- 0.05}$ & $ 11.16^{+ 0.05}_{- 0.08}$ & $
0.24^{+ 0.02}_{- 0.02}$ & $ 3.91^{+ 0.45}_{- 0.37}$ & $ -0.93^{+ 0.13}_{-
  0.14}$ & $ -0.28^{+ 0.07}_{- 0.04}$ & $ 0.41^{+ 0.07}_{- 0.06}$ & $ -0.11^{+
  0.10}_{- 0.08}$ & $ 0.03^{+ 0.02}_{- 0.08}$ & $ 1.30^{+ 0.25}_{- 0.17}$ & $
0.99^{+ 0.01}_{- 0.99}$ & 167.19 & FIT-2PCF \\

8 & & $ 10.37^{+ 0.04}_{- 0.06}$ & $ 10.94^{+ 0.13}_{- 0.08}$ & $ 0.26^{+
  0.02}_{- 0.01}$ & $ 4.82^{+ 0.61}_{- 0.76}$ & $ -1.03^{+ 0.19}_{- 0.10}$ & $
-0.26^{+ 0.06}_{- 0.04}$ & $ 0.43^{+ 0.06}_{- 0.09}$ & $ -0.04^{+ 0.11}_{-
  0.07}$ & $ 0.00^{+ 0.03}_{- 0.07}$ & $ 0.85^{+ 0.10}_{- 0.06}$ & $ 1.00^{+
  0.00}_{- 0.09}$ & 126.71 & FIT-CSMF \\

9 & WMAP5 & $ 10.56^{+ 0.03}_{- 0.14}$ & $ 11.31^{+ 0.07}_{- 0.19}$ & $
0.22^{+ 0.05}_{- 0.01}$ & $ 3.54^{+ 0.75}_{- 0.33}$ & $ -0.90^{+ 0.10}_{-
  0.16}$ & $ -0.25^{+ 0.06}_{- 0.05}$ & $ 0.42^{+ 0.07}_{- 0.07}$ & $ -0.16^{+
  0.10}_{- 0.08}$ & $ 0.03^{+ 0.03}_{- 0.06}$ & $ 1.25^{+ 0.21}_{- 0.17}$ & $
0.98^{+ 0.02}_{- 0.72}$ & 135.77 & FIT-2PCF \\

10 & & $ 10.36^{+ 0.05}_{- 0.06}$ & $ 11.06^{+ 0.08}_{- 0.15}$ & $ 0.27^{+
  0.01}_{- 0.01}$ & $ 4.34^{+ 0.96}_{- 0.52}$ & $ -0.96^{+ 0.13}_{- 0.19}$ & $
-0.23^{+ 0.05}_{- 0.06}$ & $ 0.41^{+ 0.07}_{- 0.08}$ & $ -0.11^{+ 0.11}_{-
  0.08}$ & $ 0.01^{+ 0.05}_{- 0.07}$ & $ 0.88^{+ 0.07}_{- 0.10}$ & $ 0.98^{+
  0.02}_{- 0.26}$ & 131.21 & FIT-CSMF \\

11 & Millennium & $ 10.47^{+ 0.07}_{- 0.03}$ & $ 11.19^{+ 0.12}_{- 0.05}$ & $
0.24^{+ 0.01}_{- 0.02}$ & $ 3.73^{+ 0.33}_{- 0.44}$ & $ -0.86^{+ 0.21}_{-
  0.18}$ & $ -0.19^{+ 0.06}_{- 0.07}$ & $ 0.36^{+ 0.09}_{- 0.09}$ & $ -0.10^{+
  0.05}_{- 0.10}$ & $ 0.02^{+ 0.03}_{- 0.05}$ & $ 0.97^{+ 0.19}_{- 0.12}$ & $
0.99^{+ 0.01}_{- 0.36}$ & 114.40 & FIT-2PCF \\

12 & & $ 10.37^{+ 0.05}_{- 0.06}$ & $ 11.01^{+ 0.13}_{- 0.10}$ & $ 0.26^{+
  0.02}_{- 0.01}$ & $ 4.61^{+ 0.69}_{- 0.66}$ & $ -0.84^{+ 0.19}_{- 0.20}$ & $
-0.21^{+ 0.09}_{- 0.09}$ & $ 0.29^{+ 0.11}_{- 0.07}$ & $ -0.04^{+ 0.10}_{-
  0.10}$ & $ 0.03^{+ 0.03}_{- 0.05}$ & $ 1.06^{+ 0.13}_{- 0.11}$ & $ 0.48^{+
  0.23}_{- 0.10}$ & 130.37 & FIT-CSMF \\

\enddata 

\tablecomments{The same as Table~\ref{tab:fitCSMF1}, but here 
SMF2 is used instead of SMF1.} \label{tab:fitCSMF2}
\end{deluxetable*}

\end{turnpage}


%
\begin{figure*}
\plotone{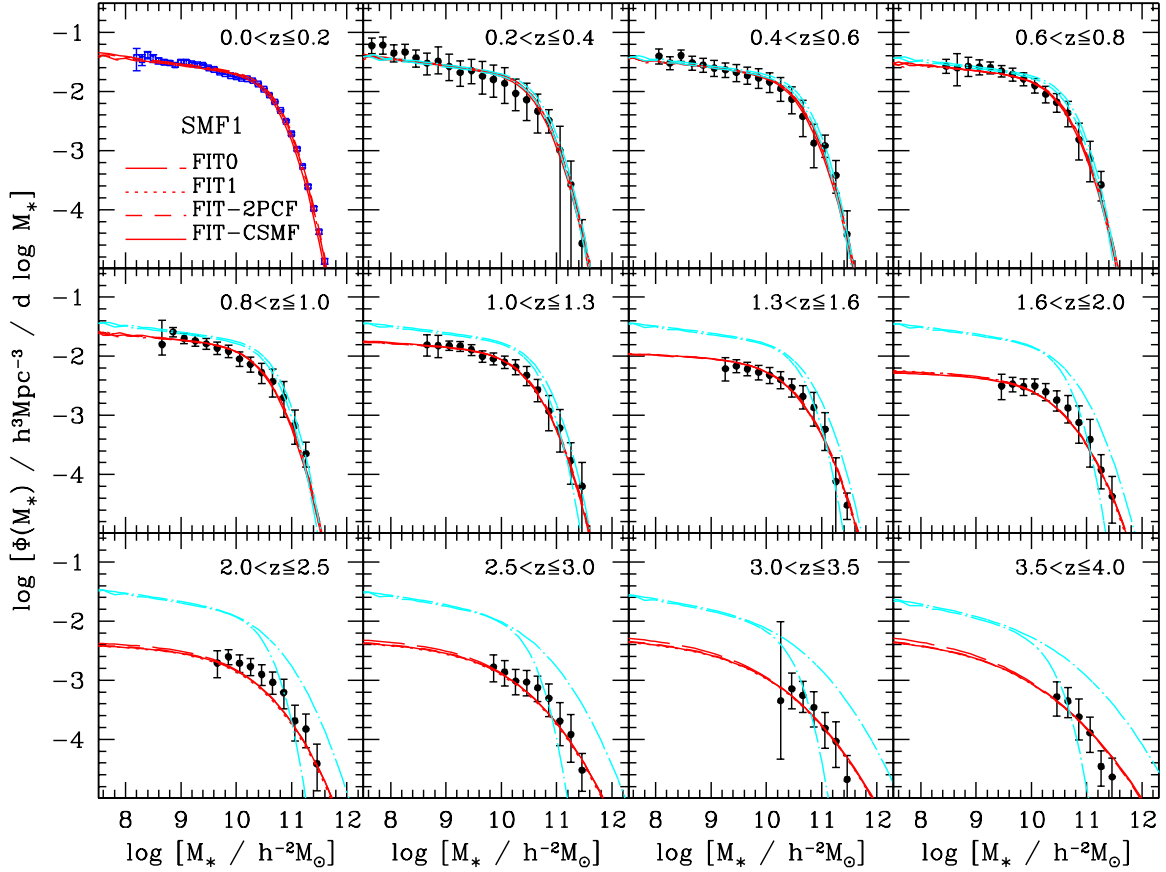}
\caption{The predicted stellar mass  functions (SMFs) of galaxies in different
  redshift bins compared  with the observed SMF of SDSS DR7  and those of PG08
  (other panels). In
  each panel, the long-dashed, dotted, dashed and solid lines, which are quite
  similar, show the  predictions of models FIT0, FIT1,  FIT-2PCF and FIT-CSMF,
  respectively, for  the case  of SMF1 and  assuming the WMAP7  cosmology.  In
  each  panel,   the  dot-long-dashed  lines  are  the   predictions  for  the
  ``no-evolution model''  in which we have  assumed that the  central galaxy -
  halo  relation does not  evolve and  is the  same as  that at  $z=0.1$. Here
  results  for $\sigma_c(z)  =  0.173$ (smaller  at  massive end  and at  high
  redshift)  and  $\sigma_c(z) =  \max[0.173,0.2z]$  are  plotted together  for
  comparison.  }
\label{fig:SMF_fit1}
\end{figure*}
\begin{figure*}
\plotone{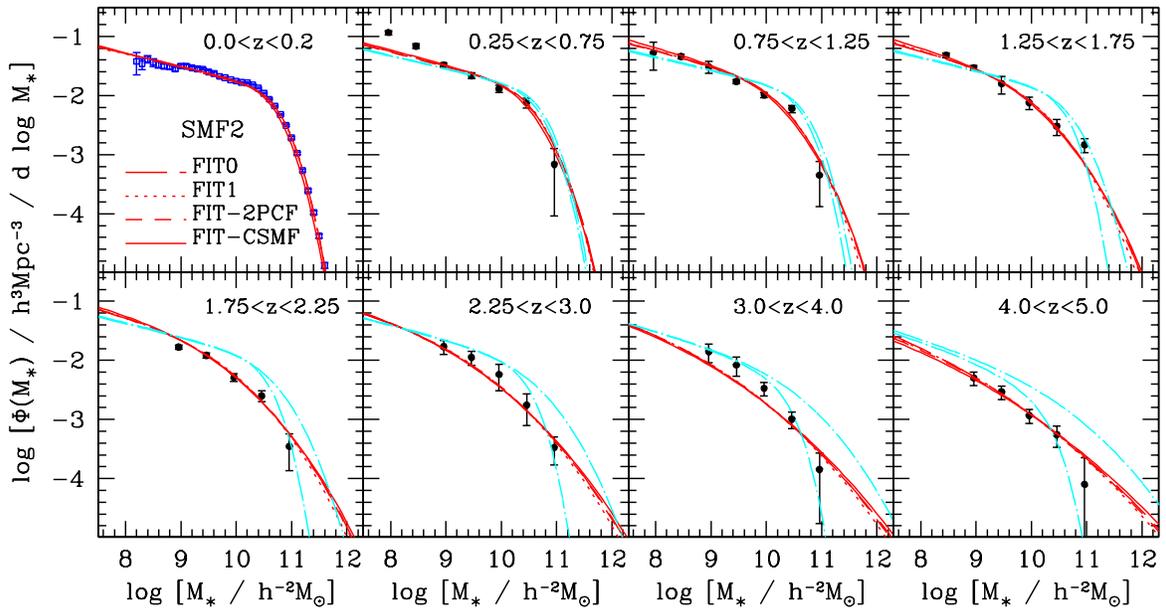}
\caption{Similar to  Fig.~\ref{fig:SMF_fit1}, but here  SMF2 has been  used as
  constraints.  The  data shown  (symbols) are the  stellar mass  functions of
  SDSS DR7 (first panel) and D05 (other panels).}
\label{fig:SMF_fit2}
\end{figure*}
\begin{figure*}
\plotone{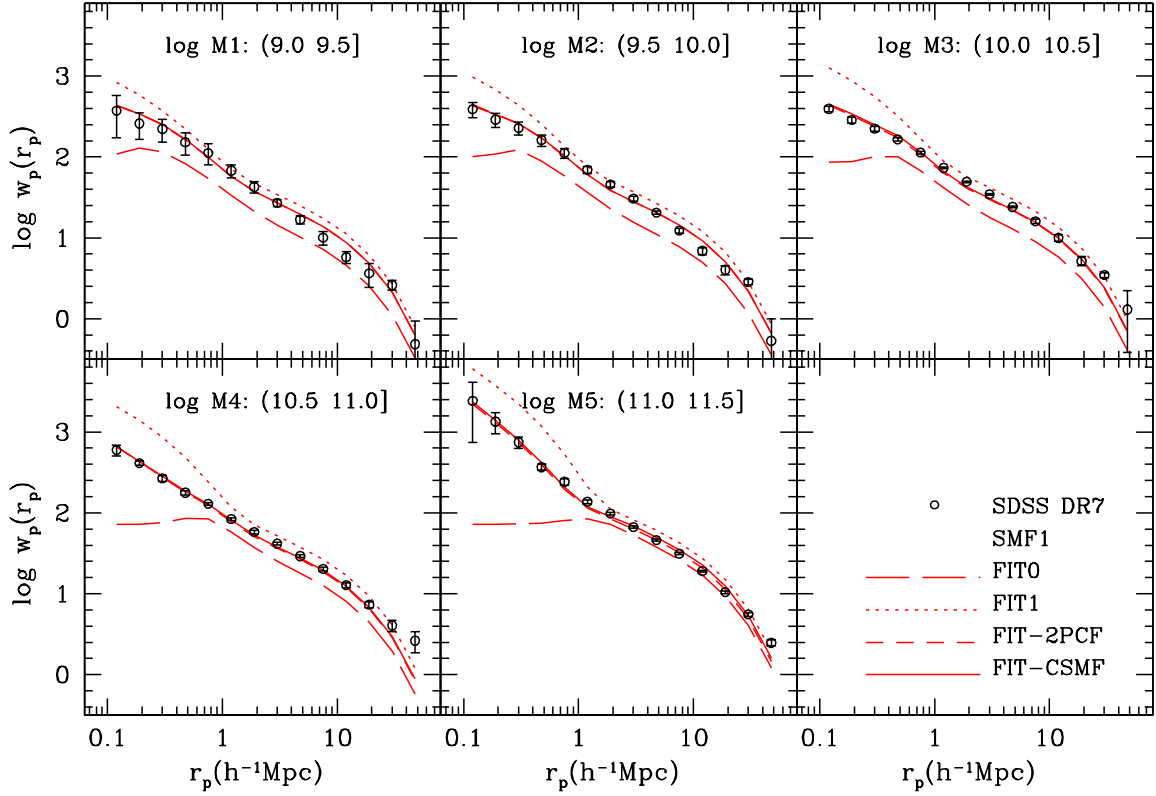}
\caption{The projected  2PCFs of  galaxies in different  stellar mass  bins as
  indicated  by  the  $\log[M_\ast/(h^{-2}\Msun)]$-values  in  brackets.  Open
  circles with error bars are the observational measurements from the SDSS DR7
  described in  Appendix~A. The  solid, dashed, dotted  and long  dashed lines
  show  the  predictions  for   models  FIT-CSMF,  FIT-2PCF,  FIT1  and  FIT0,
  respectively, where SMF1  has been used as constraints  for the stellar mass
  functions.}
\label{fig:wrp_M}
\end{figure*}
\begin{figure*}
\plotone{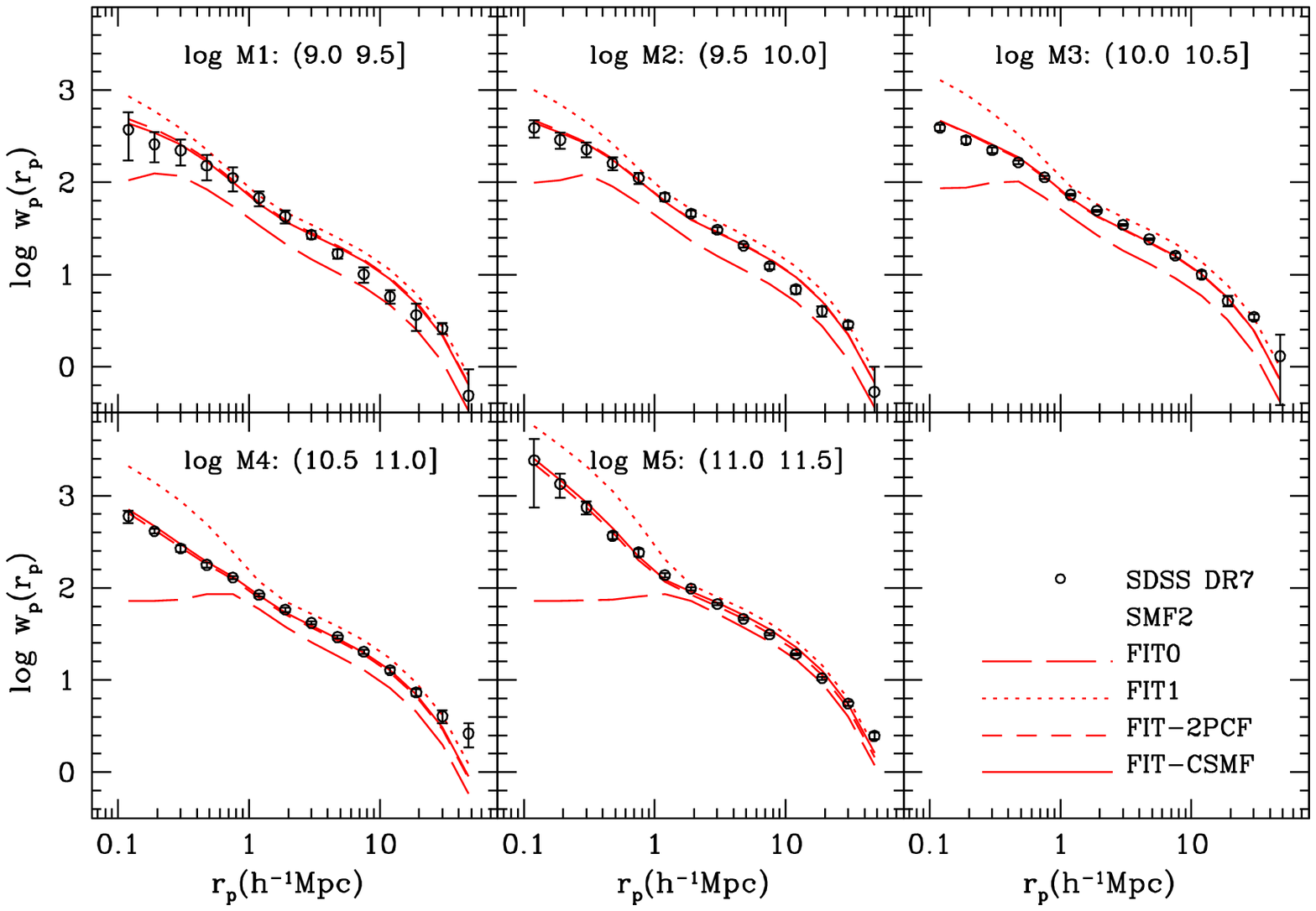}
\caption{Same as Fig.~\ref{fig:wrp_M}, except that  here SMF2 has been used as
  constraints for the stellar mass functions.}
\label{fig:wrp_Mx}
\end{figure*}
\begin{figure*}
\plotone{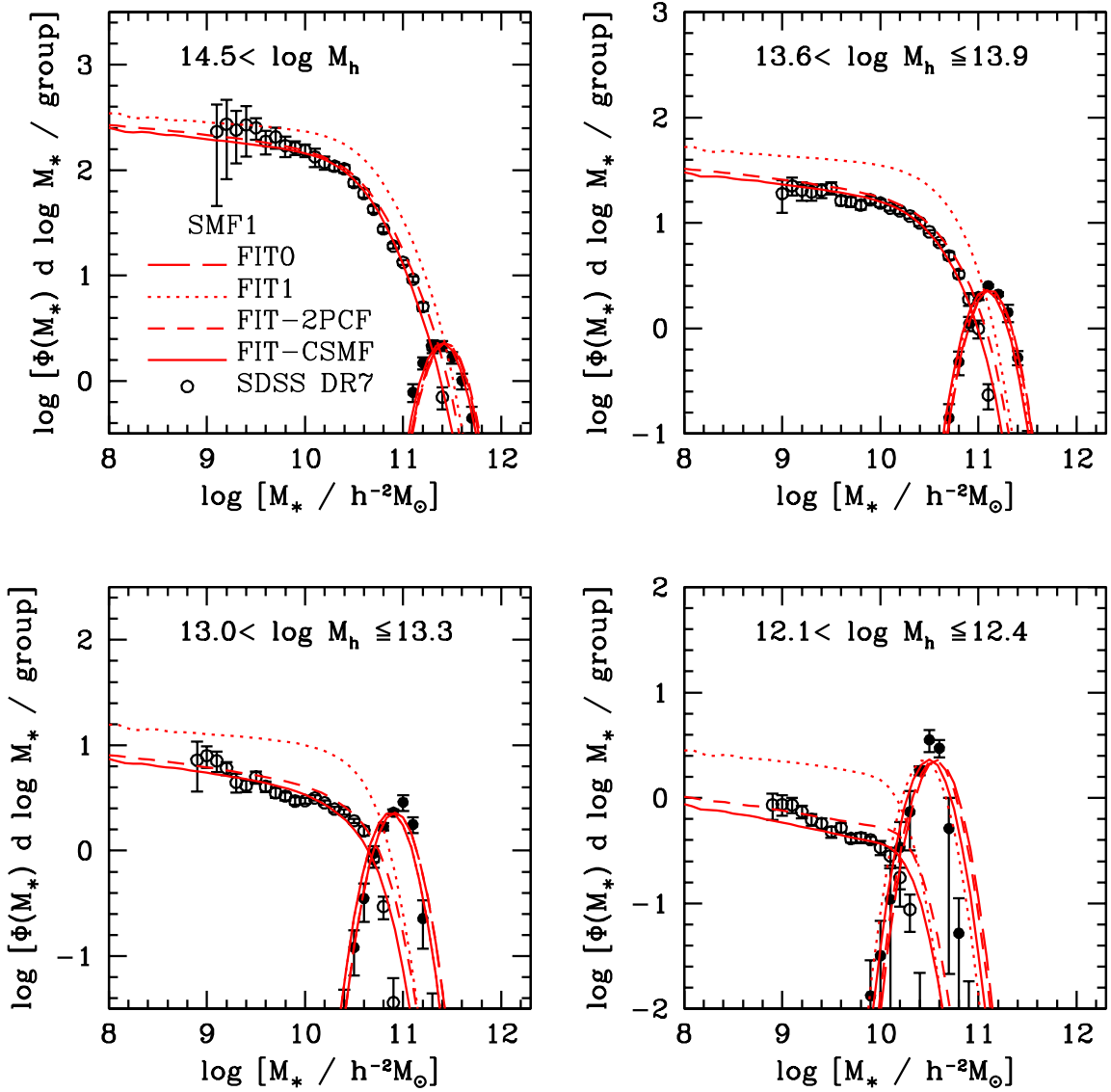}
\caption{The  conditional stellar  mass  functions for  central and  satellite
  galaxies in halos  of different masses as indicated in  each panel (the unit
  of $M_\rmh$  used is  $\msunh$). The  filled dots and  open circles  are the
  measurements from  the SDSS  DR7 group catalogue  for central  and satellite
  galaxies (see  Appendix B), respectively.  The error  bars represent scatter
  among  different bootstrap  samples.   The long-dashed,  dotted, dashed  and
  solid  lines in  each  panel show  the  predictions for  models FIT0,  FIT1,
  FIT-2PCF and  FIT-CSMF, respectively. Note  that the satellite  component is
  absent in FIT0.   Here we have adopted the WMAP7 cosmology  and used SMF1 as
  constraints for the stellar mass functions.}
\label{fig:CSMF}
\end{figure*}
\begin{figure*}
\plotone{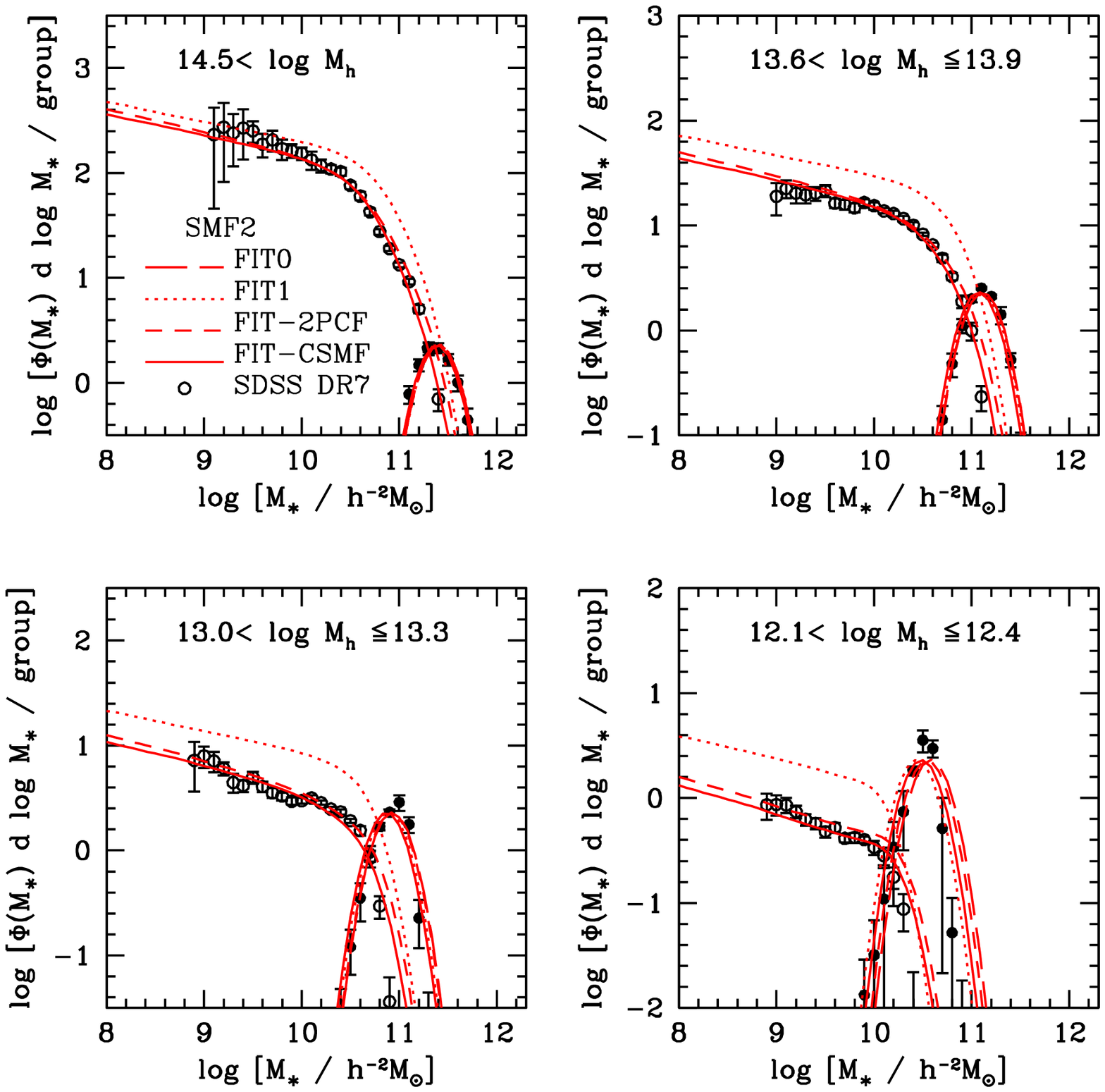}
\caption{Same as Fig.~\ref{fig:CSMF},  except that here SMF2 has  been used as
  constraints for the stellar mass functions.}
\label{fig:CSMF2}
\end{figure*}
\begin{figure*}
\plotone{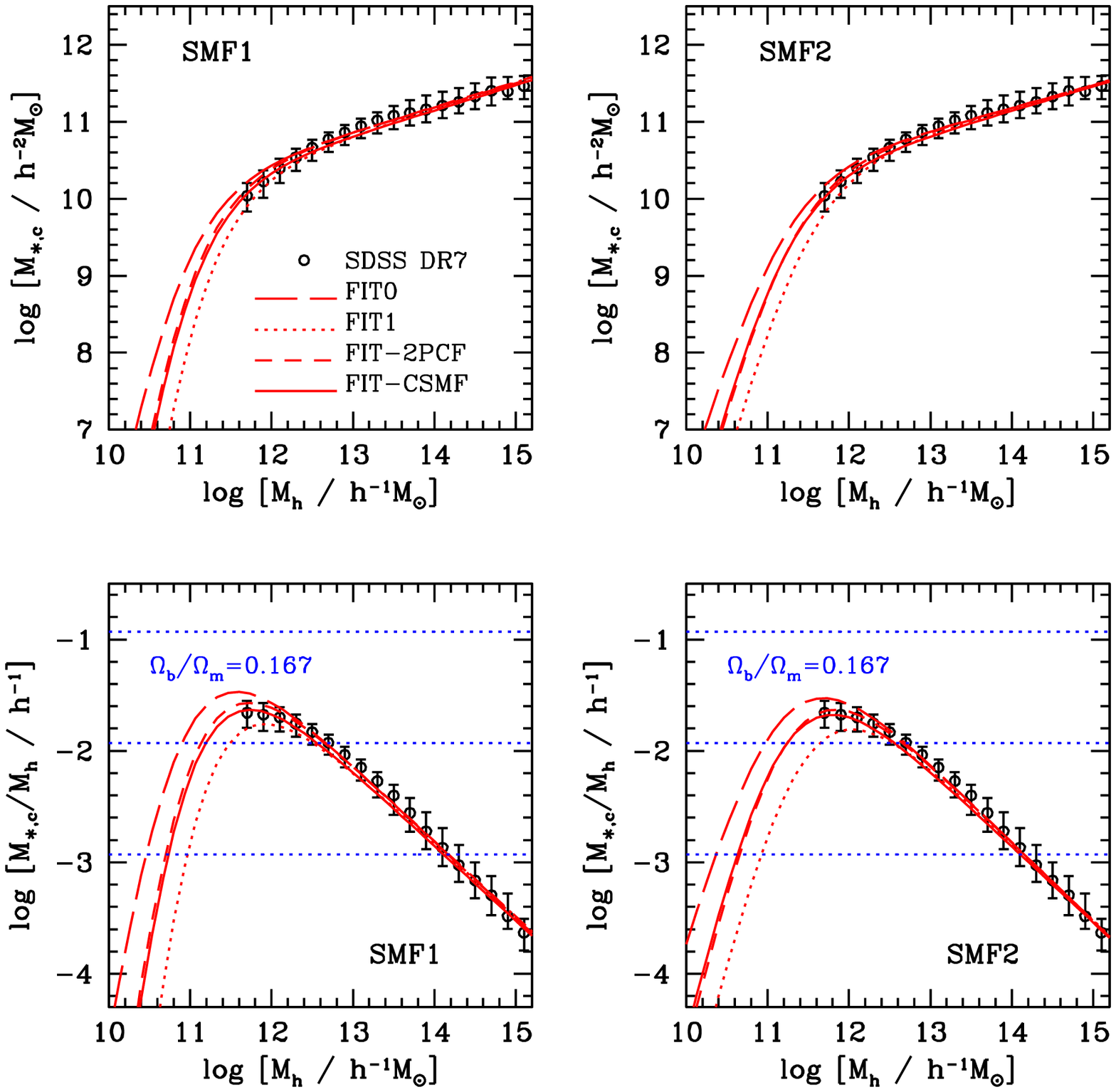}
\caption{The   $M_{\ast,c}$   -  $M$   relations   (upper   panels)  and   the
  $M_{\ast,c}/M$ ratios (lower panels) as  function of halo mass.  Results are
  shown separately  for SMF1 (left  panels) and SMF2 (right  panels), assuming
  WMAP7 cosmology.  Here the results are for galaxies at redshift $z=0.1$. The
  long-dashed, dotted, dashed  and solid lines are the  predictions for models
  FIT0,  FIT1, FIT-2PCF  and FIT-CSMF,  respectively.  The  open  circles with
  error-bars  are  the  direct  measurements  obtained  from  the  SDSS  group
  catalogue (e.g., Yang et al.  2008),  here updated to the latest DR7.  As an
  illustration, the horizontal  dotted lines in the lower  panels indicate the
  mass ratios corresponding to 100\%, 10\% and 1\% of the universal mass ratio
  between baryonic and total matter; $\Omega_\rmb/\Omega_\rmm=0.167$. }
\label{fig:lowz_MM}
\end{figure*}
%


\section{Results}
\label{sec:res}

The upper left-hand  panels of Figs.~\ref{fig:SMF_fit1} and~\ref{fig:SMF_fit2}
show  the  fits  to  the  observed  SMFs  at  low-$z$  using  SMF1  and  SMF2,
respectively.  In  both cases  we adopted the  fiducial WMAP7  cosmology.  The
long-dashed and dotted lines show the results for the two extreme models, FIT0
and FIT1, respectively.  Clearly, even these two extreme models can accurately
match  the  observed  SMFs  (as  judged by  their  reduced  $\chi^2$  values),
indicating that the  SMFs alone cannot constrain the values  of $p_t$ and $c$.
This  is  expected from  the  fact  that, as  mentioned  above,  the SMFs  are
dominated by central galaxies.  Note  that for SMF2, the $\chi^2(\Phi)$ values
are significantly larger than those for SMF1.  The main reason for this is the
much smaller  error bars in  SMF2 and the  large difference between  model and
data at  the low  mass end for  the $0.25 <  z <  0.75$ redshift bin  (see the
second panel of Fig.~\ref{fig:SMF_fit2}).

The  best  fit parameters  of  the  FIT0 and  FIT1  models  are  used to  make
predictions    for   the    projected    two-point   correlation    functions,
$w_\rmpp(r_\rmpp)$, in different stellar mass  bins, and the results are shown
in Fig.~\ref{fig:wrp_M} (for SMF1) and Fig.~\ref{fig:wrp_Mx} (for SMF2), again
as long-dashed and dotted  lines, respectively. Clearly, the model predictions
for FIT0  and FIT1 are  very different, especially  on small scales  where the
correlation function  is dominated  by the 1-halo  term.  Since FIT0  and FIT1
represent  the  minimum  and  maximum  contributions  of  satellite  galaxies,
respectively, the  correlation functions predicted by these  two models should
bracket the predictions of more realistic models. The large difference between
FIT0 and FIT1 clearly  demonstrates that the observed $w_\rmpp(r_\rmpp)$, {\it
  especially for $r_\rmpp \la 1 \mpch$}, can provide useful constraints on the
satellite population,  which in our  model is characterized by  the parameters
$p_t$ and $c$.  Note that the observed $w_\rmpp(r_\rmpp)$ falls nicely between
the predictions for  our two extreme models, suggesting  that a more realistic
treatment of satellite galaxies should be able to fit the data.

When   separated   into   central   and  satellite   components,   the   CSMF,
$\Phi(M_{\ast}|M)$,  provides  direct   information  regarding  the  satellite
fractions, and so it is also  interesting to look at the model predictions for
$\Phi(M_{\ast}|M)$.  The  long-dashed and dotted  lines in Fig.~\ref{fig:CSMF}
(for SMF1) and Fig.~\ref{fig:CSMF2} (for SMF2) show the $z=0.1$ predictions of
models  FIT0 and FIT1,  respectively.  Note  that by  definition there  are no
satellite galaxies in FIT0, so that there are no long-dashed lines present for
the satellite components.  For comparison, the solid dots and open circles are
observational   measurements    obtained   from   SDSS    DR7   (see   Section
\ref{sec:data_CSMF}).  Note how FIT0 and  FIT1 predict CSMFs for centrals that
are very  similar, and in good  agreement with the data.   However, model FIT1
clearly, and significantly, over-predicts  the CSMFs of satellites, while FIT0
also clearly fails to match the  data since it predicts zero satellites.  Once
again, the  observed CSMFs of satellites  falls in between  the predictions of
these two extreme models, suggesting that  the actual CSMF data can be used to
constrain model parameters, especially $p_t$ and $c$.

Finally, Fig.~\ref{fig:lowz_MM} shows the  $z=0.1$ relation between halo mass,
$M$, and the stellar mass  of the central galaxy, $M_{\ast,c}$ (upper panels),
and the ratio  $M_{\ast,c}/M$ (lower panels).  Here results  for SMF1 and SMF2
are shown in the left-  and right-hand panels, respectively, while long-dashed
and  dotted  lines  once again  correspond  to  the  extreme models  FIT0  and
FIT1. Note  how both models are  virtually indistinguishable at  the high mass
end, and  in good  agreement with  the results of  Yang \etal  (2008) obtained
directly  from the  SDSS group  catalog  (open circles  with errorbars),  here
updated according to the latest SDSS  DR7.  At the low mass end, however, FIT0
predicts $M_{\ast,c}$ that are significantly  higher than FIT1.  The reason is
simple: having  no satellites forces model  FIT0 to assign a  larger number of
massive galaxies, which otherwise would have been assigned to massive halos as
satellites, to relatively low-mass  halos as centrals.  Thus a self-consistent
modelling  that  takes  into  account  the subhalo  accretion  times  and  the
evolution   of  satellite   galaxies  after   accretion  mainly   impacts  the
$M_{\ast,c}$ - $M$ relation at the  low-mass end.  Such behavior has also been
noticed, e.g., in Moster et al. (2010) and Behroozi et al.  (2010).  Note also
that the  $M_{\ast,c}$ - $M$ relations  obtained from SMF1 and  SMF2 are quite
similar over the entire halo mass range.

\begin{figure*}
\plotone{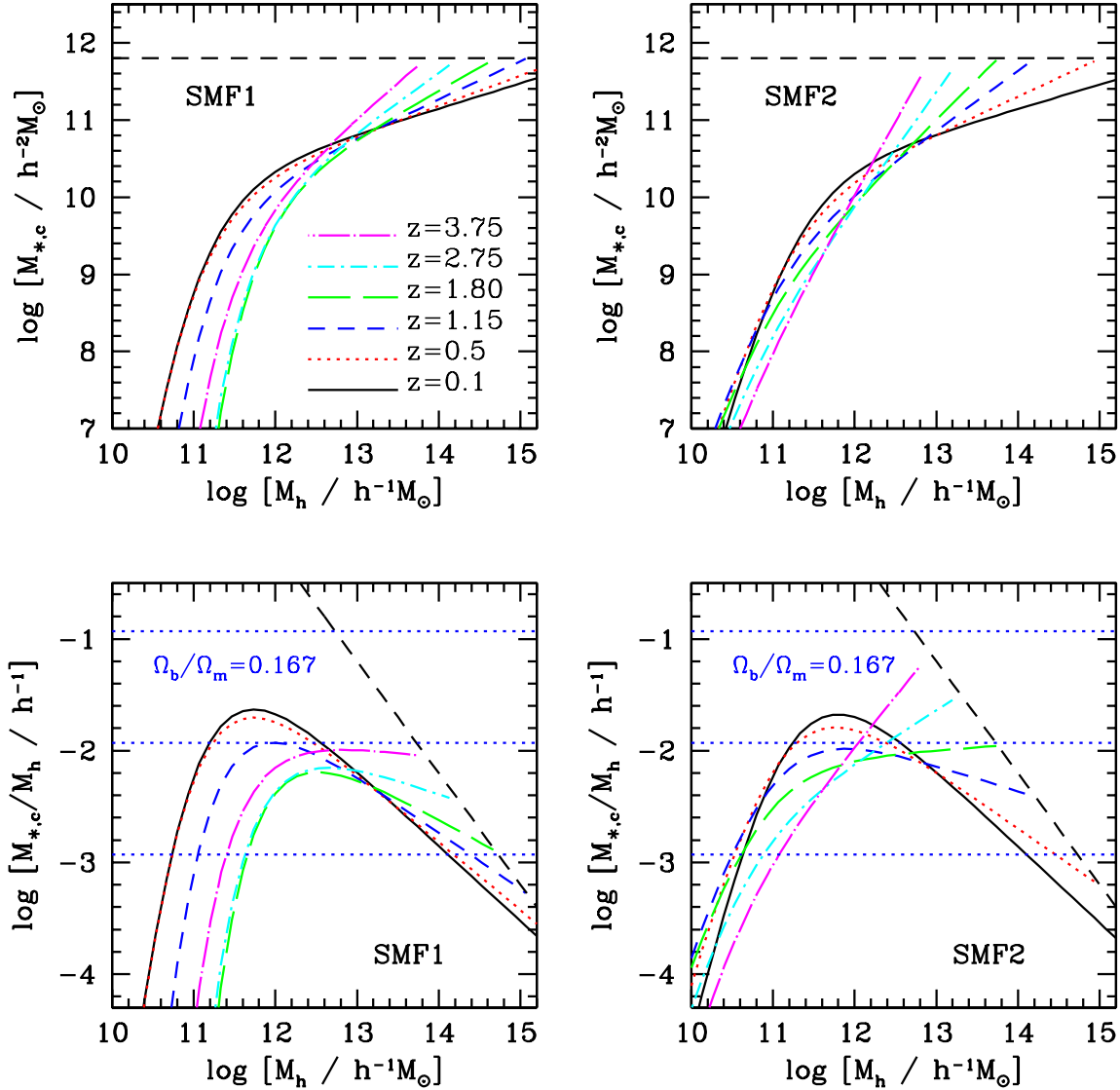}
\caption{Similar to Fig.~\ref{fig:lowz_MM}, but here we show model predictions
  at  different redshifts. Results  are shown  separately for  SMF1 (left-hand
  panels)  and SMF2  (right-hand panels).   For simplicity,  results  are only
  shown for model  FIT-CSMF, although the results for  model FIT-2PCF are very
  similar.    The   black dashed    lines   correspond   to   $M_{\ast}   =
  10^{11.8}h^{-2}\Msun$,  which  roughly reflects  the  observed stellar  mass
  limit of galaxies.}
\label{fig:highz_MM}
\end{figure*}

Next we focus on the results obtained from FIT-2PCF and FIT-CSMF, in which all
11 parameters  are kept free, and  the observed SMFs are  combined either with
the 2PCFs  (in FIT-2PCF)  or the  CSMFs (in FIT-CSMF)  to constrain  the model
parameters.   The  fitting  results   are  shown   in  Fig.~\ref{fig:SMF_fit1}
through~\ref{fig:lowz_MM} as  the dashed  and solid lines,  respectively.  The
best fit values for each of the 11 model parameters are listed under IDs 3 and
4   in  Tables~\ref{tab:fitCSMF1}   (for  SMF1)   and~\ref{tab:fitCSMF2}  (for
SMF2). There are a number of points worth noting: 
\begin{itemize}

\item  For a  given set  of SMFs  (SMF1 or  SMF2), the  results  obtained from
  FIT-2PCF and  FIT-CSMF are quite similar,  showing that the  2PCFs and CSMFs
  constrain the model in a similar way.

\item The best-fit  values for $p_t$ fall between about 0.8  and 1.3, which is
  roughly consistent  with the canonical  value, $p_t=1.0$, expected  from the
  dynamical friction time of subhalos  in numerical simulations.  The value of
  parameter $c$, which  is much larger than $0$ in  most cases, indicates that
  there is evolution  in the mass of satellite galaxies.   The mass growth for
  low-mass satellites after their accretion into the host halo is particularly
  significant,  as  indicated  by  the redshift  dependent  $M_{\ast,  c}-M_h$
  relations shown in  Fig. \ref{fig:highz_MM}.  For a given  set of SMFs (SMF1
  or  SMF2), FIT-2PCF  yields  larger $p_t$  (or  $c$) (by  $\sim 30\%$)  than
  FIT-CSMF,  because FIT-2PCF slightly  over-predicts the  CSMFs, as  shown in
  Figs.~\ref{fig:CSMF} and \ref{fig:CSMF2}.

\item Using  SMF1 or SMF2  leads to very  similar predictions for the  CSMF of
  central galaxies, but SMF2 yields larger satellite fractions at the low mass
  end (Figs.~\ref{fig:CSMF} and~\ref{fig:CSMF2}).

\item Models  FIT-2PCF and FIT-CSMF  predict $M_{\ast,c}$ - $M$  and $M_{\ast,
  c}/M$ - $M$ relations at low-$z$ that are very similar, and that lie between
  the predictions of FIT0 and  FIT1 (see Fig.\ref{fig:lowz_MM}). They are also
  in  good agreement  with the  results obtained  directly from  the  SDSS DR7
  galaxy group catalogs. These two  models predict a maximum for the $M_{\ast,
    c}/M$   ratio  of  $\sim   0.02  h^{-1}$   at  a   halo  mass   $M  \simeq
  10^{11.8}\msunh$, indicating that galaxy formation is most efficient on this
  mass scale (at least at low-$z$). Note that a value of $M_{\ast, c}/M = 0.02
  h^{-1} \simeq 0.03$ is about a factor of six lower than the average baryonic
  to total mass ratio $\Omega_\rmb  / \Omega_\rmm \simeq 0.167$ (e.g., Komatsu
  \etal 2011) of the Universe  (indicated by the upper horizontal dotted lines
  in  the lower panels  of Fig.~\ref{fig:lowz_MM}).   This indicates  that the
  overall star formation efficiency  is low.  Furthermore, since atomic and/or
  molecular  gas typically  contribute  only  a small  fraction  of the  total
  baryonic mass  of (massive) galaxies, this furthermore  underscores that the
  process of  galaxy formation has either  prevented most of  the baryons from
  cooling, or has somehow managed to eject them from the galaxy.

\item The central  stellar mass increases with halo  mass roughly as $M^{0.3}$
  at  the massive  end  and as  $M^{8}$ for  SMF1  ($M^{4}$ for  SMF2) at  the
  low-mass  end. In  low-mass halos,  the use of SMF2 leads  to more  massive
  central galaxies than the use of SMF1 (Fig.\ref{fig:lowz_MM}). This arises
  because the stellar mass functions in SMF2 are steeper at the low-mass end.

\item  Although  the parameters  in  $\sigma_c(z)$  are  not treated  as  free
  parameter  in  our  model  constraints,  its impact  is  quite  significant,
  especially at high redshift where the scatter is large. This is demonstrated
  by    the   predictions    of   the    `no-evolution'   models    shown   in
  Figs.~\ref{fig:SMF_fit1}  and~\ref{fig:SMF_fit2}. Here  the $M_{\ast,  c}$ -
  $M$ relations  at high-$z$ are  assumed to be  the same as that  at $z=0.1$.
  Two cases are  considered, one assuming $\sigma_c(z) =  0.173$ and the other
  $\sigma_c(z) = \max[0.173,0.2z]$. Note  that these two assumptions give very
  different   predictions  for  the   high-$z$  SMFs   at  the   massive  end,
  demonstrating that  $\sigma_c(z)$ needs to  be treated properly in  order to
  obtain reliable predictions for the high-$z$ SMFs at the massive end.

\item Finally,  as demonstrated by  the similar $\chi^2$ values  for different
  $\Lambda$CDM models, we  note that the clustering of  galaxies alone can not
  put stringent constraints on cosmological parameters.

\end{itemize}

These  results  clearly  demonstrate  that  the  observed  SMFs  at  different
redshifts,  and  the  CSMFs and  2PCFs  at  low-$z$  can all  be  accommodated
self-consistently within  the current $\Lambda$CDM model,  which specifies the
mass  and bias  functions of  dark matter  host halos,  the mass  function and
accretion  times of  dark  matter subhalos,  and  the orbital  decay times  of
subhalos within their hosts.

%
\begin{figure*}
\plotone{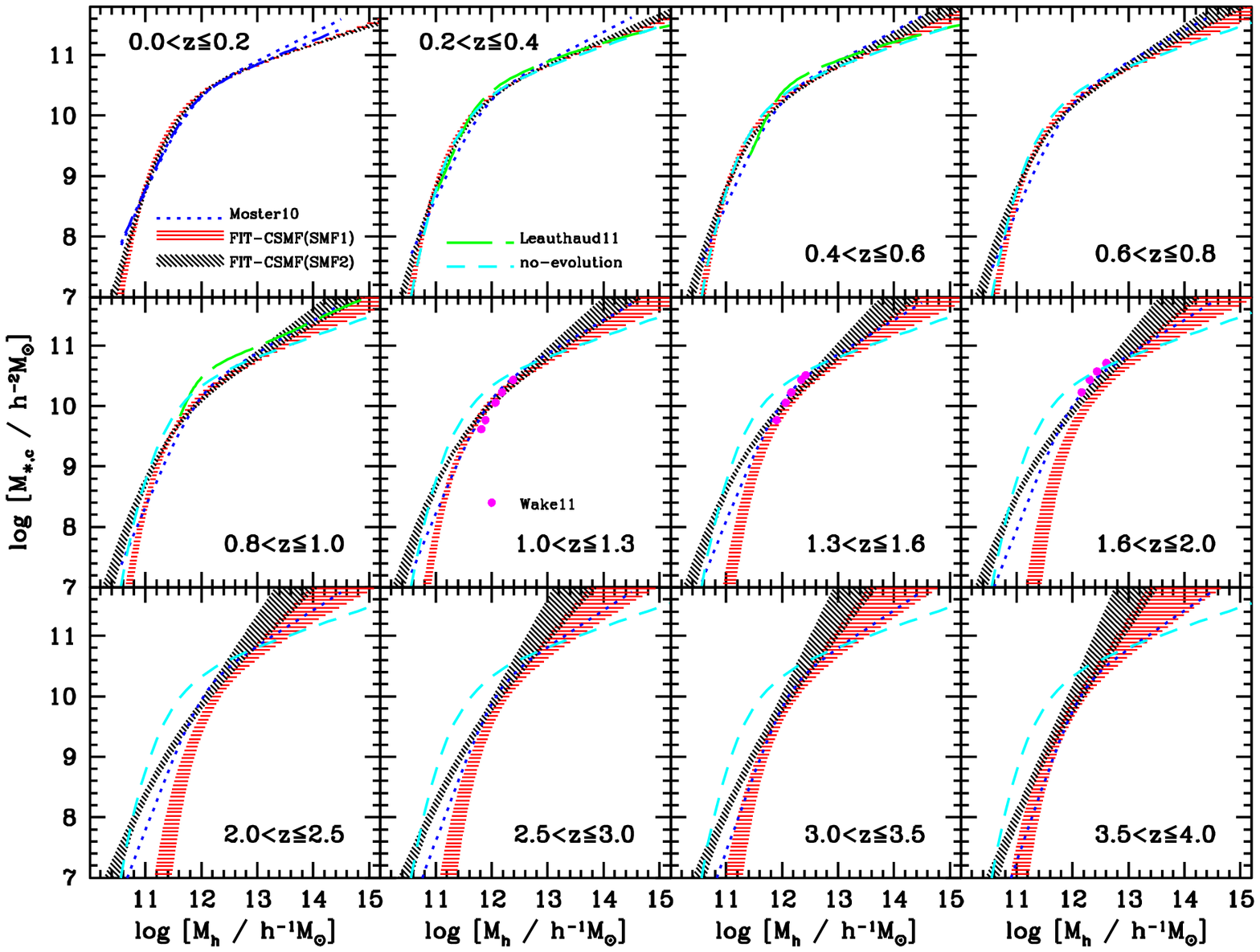}
\caption{The $M_{\ast,c}$ - $M$ relations in different redshift bins.  In each
  panel we compare the predictions of model FIT-CSMF using SMF1 and using SMF2
  (shaded areas  for the 68\%  confidence levels as  indicated), respectively.
  For  comparison,  we  also  show   in  each  panel  the  prediction  of  the
  ``no-evolution model'', which  simply is the $M_{\ast,c}$ -  $M$ relation at
  $z=0.1$ (using SMF1; dashed line), and the model predictions of Moster \etal
  (2010; dotted line).  In addition,  for three redshift bins we also indicate
  the model predictions obtained  by Leauthaud \etal (2012; long-dashed lines)
  and  Wake \etal (2011;  filled dots),  respectively.}
\label{fig:highz_MM2}
\end{figure*}
%


\section{Implications for the Assembly of Galaxies}
\label{sec:evolve}

The main ingredient of our model is a description of the relation between halo
mass and the stellar mass of its central galaxy, as a function of redshift. In
the previous section  we have seen how this relation  can be constrained using
observational data. We now investigate  what our models imply for the assembly
of galaxies.

\subsection{The Galaxy Mass - Halo Mass Relation}
\label{sec:gmhm}

The  upper panels  of Fig.~\ref{fig:highz_MM}  show  the $M_{\ast,  c}$ -  $M$
relations  at different  redshifts predicted  by model  FIT-CSMF  (results for
FIT-2PCF are fairly  similar).  The lower panels show  the same relations, but
this  time in  terms  of the  mass  ratio $M_{\ast,c}/M$.   Results are  shown
separately for models that are constrained by SMF1 (left-hand panels) and SMF2
(right-hand panels).  For comparison, in Fig.~\ref{fig:highz_MM2} we also plot
the  results obtained from  SMF1 (red  dot-dashed lines)  and SMF2  (red solid
lines) at each redshift in the  same panel, together with a no-evolution model
(cyan dashed line) and two models from the literature (Moster \etal 2010, blue
dotted lines, and Leauthaud \etal 2012, green long-dashed lines).

At  all redshifts  the central  stellar mass  increases with  halo  mass.  For
massive halos with $M\ga 10^{13}\msunh$, the (average) central stellar mass is
actually more  massive at  higher redshift. Note,  though, that this  does not
mean that the stellar mass of a particular central galaxy decreases with time.
After all, the  main branch mass of a halo also  increases with time. However,
it {\it does} mean that the central stellar mass cannot grow proportionally to
halo  mass.   The  straight   dashed  lines  in  Figs.~\ref{fig:highz_MM}  and
~\ref{fig:highz_MM2}  mark  a stellar  mass  of $10^{11.8}h^{-2}\Msun$,  which
roughly  corresponds to  the  largest stellar  mass  in the  SDSS survey  (see
Fig.~\ref{fig:redshift}). If  galaxies more massive than this  are absent from
the local  universe, it is  reasonable to assume  that there are also  no such
galaxies at higher redshifts. Note that this is consistent with the SMFs shown
in Figs.\ref{fig:SMF_fit1}  and \ref{fig:SMF_fit2}.  Unfortunately,  since the
high-mass ends  of the  high-$z$ SMFs are  uncertain (as massive  galaxies are
rare  and  current survey  volumes  are relatively  small)  and  can be  quite
significantly impacted by  the actual scatter in the  stellar mass estimations
(see   the   cyan,   dot-dashed   curves   in   Figs.~\ref{fig:SMF_fit1}   and
\ref{fig:SMF_fit2} for an  illustration), the $M_{\ast, c}$ -  $M$ relation is
not  well constrained  at the  very  massive end.   Nevertheless, despite  the
uncertainties, the observational data clearly indicate that the central galaxy
of a  high-mass halo  at high-$z$ (usually  too rare  to be observed)  is more
massive than that of a low-$z$ halo of  the same mass.  As we will see in more
detail  in Section~\ref{sec:growth},  this also  means that  massive galaxies,
currently  residing in  massive halos,  must have  formed early  in progenitor
halos with  much lower masses.  The  stellar masses of  these massive galaxies
have increased little during the  later growth of their host halos, indicating
that star  formation in these  galaxies must have  ceased at early  times, and
that they have not cannibalized too many (massive) satellites.

For low-mass  halos, especially those with  $M \la 10^{12}\msunh$,  there is a
huge increase  in the  stellar mass  for a given  halo mass  from $z\ga  2$ to
$z\sim 0$.  This increase is more than a factor of 10 for SMF1 and a factor of
more  than $3$ for  SMF2 (see  Fig.~\ref{fig:highz_MM2}).  This  has important
implications for  the star  formation efficiencies in  low mass halos  at high
$z$.   Afterall, the  $M_{\ast,c}/M$ ratio  of low  mass halos  at low  $z$ is
already  far  below  the  universal  baryon  fraction  (see  lower  panels  of
Fig.~\ref{fig:highz_MM}). If  this ratio is a  factor 3 to 10  lower at higher
redshifts, it indicates  that star formation in these  halos must be extremely
inefficient. As  to be discussed in more  detail in Section~\ref{sec:summary},
this   poses  a   serious  challenge   to   our  current   theory  of   galaxy
formation. Although it is standard  practice to invoke supernova feedback as a
feedback mechanism to suppress star formation in low mass haloes, this process
is not  expected to be very efficient  in haloes with $M  \gta 10^{11} \msunh$
(e.g.,  Dekel \& Silk  1986), and  it appears  that additional  mechanisms are
required to explain the low  inferred star formation efficiencies at high-$z$.
Alternatively,  one might  question whether  the  data for  faint galaxies  is
sufficiently reliable.  After  all, the SMFs at the  low-mass ends obtained by
D05 are  significantly higher than those  by PG08.  However, as  one can infer
from a comparison  between the data points and the  cyan, dot-dashed curves in
Figs.~\ref{fig:SMF_fit1} and~\ref{fig:SMF_fit2}, the `no-evolution' assumption
(assuming the $M_{\ast, c}$ - $M$ relations at high-$z$ to be the same as that
at $z=0.1$) over-predicts the high-$z$  stellar mass functions at the low-mass
ends by very large amounts, about a factor of 5 for SMF1 and about a factor of
2  for  SMF2  at $\log (M_{\ast}/\msunhh)\sim  10$  
at  redshift  $z\sim 2.0$.   Such  large
discrepancies are  difficult to  be explained by  observational uncertainties.
Indeed, as shown  in Marchesini \etal (2009), the SMFs of  D05 at the low-mass
ends  are already  the highest  among a  handful of  independent observations.
Hence,  the implied  low  star  formation efficiencies  in  low-mass halos  at
high-$z$ have to be taken serious.

\subsection{Comparison with Previous Studies}
\label{sec:comparison}

For  completeness, we  compare  our  results regarding  the  evolution of  the
$M_{\ast, c}$-$M$  relation to a few  previous studies. Note that  in order to
carry  out  proper  comparisons,  one  needs  to  take  into  account  various
systematics  among different analyses.   First, most  of the  analyses, except
perhaps those based  on gravitational lensing, have to  assume a cosmology, so
that the  result is  cosmology dependent  (see van den  Bosch et  al.  2003b).
Second, different  SED (Spectral Energy Distribution) fitting  codes may yield
significantly  different   stellar  masses  even   with  the  same   IMF  (see
e.g. Behroozi et al. 2010). Finally, the uncertainty in the stellar masses may
be redshift dependent (e.g.  Drory et al. 2005).

At low redshift,  the $M_{\ast, c}$-$M$ relation has been  studied not only in
the framework of HOD/CLF, but also with the use of weak gravitational lensing,
X-ray observations, and satellite kinematics.   As shown in More et al. (2009)
and Leauthaud et al. (2012),  these measurements are in general agreement with
each other.  As we have  shown above, our  results at redshift $z=0.1$  are in
good  agreement with  those  of  Yang et  al.  (2008), and  thus  are also  in
agreement with other  results (see e.g., More et al. 2009).   Here we focus on
comparison of results at high redshift.

Moster \etal  (2010) used a method  that is very similar  to subhalo abundance
matching, and stellar  mass functions of Fontana \etal  (2006) covering a wide
range in  redshifts, to  constrain the galaxy-dark  matter connection  from $z
\sim 4$ to  the present. The resulting $M_{\ast,c}$ -  $M$ relations are shown
as blue,  dotted lines in  Fig.~\ref{fig:highz_MM2}.  At low  redshift, Moster
\etal  used the  stellar mass  function  of Panter  et al.   (2007), which  is
actually very similar to  the SMF (for SDSS DR7) we adopt  (see Fig.  2).  The
significantly higher $M_{\ast,c}$ at the massive end obtained by Moster et al.
(2010) is likely due to the  following two factors. First, the model of Moster
et al. (2010)  assumes zero scatter in the $M_{\ast,c}$ -  $M$ relation.  As a
demonstration, we show their  model predition assuming $\sigma_c=0.15$ instead
of  $\sigma_c=0$ in  the  top-left panel  of  Fig.~\ref{fig:highz_MM2} as  the
dot-dashed  line.   As  one  can  see, assuming  a  finite  dispersion  indeed
significantly suppresses the  slope of the $M_{\ast,c}$ -  $M$ relation at the
high mass end. Second, Moster et al.'s results are based on abundance matching
using a  particular $N$-body simulation of  a box size  of $L=100\,{\rm Mpc}$,
and may  be affected by box-size  effect, particularly for  massive halos.  It
might be that massive halos are under-represented in their simulation, so that
their  model has  to  assign massive  galaxies  to halos  with relatively  low
masses.  At high redshift, Moster \etal used a set of SMFs obtained by Fontana
\etal (2006),  where uncertainties  in the estimated  stellar masses  might be
significant.  Nevertheless,  if we  use the difference  in the  predictions of
SMF1 and  SMF2 as an indication  of the observational  uncertainty, the Moster
\etal results are consistent with ours.

We also compare  our results to those of Leauthaud \etal  (2012), at least for
the three  redshift bins  $0.2<z\le 0.4$, $0.4<z\le  0.6$, and  $0.8<z\le 1.0$
(green  long-dashed curves  in Fig.~\ref{fig:highz_MM2}).   Rather  than using
subhalo  abundance  matching, Leauthaud  \etal  used  galaxy number  densities
(SMFs),  galaxy clustering  and  galaxy-galaxy lensing  in  the COSMOS  survey
(Scoville \etal 2007), in order to constrain the galaxy dark matter connection
from  $z  \simeq  0.2$  to  $z  \simeq 1.0$.   Since  they  parameterized  the
occupation statistics of centrals  and satellites separately, they effectively
allow for  a dependence on subhalo  accretion times, and for  evolution in the
satellite population  after accretion.  A  comparison of their  $M_{\ast,c}$ -
$M$ relations to ours shows that  their stellar masses for central galaxies at
a given halo  mass are generally very similar to ours,  especially for the two
low  redshift  bins.   For  the  highest redshift  bin,  there  is  noticeable
discrepancy  in   the  intermediate  halo  mass   range  ($M\sim  10^{12.0}$).
Unfortunately the origin of the discrepancy is not clear.

Finally, we also compare our results to those obtained by Wake \etal (2011) in
three redshift bins: $1.0 < z \leq 1.2$, $1.2 < z \leq 1.6$, and $1.6 < z \leq
2.0$ (filled dots  in Fig. 12). These data are obtained  by fitting HOD models
to  the abundance and  clustering of  galaxies, as  function of  stellar mass,
obtained  from the  NEWFIRM Medium  Band Survey  (van Dokkum  \etal  2009).  A
comparison with our results shows  that their $M_{\ast,c}$ - $M$ relations are
in quite good  agreement with our results.  Note however, as  pointed out in a
recent paper by Leauthaud et  al.  (2011), that using different functional form
for the HOD model might impact the average halo masses that host given stellar
mass galaxies as well.

\begin{figure*}\plotone{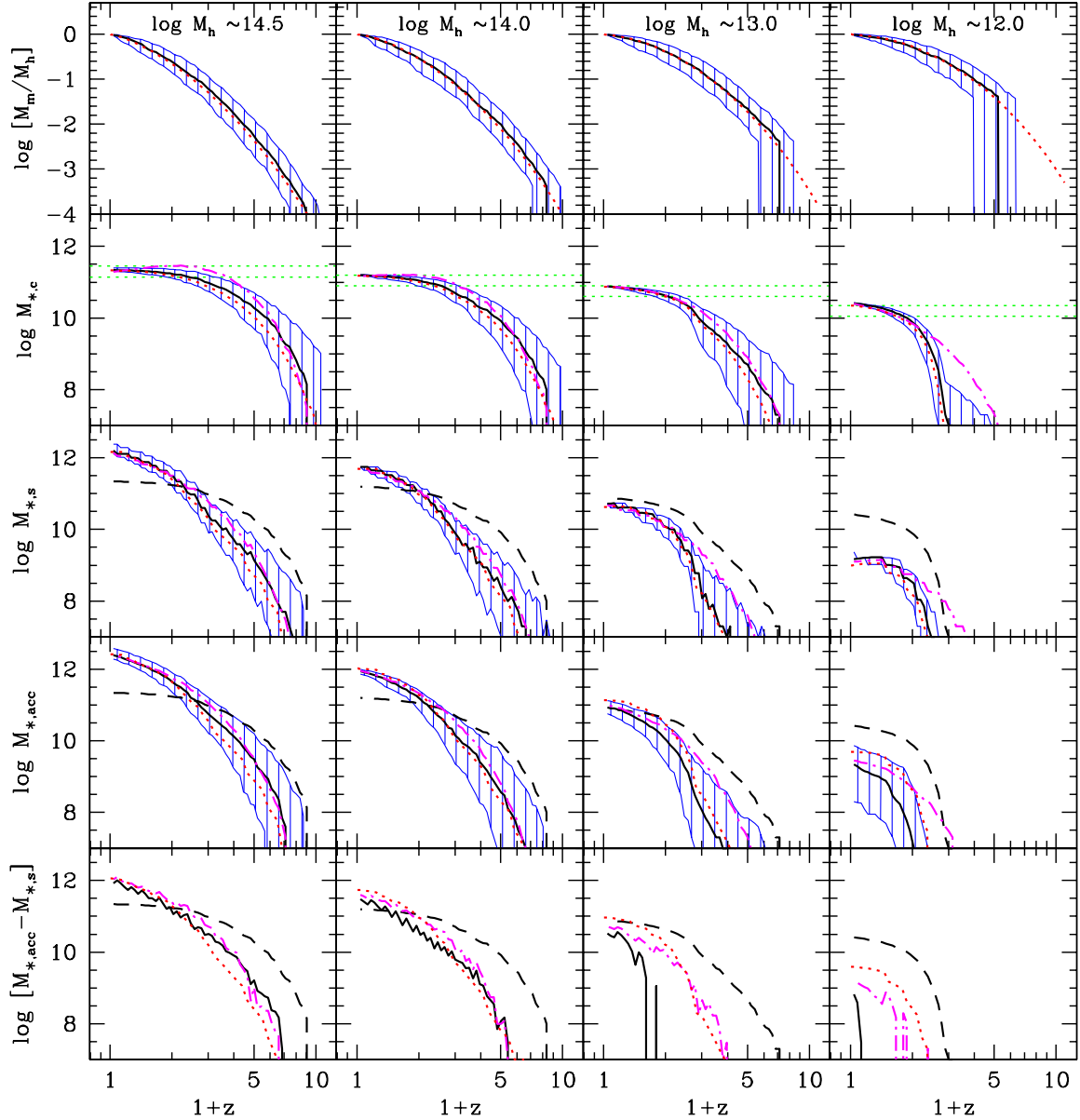}
\caption{The  growth of host  (main branch)  halo mass  (top row  panels), the
  stellar mass  of central  galaxies (second row),  the total stellar  mass in
  surviving  satellite galaxies (third  row), the  total stellar  mass brought
  into the main branch by sub-halos  (fourth row), and the stellar mass of the
  ``disrupted''  component  (bottom row,  see  text  for  definition), all  as
  function  of  redshift.  These  results  are  obtained  using  the  best-fit
  parameters of  model ID=10, listed in  Table~\ref{tab:fitCSMF1} (using WMAP5
  cosmology and SMF1).  Results in  different columns correspond to host halos
  with masses centered at $\log (M/\msunh)  = 14.5$, 14.0, 13.0 and 12.0, from
  left to  right.  The solid line and  shaded area in each  panel indicate the
  median and  68\% percentile obtained  from 200 simulated halo  merger trees.
  The dotted  line is  the prediction obtained  using the analytical  model of
  Y11.  The two horizontal dotted lines in the second row indicate the maximum
  and half-maximum values of the  median stellar mass of the central galaxies.
  The dashed lines  in the third, fourth  and bottom rows are the  same as the
  solid  lines in  the top  panels.  Finally,  for comparison,  the dot-dashed
  lines (in all rows except the  top one) are model predictions obtained using
  the best  fit parameters of  model ID=10 listed  in Table~\ref{tab:fitCSMF2}
  (using WMAP5 cosmology and SMF2).}
\label{fig:merger}
\end{figure*}

\begin{figure*}\plotone{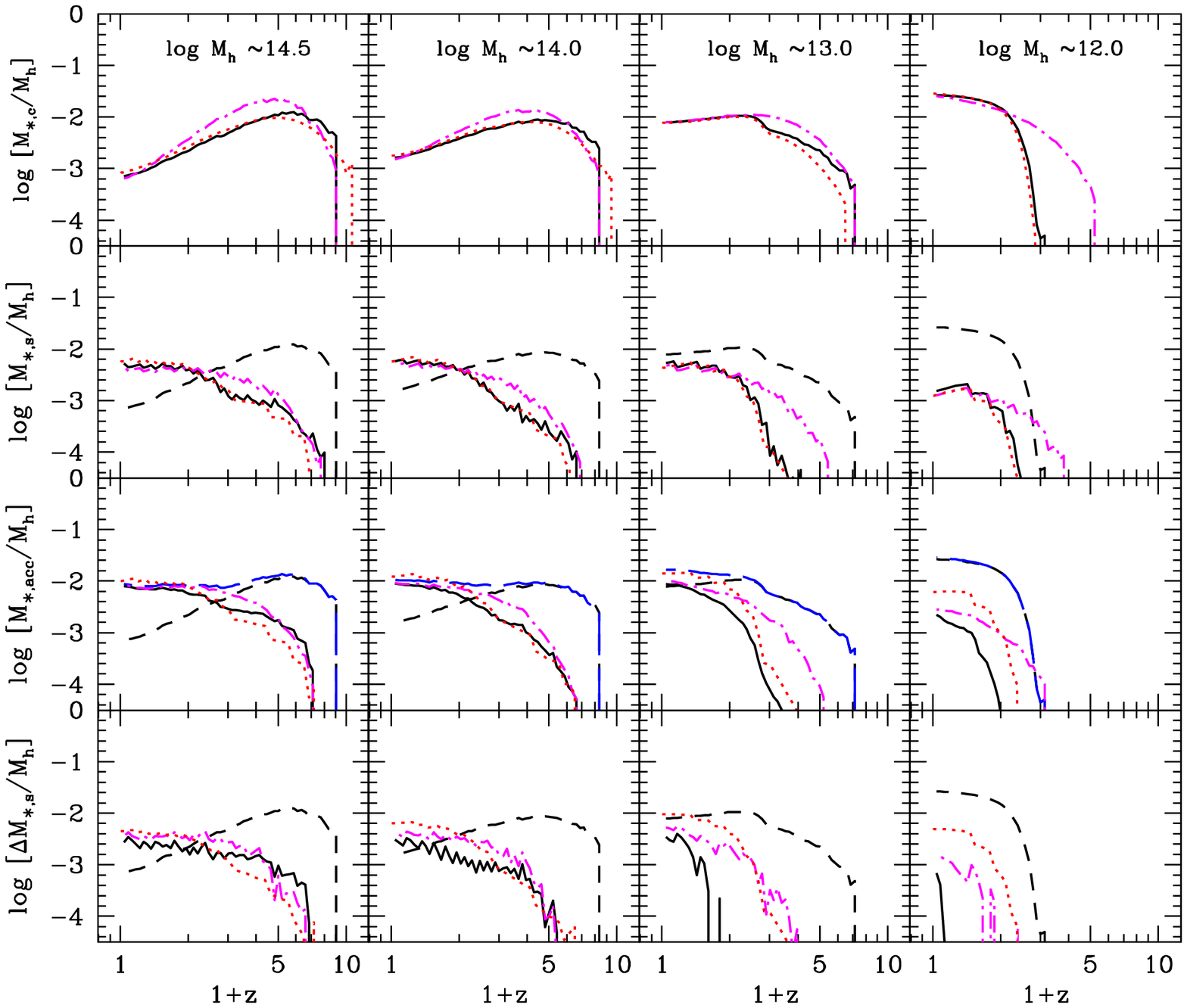}
\caption{Similar to the lower four rows of Fig.~\ref{fig:merger}, but this
  time we have normalized all stellar masses by the median, instantaneous
  main-branch halo mass, $M_\rmh$ (the solid lines in the top panels of
  Fig.~\ref{fig:merger}); i.e., the various lines indicate the {\it median}
  stellar-to-halo mass ratios (in units of $h^{-1}$). From the top row to the
  bottom row, the results are shown for the stellar mass in central galaxies
  (first row), the total stellar mass in surviving satellite galaxies (second
  row), the total stellar mass brought into the main branch by sub-halos
  (third row), and the stellar mass of the ``disrupted'' component (bottom
  row).  Line styles are the same as in Fig.~\ref{fig:merger}.}
\label{fig:ratios}
\end{figure*}

\subsection{The Growth of Stellar Components in Dark Matter Halos}
\label{sec:growth}

Using the redshift-dependent  CSMFs obtained above, we can  predict the growth
of the  stellar masses of both  central and satellite galaxies  along the main
branch of their dark matter halos.  The median stellar mass at redshift $z$ of
a central galaxy  that at redshift $z_0 \leq  z$ is located in a  host halo of
mass $M_0$ can be written as 
\begin{equation}\label{cengrowth}
M_{\ast, c} (z|M_0,z_0) = M_{\ast, c} (M_a,z)\,.
\end{equation}
Here, as before, $M_a$ is the median mass at redshift $z \geq z_0$ of the main
progenitor of  a host halo of mass  $M_0$ at redshift $z_0$.  The median, {\it
  total} stellar mass  of surviving satellite galaxies in  the main branch can
be obtained by integrating the CSMF of satellite galaxies: 
\begin{equation}\label{satgrowth}
 m_{\ast, s} (z|M_0,z_0) = \int \rmd \log m_{\ast} \, m_{\ast} \,
 \Phi_\rms(m_{\ast}|M_a,z) \,.
\end{equation}
Thus,  once  we know  the  assembly  history of  a  dark  matter  halo, it  is
straightforward  to  use  our  model  to  obtain  the  corresponding  assembly
histories of the stellar components  of its central and satellite galaxies. In
addition,  we also  examine what  the  models predict  for the  total mass  of
satellite  galaxies  that have  been  accreted  onto  the main  branch,  which
includes the surviving satellites as well as those that have been cannibalized
by the central or  disrupted by the tidal field. Note that  here we have taken
into account  the evolution of  the satellite galaxies after  their accretion,
e.g.  the addtional star formation  and passive evolution, using the parameter
$c$. This component is thus given by
\begin{eqnarray}\label{satacc}
m_{\ast, {\rm acc}} (z|M_0,z_0) = \int_{z_0}^{z}
{\rmd z_a \over (1+z_a)} \int \rmd \log m_{\ast} \, m_{\ast} ~~~~~~~~~~&&\\
\Phi_\rme(m_{\ast}|m_a,z_a,z)  \,
n_{\rm sub}(m_a,z_a|M_0,z_0)\,.~~~~~~~~~~~~~~\nonumber
&&
\end{eqnarray}

Here we adopt  two different methods to obtain the  assembly histories of dark
matter halos. The first is based  on halo merger trees extracted directly from
a large $N$-body simulation. The second  uses the analytical model of Y11 (see
also Section \ref{ssec:subhalo}), which relies  on the fitting formula of Zhao
\etal (2009) to describe the mean  assembly history for halos of a given final
mass and  adopts a  lognormal dispersion around  the mean main-branch  mass at
each  time. As  shown in  Y11, this  analytical model  accurately  matches the
simulation results averaged over halos of a given final mass. However, it does
not fully take into account the  variance among different halos, as it ignores
the  covariance  of  the  masses  at different  times  along  individual  main
branches. Using the merger trees extracted from the numerical simulations does
not suffer  from this shortcoming and  therefore allows us to  get an estimate
for the  halo-to-halo scatter,  taking full account  of the variance  in their
assembly histories.
 
The simulation  used here is the  $300\mpch$ simulation box  described in Y11,
which  evolves  $1024^3$ dark  matter  particles  in  a WMAP5  cosmology  from
redshift $z=100$  to the  present epoch ($z=0$)  using the  massively parallel
GADGET2 code  (Springel et al. 2001;  2005).  The particle  mass and softening
length are  $1.87\times10^9\msunh$ and  $6.75\kpch$, respectively. A  total of
100  outputs  are  available  between   $z=50$  and  $z=0$,  spaced  in  equal
$\log(1+z)$ intervals.  Dark matter halos were identified at each output using
the standard  friends-of-friends (FOF)  algorithm (Davis et  al. 1985)  with a
linking length of  $0.2$ times the mean interparticle  separation.  Only halos
with at  least 20 particles  are used.  Based  on halos at  different outputs,
halo merger trees were constructed.  A halo in an earlier output is considered
to be a progenitor of the present  halo if more than half of its particles are
found in  the present halo.  The  main branch of  a merger tree is  defined to
consist  of all the  progenitors as  one climbs  from the  bottom to  the top,
choosing  always the  most massive  branch  at every  branching point.   These
progenitors are  the main-branch progenitors,  and the time dependence  of the
main branch mass is the assembly history of the halo.  We extracted 200 merger
trees  for present-day  halos  in each  of  the mass  bins  centered at  $\log
(M/\msunh) = 12.0$, 13.0, 14.0 and 14.5.  The median and the 68\% range of the
distribution of assembly histories thus  obtained are indicated as solid lines
and  hatched   areas,  respectively,  in  the   panels  in  the   top  row  of
Fig.~\ref{fig:merger}.   For comparison,  the dotted  lines indicate  the mean
assembly histories obtained  using the analytical model of  Zhao \etal (2009),
which  are in  excellent agreement  with the  solid lines.   In  general, more
massive  halos   assemble  later  than   smaller  ones,  which   reflects  the
hierarchical nature  of structure  formation in the  CDM model (e.g.,  van den
Bosch 2002;  Wechsler \etal 2002).  Note that, as  we have tested  our results
using another  set of  halo merger trees  constructed with a  different method
from simulations  of the WMAP1 cosmology  (Han \etal 2011), we  found that our
results  are robust  against the  change  in cosmology  and in  the method  of
merger-tree construction.

In the following subsections, we  use these halo assembly histories to examine
in detail how  the different stellar components in  dark matter halos (central
galaxies, satellite galaxies, and halo stars) grow with time.

\subsubsection {Central Galaxies}

Using   model   FIT-CSMF   for   the   WMAP5  cosmology   (ID=10   in   Tables
\ref{tab:fitCSMF1} and~\ref{tab:fitCSMF2}), we populate  the host halos in the
simulation  outputs  with  central  galaxies\footnote{We have  also  used  the
  FIT-2PCF models (ID=9) instead, and found very similar results. For the sake
  of brevity, we  will only show results for model  FIT-CSMF.}. Here we ignore
the scatter in  the central stellar mass - halo  mass relation, simply because
it is  relatively small and  because it is  difficult to take it  into account
properly owing to possible covariances  between the stellar masses in the main
branch of a halo  at different times. The results are shown  in the second row
of  Fig.~\ref{fig:merger},  where  different  panels correspond  to  different
present-day halo masses, as indicated in  the top panels.  The solid lines and
shaded   areas  indicate   the  median   and  68   percentiles   of  $M_{\ast,
  c}(z|M_0,z_0)$ obtained from 200 halo merger trees. For comparison, the red,
dotted   curves  show   the   $  M_{\ast,   c}   (z|M_0,z_0)$  obtained   from
Eq.~(\ref{cengrowth}). Note  that these agree extremely well  with the results
obtained from the merger trees in the numerical simulations.

In each panel, the two horizontal doted lines indicate the maximum and half of
the maximum of the median stellar mass of the central galaxies at $z=0$. These
can be used  to read off when the central galaxies,  on average, acquired half
their present day  stellar masses; for centrals in  present-day halos of $\log
(M/\msunh) \sim 12.0$, 13.0, 14.0 and  14.5 this occurs at $z\sim 0.5$, $1.0$,
$1.5$  and $2.0$,  respectively.  Thus,  although more  massive  halos assembe
their masses  later, their central  galaxies actually assembled  {\it earlier}
than centrals in lower-mass halos (see also Conroy \& Wechsler 2009). Note how
the stellar  mass assembly  histories of central  galaxies reach a  plateau in
halos with $M \ga 10^{13}\msunh$,  indicating that central galaxies in massive
halos  grow very  little in  stellar mass  at low  redshifts.   In particular,
central galaxies in clusters with $M  \sim 10^{14.5} \msunh$ have not grown in
mass  (significantly) since  $z\sim  1.0$. In  contrast,  central galaxies  in
Milky-Way sized halos with $M  \sim 10^{12}\msunh$ continue to grow in stellar
mass down to the present day.   The slow growth of central galaxies in massive
halos suggests not  only that their star formation must  have been quenched by
some processes,  but also  that their  increase in stellar  mass due  to the
accretion of satellite galaxies cannot  have been signficant.  As discussed in
Section~\ref{sec:summary},   this   has   important   implications   for   our
understanding of galaxy formation and evolution.

Another interesting  way to depict the  growth of the stellar  mass of central
galaxies is  to show the  ratio between stellar  mass and host halo  mass (the
main branch mass) at the redshift  in question.  These curves are shown in the
top  panels  of  Fig.~\ref{fig:ratios}  using  the  same  line  styles  as  in
Fig.~\ref{fig:merger}.   As one  can see,  central galaxies  in host  halos of
different final masses reach a maximum of about $(0.015 - 0.03) h^{-1}$, quite
independent of  the final host halo  mass. However, the maximum  is reached at
very different redshifts of $z\sim 5$, $4$, $1.5$ and $0$ for $\log (M/\msunh)
\sim 14.5$, $14.0$, $13.0$, and $12.0$, respectively. Interestingly, all these
redshifts  roughly  correspond  to  a   similar  main  branch  mass  of  $\sim
10^{12}\msunh$.
 
Finally,  in order  to  illustrate the  effects  of the  uncertainties in  the
high-$z$    SMF    measurements,   the    magenta    dot-dashed   curves    in
Figs.~\ref{fig:merger}  and~\ref{fig:ratios} show  the  median stellar  masses
obtained using the  D05 SMFs (case SMF2, the best-fit  parameters of which are
listed under ID=10 in Table~\ref{tab:fitCSMF2}). The results are qualitatively
the same as  those obtained from SMF1. However, the  stellar masses of central
galaxies in  low-mass halos  at high-$z$ are  higher than those  obtained from
SMF1. This  simply reflects  that the high-$z$  SMFs of D05  are significantly
steeper at the low mass end than the PG08 SMFs (see Fig.~\ref{fig:highz_SMF}).
\begin{figure*}
\plotone{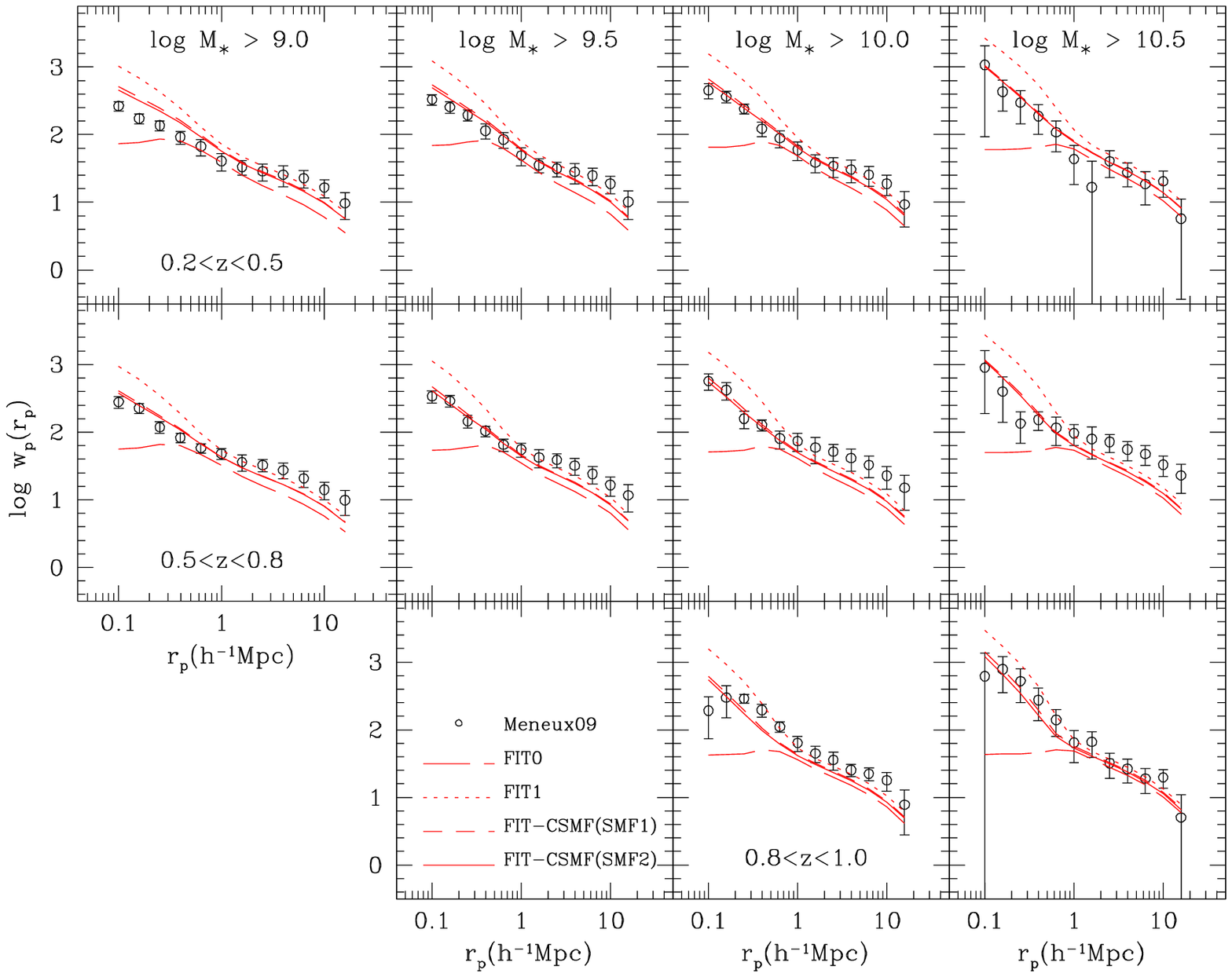}
\caption{The  projected  2PCFs for  galaxies  in  different  stellar mass  and
  redshift bins.  Here we compare  model predictions (various lines)  with the
  observational  measurements obtained  by  Meneux \etal  (2009) from  zCOSMOS
  (open circles  with error-bars). The  long-dashed, dotted, and  dashed lines
  show the predictions  of models FIT0, FIT1, and  FIT-CSMF, respectively, all
  of which have used SMF1 as  constraints for the stellar mass functions.  For
  comparison, the  solid lines show  the predictions for model  FIT-CSMF using
  SMF2.}
\label{fig:highz_wrp}
\end{figure*}
\begin{figure*}
\plotone{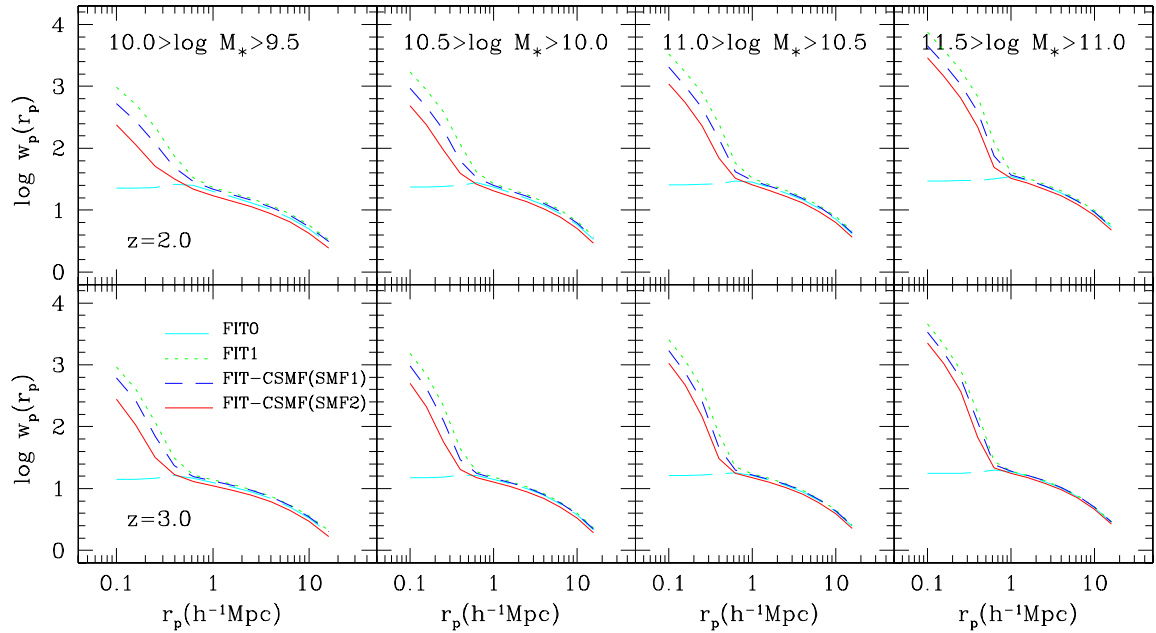}
\caption{Predictions for the projected 2PCFs in different stellar mass bins at
  redshifts $z=2.0$  (upper panels) and  $z=3.0$ (lower panels).   Results are
  shown for  the same  four models as  in Fig.~\ref{fig:highz_wrp},  using the
  same line-styles, as indicated.}
\label{fig:highz_wrpp}
\end{figure*}

\subsubsection {Satellite Galaxies}

Next we  use the  merger trees in  our simulation  box to investigate  how the
population of surviving satellite galaxies grows in stellar mass.  The results
are shown in the third row  of panels in Fig.~\ref{fig:merger}. Once again the
solid  lines  and shaded  areas  indicate the  median  and  68 percentiles  of
$m_{\ast, s}(z|M_0,z_0)$  obtained from 200  halo merger trees, while  the red
dotted  lines indicate  the $\langle  m_{\ast,  s}\rangle(z|M_0,z_0)$ obtained
from  Eq.~(\ref{satgrowth}) using  the analytical  model of  Y11 for  the halo
assembly history.  The black, dashed  lines indicate the median of the stellar
mass of  central galaxies, taken  from the panels  in the second row,  and are
shown  for comparison.   As one  can see,  the total  mass of  satellites only
becomes  larger than  that  of the  central  galaxy in  massive  halos at  low
redshifts.  In  Milky-Way sized halos  the total satellite mass  never exceeds
$\sim 10\%$  of the central mass.   For all halos,  satellite galaxies contain
less than a few  percent of the total stellar mass at  all $z\ga 3$. The total
satellite mass in a massive halo continues  to grow all the way to the present
time, whereas  the central galaxy  ceases to grow  once the halo  mass reaches
$\sim 10^{12} \msunh$.  As shown by  the magenta, dot-dashed lines, the use of
SMF2  predicts higher  satellite  masses at  high-$z$  than the  use of  SMF1,
especially  in low-mass  halos. Once  again this  reflects the  fact  that the
high-$z$  SMFs  of D05  are  steeper  at the  low  mass  end  than their  PG08
counterparts.

The ratios between the total satellite  mass and the main branch halo mass are
shown in the second row  of Fig.~\ref{fig:ratios}.  For low-mass halos and for
massive halos at early times, this ratio increases rapidly with time, implying
that the total satellite mass grows  faster than the halo mass. The reason for
this is that  for a given halo  mass, the stellar mass of  central galaxies in
low-mass   halos  increases   rapidly   with  time   (see   upper  panels   of
Fig.~\ref{fig:ratios}), and  so the  ratio between the  stellar mass  and dark
matter mass of  the accreted matter increases rapidly  with time.  For massive
halos,  the  ratio  flattens  at   low-$z$  to  a  value  $\sim  0.002h^{-1}$,
corresponding to a  halo mass to stellar  mass ratio of $\sim 200  h$ which is
close to the universal mass-to-light ratio.

\subsubsection{Halo Stars and Star Formation in Central Galaxies}
\label{sec:SF}

The third component we consider is the total stellar mass that is brought into
the main branch by all satellite galaxies at the time of their accretion. Here
we  have  taken into  account  the later  evolution  effect  of the  satellite
galaxies, such as the  addtional star formation and passive  evolution. This
mass contains  not only  the mass of  the `surviving' satellite  galaxies, but
also that of the satellites that have been cannibalized or disrupted. The mass
growth  histories of  this  mass component  are  shown in  the  fourth row  of
Fig.~\ref{fig:merger}, where as before the solid lines and shaded areas depict
the medians and  68 percentiles obtained from 200  merger trees, respectively.
The    dotted   lines    show    the   results    obtained   by    integrating
Eq.~(\ref{satacc}). The  corresponding ratios between  this mass  component and
the  main  branch  halo  mass  are  shown  in  the  third  row  of  panels  of
Fig.~\ref{fig:ratios}. In halos with $M \la 10^{13}\msunh$, the total accreted
stellar  mass  is   roughly  similar  to  that  of   the  surviving  satellite
population. In halos  with $M \ga 10^{14}\msunh$, however,  the total accreted
stellar masses are significantly larger  than those in the surviving satellite
population, especially at low redshift ($z\la 1.0$). This is more clearly seen
in the  bottom rows of Fig.~\ref{fig:merger}  and~\ref{fig:ratios}, which show
the differences  between the median accreted  mass and the median  mass of the
surviving  satellites. This  indicates  that in  massive  halos a  significant
fraction of  the total  stellar mass  brought in by  subhalos is  actually not
associated  with  surviving  satellite   galaxies  and  must  exist  in  other
forms. For  convenience, we will refer  to this stellar mass  component as the
``disrupted'' component.

There  are  two possible  fates  for the  disrupted  component:  either it  is
accreted by the central galaxy or it  gives rise to a population of halo stars
(also  referred to  as intracluster  stars,  or ICS,  in the  case of  massive
halos).  Unfortunately,  our model cannot uniquely  discriminate between these
two components, as  the mass of the central galaxy can  also grow through star
formation.   However, we  can obtain  important constraints  by  comparing the
growth  of the disrupted  component to  that of  the central  galaxies (black,
dashed  lines).   This  shows that  central  galaxies  in  halos with  $M  \la
10^{14}\msunh$ must  have grown mainly  through {\it in situ}  star formation,
simply because the total mass in  the disrupted component is much smaller than
that of the  central. For more massive halos, however,  the situation could be
very  different. Here the  total stellar  mass in  the disrupted  component is
actually larger than  that of the central galaxy. Hence,  the existence of ICS
is inevitable;  the total stellar  mass locked up  in the stellar halo  can be
more than 2  times as massive as the central galaxy  at redshifts $z<1.5$ (see
also Purcell  \etal 2007).  As  one can  see from the  left two panels  in the
bottom row of Fig.~\ref{fig:merger}, although  the total mass of the disrupted
fraction is sufficient to form the central galaxy in massive halos, in reality
central galaxies in  these halos must have acquired  their stellar mass mainly
through {\it  in situ}  star formation, rather  than through the  accretion of
satellites  (see also results along  this line  
obtained from hydro-dynamical  simulations by
e.g., Naab et al. 2009; Oser et  al. 2010; 2012; Hrischmann et al. 2012). This
can be inferred  from the fact that these central  galaxies have already build
up  most of  their  stellar  mass at  high  redshifts, when  the  mass in  the
disrupted component  is too small  to make a dominant  contribution. Accretion
can  only be  significant at  $z\la 2$  in massive  halos, when  the disrupted
component  is  sufficiently massive  compared  to  the  mass of  the  central.
However, at these  low redshifts, the central galaxies  in these massive halos
no  longer seem  to grow  much in stellar mass.  
Hence, even  central galaxies  in
massive halos must have mainly grown via {\it in situ} star formation.
         
Finally, we point  out that the use  of SMF2 leads to masses  in the disrupted
component that  are quite  similar to those  obtained using SMF1.   The higher
total  accreted mass predicted  with SMF2  for low-mass  halos at  high-$z$ is
largely due to the higher mass  in the survived satellites. Note also that the
analytical  model predicts larger  disrupted components  in low-mass  halos at
high-$z$ than  using the  merger trees in  the numerical simulation  (cf solid
black   lines   and    dotted   red   lines   in   the    bottom   panels   of
Figs.~\ref{fig:merger} and~\ref{fig:ratios}).  This is mainly due  to the fact
that small  progenitors are  not well resolved  in the simulation  at high-$z$
(see the top-right panel of Fig.~\ref{fig:merger}).

\subsection{Galaxy Clustering at High Redshift}

With  the model  described  in  Section~\ref{sec:2pcf}, we  are  also able  to
predict   the   correlation   functions   of   galaxies   at   high-$z$.    In
Fig.~\ref{fig:highz_wrp}  we show the  model predictions  for FIT0,  FIT1, and
FIT-CSMF using  SMF1 as well  as for FIT-CSMF  using SMF2 (the  predictions of
FIT-2PCF are similar to those of FIT-CSMF). The model predictions are compared
with the observational results of Meneux \etal (2009), which are shown as open
circles  with errorbars.  As  one can  see, all  our models  under-predict the
correlation  function  on  large  scales,  particularly for  galaxies  in  the
redshift range  $0.5<z<0.8$.  As  pointed out in  Meneux \etal (2009),  and as
briefly mentioned in Section~\ref{sec:gmhm} above, these enhanced correlations
on large scales are  likely due to the fact that the  zCOSMOS field happens to
correspond  to a  high  density  region of  the  Universe.  Nevertheless,  the
results  show that  different models  predict very  different  correlations on
small  scale ($r_\rmpp \la  1\mpch$), once  again demonstrating  that accurate
clustering measurements  on small scales can provide  important constraints on
the  galaxy-dark  matter connection  (in  particular  regarding the  dynamical
evolution of satellite galaxies).

In Fig.~\ref{fig:highz_wrpp} we show model predictions for the projected 2PCFs
at $z=2.0$ (upper panels) and  $z=3.0$ (lower panels).  Here again results are
shown for FIT0, FIT1, and FIT-CSMF using SMF1, as well as FIT-CSMF using SMF2.
Different panels correspond  to galaxies in different stellar  mass ranges, as
indicated.  The projected 2PCFs on  large scales decrease with both increasing
redshift  and decreasing  stellar mass.  Note that  the differences  among the
different models  are extremely small  at large scale ($r_\rmpp  \ga 1\mpch$).
On  small  scales our  models  predict that  the  2PCFs  are greatly  enhanced
relative  to a simple  extrapolation from  large scales,  except for  the FIT0
models which  has zero satellites, and  therefore no one-halo  term. Note that
this enhancement is stronger for more  massive galaxies, and that it is fairly
different  for the different  models.  Hence,  as for  the low-$z$  2PCFs, the
clustering strength  on large  scales does not  provide much  more information
than the  SMFs themselves (see  also Moster \etal  2010), while that  on small
scales does.
   
\section{Summary and Discussion} 
\label{sec:summary}

In this paper we have developed a new and self-consistent model that describes
the  galaxy-dark matter connection  over cosmic  history.  Unlike  the popular
abundance  matching  technique, our  model  takes  account  of the  fact  that
sub-halos in a host halo are  accreted at different times, so that the stellar
masses associated with them are likely  to depend on both their halo masses at
accretion  and  their accretion  times.  In  addition,  our model  allows  for
dynamical  evolution  of  the  satellite  population  (mass  stripping,  tidal
disruption, cannibalism).

We have used  galaxy stellar mass functions observed in  the redshift range $0
\leq  z  \la 4$,  together  with the  conditional  stellar  mass function  and
two-point correlation  functions (both  at $z \simeq  0.1$, and  both obtained
from SDSS),  to constrain the  evolution of the galaxy-dark  matter connection
from $z\sim 4$ to the present. The relation between halo mass and stellar mass
thus obtained is  used, together with simulated and  theoretical halo assembly
histories, to predict  how the masses of different  stellar components in dark
matter halos (central galaxies, satellite  galaxies, and halo stars) evolve as
function of  time. We  also used our  model to  predict the 2PCFs  of high-$z$
galaxies  as a  function of  their stellar  masses. Our  main findings  can be
summarized as follows.  
\begin{enumerate}

\item  Our  model  provides a  reasonable  fit  to  all  the data  within  the
  observational uncertainties,  which indicates that  the current $\Lambda$CDM
  model  is  consistent with  a  wide variety  of  data  regarding the  galaxy
  population across cosmic time.

\item  At  low-$z$,  the  stellar  mass  of  central  galaxies,  $M_{\ast,c}$,
  increases with halo  mass as $M^{0.3}$ and $M^{\ga 4.0}$  at the massive and
  low-mass ends,  respectively. The ratio $M_{\ast,c}/M$ reveals  a maximum of
  $\sim 0.03$  at a  halo mass  $M \sim 10^{11.8}\msunh$.  The fact  that this
  maximum value is much lower than the universal baryon fraction ($\sim 0.17$)
  reflects the overall inefficiency of star formation. At higher redshifts the
  maximum in $M_{\ast,c}/M$ remains close to $\sim 0.03$, but shifts to higher
  halo mass (see also Wake \etal 2011).

\item  The $M_{\ast,c}$-$M$ relations  obtained here are roughly in
  agreement with previous studies (e.g., Moster \etal 2010; Wang \& Jing 2010;
  Leauthaud \etal  2012).  However, our model allows us to put 
  constraints on the satellite population so that we can interpret the
  results in terms of the evolution of galaxies in dark halos. We find
  that low-mass satellite galaxies can significantly 
  increase their stellar masses after accretions by their host halos.  
   
\item The  time-scale for  the disruption of  satellite galaxies is  about the
  same  as the dynamical  friction time  scale of  their subhalos  obtained in
  $N$-body simulations, suggesting that most
  disruption  occurs once  the galaxy  has `depleted'  its orbital  energy, at
  which point it is either accreted by the central galaxy or torn apart by the
  strong tidal field.

\item The  stellar mass  assembly history for  central galaxies  is completely
  decoupled from the  assembly history of its host  halo.  Initially the ratio
  $M_{\ast,c}/M$  increases rapidly  with time,  until the  halo  mass reaches
  $\sim 10^{12} \msunh$, at which point $M_{\ast,c}/M \sim 0.03$. Once $M \gta
  10^{12} \msunh$,  there is little growth in $M_{\ast,c}$,  causing the
  ratio $M_{\ast,c}/M$ to decline.

\item Most  massive centrals  assemble their stellar  mass earlier  than their
  less massive counterparts, even though  their host haloes assemble later. In
  Milky-Way  sized  halos  more than  half  of  the  central stellar  mass  is
  assembled at $z\la 0.5$, while  brightest cluster galaxies had already grown
  to half their present day stellar mass by $z \simeq 2.0$.

\item The accretion of satellite  galaxies contributes little to the formation
  of central  galaxies in small  halos. Hence, most  of their stars  must have
  formed {\it in situ} so that their assembly histories closely resemble their
  star  formation histories.  In massive  halos more  than a half  of the
  stellar mass of the  central galaxy has to be formed {\it  in situ} at $z\ga
  2$.   The accretion  of stars  in satellites  by the  centrals  only becomes
  significant at  $z\la 2$,  when the total  mass available for  the accretion
  reaches a significant fraction of the mass of the central galaxies.

\item In massive  halos, the total mass  in halo stars is more  than two times
  that of the central galaxy, while  in Milky-Way sized halos the mass in halo
  stars is much smaller ($\lta 10\%$) than that of the central galaxy.

\item  The  2PCFs  of galaxies  on  large  scales  do  not provide  much  more
  constraints than  those already  provided by the  stellar mass  functions of
  galaxies (see also  Moster \etal 2010). On small  scales, however, the 2PCFs
  provides important  constraint on how satellite galaxies  evolve within dark
  matter halos.  Future observations of the small scale clustering strength of
  galaxies at  high-$z$ will therefore  be important for our  understanding of
  galaxy formation and evolution.  Our  models predict that the high-$z$ 2PCFs
  on  small scales  are  much steeper  than  at low-$z$,  especially for  more
  massive galaxies.

\end{enumerate}

These  results have  important implications  for our  understanding  of galaxy
formation and evolution. Here we highlight a few issues which we believe to be
particularly interesting.

One of  the implications  of our results  is that  star formation in  low mass
halos at high-$z$ must be  extremely ineffecient.  This basically follows from
the strong  evolution in the  $M_{\ast,c}$-$M$ relation that we  infer between
high-$z$ and  the present at the  low mass end: halos  of $10^{11}\msunh$ host
central galaxies at $z \sim 4$  that are orders of magnitude less massive than
centrals in  halos of  the same mass  at $z=0$  (see Fig.~\ref{fig:highz_MM}).
Note that  even at $z=0$  central galaxies in  halos with $M  = 10^{11}\msunh$
have  a  stellar  mass that  is  only  2-3  percent  of the  universal  baryon
fraction. Hence,  only a  minute fraction of  the baryons associated  with low
mass halos  at high-$z$ are turned  into stars.  It is  typically assumed that
star  formation is  suppressed in  low-mass  halos due  to supernova  feedback
and/or photo-ionization.   However, it has  become clear in recent  years that
either  the standard  treatment  of  these processes  is  inadequate, or  that
additional mechanisms are  needed to help suppress star  formation in low mass
halos.  In  particular, recent observations  have indicated that  the specific
star formation rates (SSFRs) of  star forming galaxies, at fixed stellar mass,
increase with  redshift from $z=0$  until they reach  a plateau at $z  \gta 2$
(see e.g., Stark \etal 2009; Labb\'e et al 2010a,b; Gonz\'alez \etal 2010). As
pointed out  by several studies, this  constancy of the  SSFRs is inconsistent
with  predictions of  `standard' models  for galaxy  formation,  which instead
predict SSFRs  that continue  to increase with  redshift (Lo Faro  \etal 2009;
Dutton, van den Bosch \& Dekel 2010; Bouche \etal 2010; Firmani \& Avila-Reese
2010; Weinmann, Neistein \& Dekel  2011; Lacey \etal 2011).  Simply increasing
the supernova feedback  efficiency in order to lower  the star formation rates
at high-$z$ is  not a solution, because it  has no impact on the  SSFRs, as it
changes both  the SFR {\it  and} the stellar  mass (Dutton \etal  2010; Bouche
\etal 2010).   In addition, it will result  in too little stellar  mass at low
and  intermediate  redshifts, not  to  mention  the  fact that  the  supernova
feedback efficiencies  that are typically invoked  are already unrealistically
high  (e.g.,  Benson \etal  2003).   What  seems to  be  needed  instead is  a
modification of the  `standard' model that (i) causes  a larger suppression of
star  formation in  low  mass halos  at high-$z$,  and  (ii) a  boost in  star
formation at intermediate redshifts ($z \sim 1-2$). Although there has been no
shortage of ideas (e.g., Mo \etal 2005; Lo Faro \etal 2009; Bouche \etal 2010;
Lacey \etal  2011; Krumholz \& Dekel  2011; Weinmann, Neistein  \& Dekel 2011;
Wang, Weinmann  \& Neistein 2011; Avila-Reese  et al. 2011), it  is clear that
much more work is required before this outstanding problem in galaxy formation
and evolution is adequately addressed.  We believe that our constraints on the
stellar assembly histories of (central) galaxies may play an important role in
testing and calibrating  these new models.  In particular,  the rapid increase
in  $M_{\ast,c}/M$ at high-$z$  (before the  host halo  reaches a  mass $M\sim
10^{12}\msunh$) indicates  that the specific star formation  rate of (central)
galaxies is much higher than the specific growth rate of its dark matter halo,
which may hold important clues.

Another intriguing  feature from  the stellar mass  assembly histories  is the
fact that $M_{\ast,c}/M$  reaches a maximum of $\sim 0.03$  once $M$ reaches a
mass of $\sim  10^{12} \msunh$. This feature seems to  hold independent of the
final  host halo mass  of the  central galaxy.  Since, as  we have  shown, the
stellar mass growth in host halos with $M \lta 10^{12} \msunh$ is dominated by
{\it  in situ}  star formation  (rather  than accretion),  this suggests  that
something  quenches star  formation in  central  galaxies once  its halo  mass
reaches $10^{12}\msunh$.   Interestingly, this mass  scale is very  similar to
the cold-mode to hot-mode transition  scale (Birnboim \& Dekel 2003; Kere\v{s}
\etal 2005), suggesting that the  quenching of central galaxies coincides with
the formation of a hot gaseous halo.  This is indeed what seems to be required
in order  to explain  the observed bimodality  in colors and/or  specific star
formation rates of the galaxy  population (e.g., Cattaneo \etal 2006; Birnboim
\etal  2007).  Different  mechanisms have  been  invoked to  explain why  this
transition  is associated with  a shutdown  of star  formation in  the central
galaxies, ranging  from AGN  feedback (e.g., Tabor  \& Binney 1993;  Ciotti \&
Ostriker 1997;  Croton \etal 2006; Bower  \etal 2006; Hopkins  \etal 2006) and
gravitational heating (e.g., Fabian 2003;  Khochfar \& Ostriker 2008; Dekel \&
Birnboim 2008;  Birnboim \&  Dekel 2011) to  thermal conduction (e.g.,  Kim \&
Narayan 2003)  and turbulence  (e.g., Zhu, Feng  \& Fang 2011).   Although our
results  are  unable  to  provide  direct insight  into  the  exact  quenching
mechanism, it is fascinating (or at least reassuring) that our analysis, which
uses no (direct)  data on star formation rates or  the color-bimodality of the
galaxy population,  also comes  to the conclusion  that galaxies  are quenched
once their halo mass reaches $\sim 10^{12}\msunh$.

Another important  result regards the  implied build-up of stellar  halos.  We
have argued  in Section~\ref{sec:SF} that  central galaxies build up  the vast
majority of  their stellar  mass via  {\it in situ}  star formation  (the only
possible exception are  the most massive centrals, which may  have build up as
much as roughly half their stellar  mass via accretion). This implies that the
difference  between  the  total  mass  in accreted  satellites  and  surviving
satellites,    what    we    called    the    ``disrupted''    component    in
Section~\ref{sec:SF}, must have given rise  to a stellar halo.  This result is
consistent with a number of  recent studies which have argued that reconciling
halo occupation statistics with halo  merger rates requires that a significant
fraction of satellite  galaxies is indeed tidally disrupted  (e.g., Conroy, Ho
\& White 2007; Conroy, Wechsler \&  Kravtsov 2007; Kang \& van den Bosch 2008;
Yang, Mo  \& van  den Bosch  2009a). As these  studies have  argued, satellite
disruption is an important ingredient  of galaxy formation, which has hitherto
been  largely  ignored  in  semi-analytical models.  Properly  accounting  for
satellite disruption  may alleviate the  problem with the excessive  growth of
massive galaxies  (e.g., Monaco \etal 2006; Conroy, Wechsler  \& Kravtsov 2007,
Brown  \etal  2008), with  the  overabundance  of  satellite galaxies  in  the
semi-analytical  models  (Liu \etal  2010),  and  may  even be  important  for
understanding the observed metallicities of satellite galaxies (Pasquali \etal
2010). Note that our model predicts that the fraction of stars associated with
the stellar  halo is an  increasing function of  halo mass (see  also Purcell,
Bullock \& Zentner 2007 and Henriques, Bertone \& Thomas 2008), in qualitative
agreement with observations (Gonzalez, Zaritsky \& Zabludoff 2007).

Finally we point out that our  results may have important implications for the
`missing  satellite problem',  which refers  to  the fact  that the  predicted
subhalo count  in a  Milky-Way (MW)  sized halo vastly  exceeds the  number of
observed  satellite  galaxies  in  the  MW (Moore  \etal  1999;  Klypin  \etal
1999). Although  theoretical progress combined  with the discovery of  a large
population of new (typically ultra-faint)  satellite galaxies in the Milky Way
(e.g., Belokurov  \etal 2007) has  at least partially alleviated  this concern
about  a  mismatch between  the  numbers of  low-mass  subhalos  and faint  MW
satellites, it has become clear that it is difficult to reconcile the observed
abundance   of  satellite   galaxies  with   their  kinematics:   whereas  the
line-of-sight velocity dispersions of  the satellites suggest that they reside
in subhalos  of relatively low mass,  their abundance suggests  they reside in
much   more  massive   subhalos  (e.g.,   Madau,  Diemand   \&   Kuhlen  2008;
Boylan-Kolchin,  Bullock \&  Kaplinghat 2011).   This problem  is particularly
acute if  one uses  subhalo abundance matching  techniques to  link satellite
galaxies to  subhalos (e.g., Boylan-Kolchin  \etal 2011, see also  Busha \etal
2011). As we have argued, subhalo abundance matching is not self-consistent in
that it  does not account for dependence on subhalo  accretion times. Since
our  model shows  that  there is  very strong  redshift  dependence in  the
$M_{\ast,c}$-$M$ relation at  the low mass end, with low  mass halos at higher
$z$ hosting central galaxies that are  much less massive, the large scatter in
subhalo accretion times implies a huge  amount of scatter in the ratio between
the stellar mass  of satellite galaxies and their halo mass  at infall. Such a
large  scatter  may  be exactly  what  is  needed  to reconcile  the  observed
satellite population of the  MW with predictions from $\Lambda$CDM cosmologies
(e.g,., Madau \etal 2008), and may already have observational support from the
fact that the  ultrafaint dwarfs appear to reside in halos  of similar mass as
the much  more massive `classical'  dwarfs (Strigari \etal 2008;  Walker \etal
2009). We intend to return to this intriguing issue in more detail in a future
paper.

                                                                           
\acknowledgments We are grateful  to Mike Boylan-Kolchin, Aaron Dutton, Alexie
Leauthaud, Cheng Li,  Yu Lu and Jeremy Tinker for  useful discussions. We also
thank the  anonymous referee  for helpful comments  that greatly  improved the
presentation of  this paper.  This work  is supported by the  grants from NSFC
(Nos.  10925314, 11128306, 11121062) and  CAS/SAFEA International Partnership
Program  for  Creative  Research  Teams  (KJCX2-YW-T23).  HJM  would  like  to
acknowledge the support of NSF AST-1109354 and NSF AST-0908334


\appendix

\section{A. The projected two-point correlation function 
         of galaxies and its dependence on stellar mass from SDSS DR7}

\begin{figure*}
\plotone{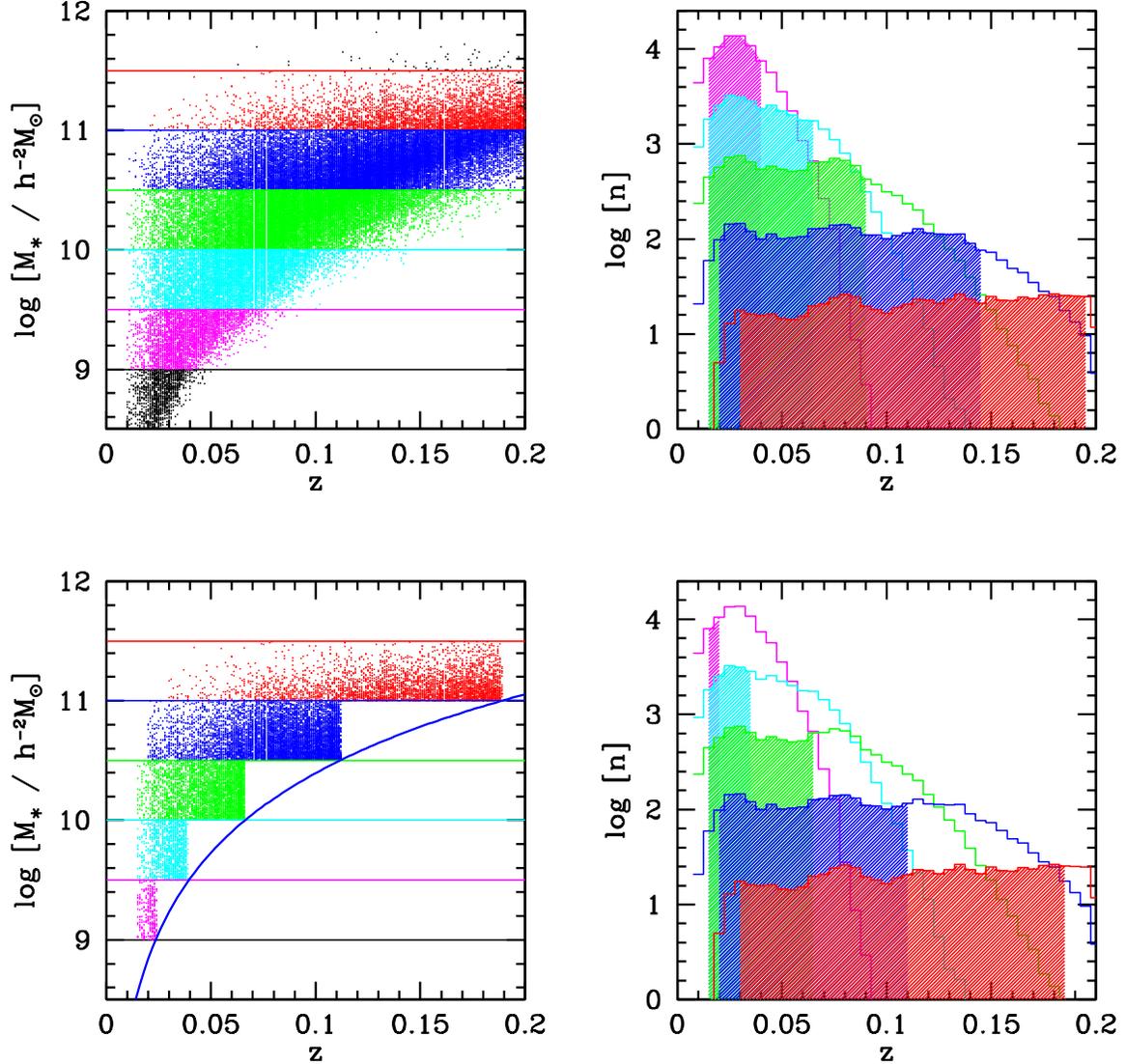}
\caption{The distributions of stellar mass versus redshift for galaxies in the
  SDSS DR7  (left-hand panels), and  the number density distribution  of these
  galaxies as a function of redshift (right-hand panels). Galaxies within five
  stellar  mass bins  are  selected.  The selection  criteria  for these  five
  samples are specified using  their redshift distributions (histograms and/or
  the shaded areas). See text for details regarding the constructions of these
  samples.  }
\label{fig:redshift}
\end{figure*}
\begin{figure*}
\plotone{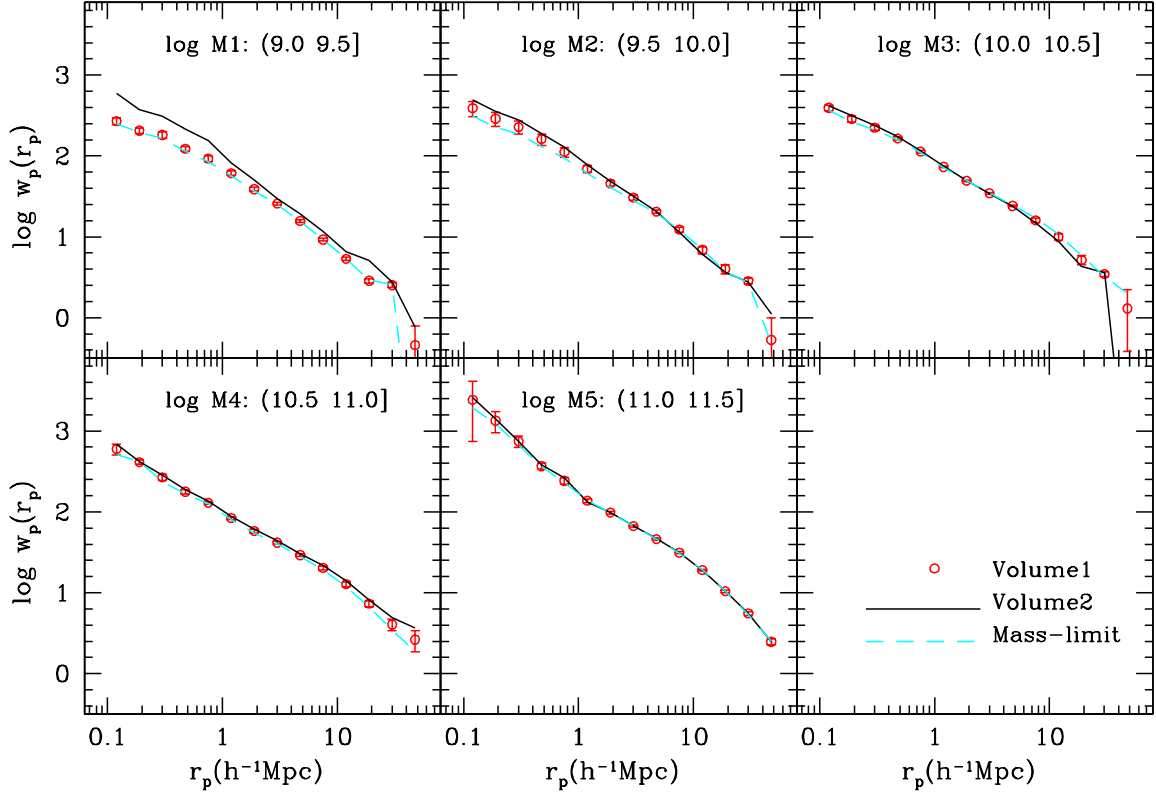}
\caption{The projected  2PCFs of galaxies  in different stellar mass  bins, as
  indicated  by  the  $\log[M_\ast/(h^{-2}\Msun)]$-values  in  brackets.  Open
  circles with error  bars are the results for the  Volume1 samples. Solid and
  dashed  lines show  the results  for  the Volume2  samples and  Mass-limited
  samples, respectively.  }
\label{fig:wrp_SDSS}
\end{figure*}
\begin{deluxetable*}{lrrrrrrrrrrr}
\tabletypesize{\scriptsize} \tablecaption{The projected 2PCFs
    measured from the SDSS DR7  within different stellar mass bins.}
\tablewidth{0pt} 
\tablehead{ID & $r_\rmpp$ & $w_\rmpp(r_\rmpp)$ & 
$\sigma_{w_\rmpp}$ & $w_\rmpp(r_\rmpp)$ & $\sigma_{w_\rmpp}$ & $w_\rmpp(r_\rmpp)$ & 
$\sigma_{w_\rmpp}$ & $w_\rmpp(r_\rmpp)$ & $\sigma_{w_\rmpp}$ & $w_\rmpp(r_\rmpp)$ &
$\sigma_{w_\rmpp}$\\ 
\cline{1-12} \\ (1) & (2) & (3) & (4) & (5) & (6) &
              (7) & (8) & (9) & (10) & (11) & (12)\\ \\ & & 
\multicolumn{2}{c}{[9.0, 9.5]} & \multicolumn{2}{c}{[9.5, 10.0]} & 
\multicolumn{2}{c}{[10.0, 10.5]} & \multicolumn{2}{c}{[10.5, 11.0]} & 
\multicolumn{2}{c}{[11.0,11.5]} }

\startdata
 1 &   0.12 &  373.399 &  200.617 &  388.815 &   83.198 &  393.234 &   30.913 &  597.466 &   90.192 & 2423.422 & 1684.429  \\
 2 &   0.19 &  259.170 &   92.989 &  288.384 &   55.778 &  287.154 &   25.668 &  411.860 &   24.702 & 1340.480 &  391.143  \\
 3 &   0.30 &  221.851 &   69.188 &  228.047 &   40.157 &  223.528 &   14.039 &  267.002 &   23.886 &  746.369 &  123.150  \\
 4 &   0.48 &  152.048 &   46.861 &  161.172 &   25.446 &  164.613 &    7.819 &  177.249 &   10.115 &  366.864 &   30.737  \\
 5 &   0.75 &  112.231 &   32.711 &  111.791 &   14.929 &  113.182 &    3.683 &  129.185 &    5.566 &  241.362 &   21.748  \\
 6 &   1.20 &   67.752 &   12.318 &   69.407 &    7.387 &   73.073 &    1.938 &   83.692 &    3.573 &  137.289 &    7.388  \\
 7 &   1.90 &   42.485 &    6.752 &   45.751 &    3.630 &   49.318 &    0.960 &   57.954 &    2.935 &   98.301 &    3.801  \\
 8 &   3.01 &   27.063 &    2.678 &   30.453 &    2.027 &   34.459 &    0.776 &   41.691 &    1.673 &   66.697 &    2.229  \\
 9 &   4.77 &   16.820 &    1.987 &   20.490 &    0.995 &   24.065 &    0.617 &   29.157 &    1.126 &   45.976 &    1.532  \\
10 &   7.55 &   10.066 &    1.942 &   12.326 &    0.888 &   15.995 &    0.983 &   20.199 &    1.041 &   31.164 &    0.951  \\
11 &  11.97 &    5.762 &    0.959 &    6.884 &    0.677 &    9.983 &    0.895 &   12.782 &    0.942 &   18.996 &    0.569  \\
12 &  18.97 &    3.637 &    1.183 &    4.021 &    0.539 &    5.171 &    0.673 &    7.316 &    0.618 &   10.485 &    0.414  \\
13 &  30.07 &    2.613 &    0.369 &    2.850 &    0.229 &    3.470 &    0.229 &    4.050 &    0.647 &    5.545 &    0.297  \\
14 &  47.66 &    0.486 &    0.448 &    0.532 &    0.463 &    1.306 &    0.922 &    2.619 &    0.762 &    2.473 &    0.213  \\
\enddata

\tablecomments{ Column (1): a counting ID.  Column (2): the projected comoving
  distances  in unit of  $\mpch$. Column  (3-12): the  projected 2PCFs,$w_{\rm
    p}(r_{\rm  p})$,  and their  errors,  $\sigma_{w_{\rm  p}}$, for  galaxies
  in  different  stellar  mass  bins,  as  indicated  (in  terms  of  $\log
  (M_{\ast}/ \msunhh)$).  } \label{tab:wrp}
\end{deluxetable*}

In  this Appendix  we  present our  measurements  of the  projected 2PCFs  for
galaxies  in different  stellar mass  bins.  We  use the  New  York University
Value-Added Galaxy Catalogue (NYU-VAGC; Blanton \etal 2005), which is based on
SDSS DR7 (Abazajian  \etal 2009) but with an  independent set of significantly
improved reductions.   From the NYU-VAGC, we  select all galaxies  in the Main
Galaxy Sample  with an extinction  corrected apparent magnitude  brighter than
$r=17.72$, with  redshifts in  the range $0.01  \leq z  \leq 0.20$ and  with a
redshift completeness  ${\cal C}_z > 0.7$.   This gives a  sample of $639,359$
galaxies  with a  sky coverage  of 7748  square degrees.  For each  galaxy, we
estimate its stellar mass using the fitting formula of Bell \etal (2003).

\subsection{A.1. Galaxy samples} 

Galaxies are  separated into  five stellar mass  bins, with $\log  [M_{\ast} /
  \msunhh]  =[9.0, 9.5]$, $[9.5,  10.0]$, $[10.0,  10.5]$, $[10.5,  11.0]$ and
$[11.0, 11.5]$, respectively. These  samples are referred to as `Mass-limited'
samples, and are  indicated in the stellar mass versus  redshift plot shown in
the upper left-hand panel of Fig.~\ref{fig:redshift}.  The number densities of
galaxies in  the five stellar mass bins  as function of redshift  are shown as
histograms  in the  upper  right-hand panel  of Fig.~\ref{fig:redshift}.   For
clarity the number densities are scaled by constant factors.  The advantage of
using these `Mass-limited' samples is that it maximizes the number of galaxies
in  each  mass bin.  However,  the  drawback is  that  these  samples are  not
homogeneous  in the  radial direction,  and therefore  not  straightforward to
compare with  model predictions.  We  therefore also construct another  set of
samples as follows: For each of the  five stellar mass bins, we first find the
maximum number density  of galaxies (at some redshift).  Next we determine the
minimum and maximum redshifts at which the number density of galaxies (in that
stellar mass  bin) drops  to {\it  half} of the  maximum value.  Only galaxies
within these  two redshift limits are kept  in the sample. In  what follows we
refer  to  the  five samples  thus  selected  as  `Volume1' samples,  and  the
corresponding number densities as function of redshift are shown as the shaded
areas in the upper right-hand panel of Fig.~\ref{fig:redshift}.

Since the  SDSS galaxy sample  is flux limited,  red and blue galaxies  of the
same stellar  mass suffer  from different selection  effects because  of their
different  stellar mass-to-light  ratios. Using  conservative limits,  van den
Bosch \etal  (2008) have shown that  the stellar mass completeness  limit as a
function of redshift for the SDSS catalogue is given by 
\begin{equation} \label{eq:mstarlim}
\log[M_{*,{\rm lim}}/(h^{-2}\Msun)] = {4.852 + 2.246 \log D_L(z) +
  1.123 \log(1+z) - 1.186 z \over 1 - 0.067 z} \,.
\end{equation}
Using this  limit we also  construct five samples  in stellar mass  that are
`complete'.  The  corresponding selection  criteria  are  shown  in the  lower
left-hand  panel of  Fig.~\ref{fig:redshift}, while  the shaded  areas  in the
lower right-hand panel indicate the corresponding number densities as function
of  redshift. In  what follows  we refer  to these  five samples  as `Volume2'
samples.

To estimate the two-point correlation functions, one needs to construct random
samples to  normalize the  galaxy-pair counts (see  below).  In order  to take
account  of the  overall selection  effects in  the SDSS  catalogue,  we first
construct a  random sample using the  SDSS luminosity function  and then apply
various   observational   selection   effects  (magnitude   limits,   redshift
completeness, sky boundary) according to  the SDSS survey mask. Since galaxies
in different stellar  mass bin do not follow  the overall luminosity function,
we re-assign a redshift to each mock galaxy by randomly sampling the redshifts
of the SDSS galaxies in the  corresponding stellar mass bin. Hence, the random
samples have  exactly the same  redshift distributions as  their corresponding
SDSS  samples.  By  applying  the same  redshift  cuts, we  obtain the  random
samples  for  the  `Mass-limited',  `Volume1' and  `Volume2'  galaxy  samples,
respectively.

\subsection {A.2. The estimator of the two-point 
    correlation function and the results}

We   estimate  the  two-point   correlation  function   (2PCF),  $\xi(r_\rmpp,
r_{\pi})$, for galaxies in each sample using the following estimator, 
\begin{equation}
\label{eq:tpcfest}
\xi(r_\rmpp,r_{\pi}) = {\langle RR \rangle \, \langle DD \rangle \over
\langle DR \rangle^2} - 1\,,
\end{equation}
where $\langle  DD \rangle$,  $\langle RR \rangle$,  and $\langle  DR \rangle$
are,   respectively,   the  number   of   galaxy-galaxy,  random-random,   and
galaxy-random pairs with  separation $(r_\rmpp,r_{\pi})$ (Hamilton 1993).  The
variables $r_\rmpp$  and $r_{\pi}$ are the pair  separations perpendicular and
parallel  to  the  line-of-sight,  respectively.  Explicitly, for  a  pair  of
galaxies, one located  at ${\bf s_1}$ and the other at  ${\bf s_2}$ with ${\bf
  s_i} = c z_i {\bf \hat{r}_i}/H_0$, we define 
\begin{equation} \label{rppi}
r_{\pi} = {{\bf s} \cdot {\bf l} \over \vert {\bf l} \vert} \; ,
\;\;\;\;\;\;\;\;\;\;\;\;\;\;
r_\rmpp = \sqrt{{\bf s} \cdot {\bf s} - \pi^2}\,,
\end{equation}
where ${\bf l} = ({\bf s_1} +  {\bf s_2})/2$ is the line of sight intersecting
the  pair, and  ${\bf  s} =  {\bf  s_1} -  {\bf  s_2}$.  Since  redshift-space
distortions  only affect $r_{\pi}$,  the projection  of $\xi(r_\rmpp,r_{\pi})$
along the $r_{\pi}$ axis is not sensitive to peculiar velocities, and directly
related  to  the  real-space   correlation  function.   This  projected  2PCF,
$w_\rmpp(r_\rmpp)$, is estimated using 
\begin{equation}
\label{abel}
w_\rmpp(r_\rmpp) = \int_{-\infty}^{\infty} \xi(r_\rmpp,r_{\pi}) \, \rmd r_{\pi}
= 2 \sum_{k} \xi (r_\rmpp,r_{\pi}) \, \Delta r_{\pi} \,
\end{equation}
(Davis \& Peebles 1983).  In our analysis, the summation is made over $k=1$ to
$40$,  which,  for  our  adopted  bin  width of  $\Delta  r_{\pi}  =  1\mpch$,
corresponds to an integration from $r_{\pi}= 0 \mpch$ to $r_{\pi} = 40 \mpch$.
In  order  to obtain  the  error  bars and  the  covariance  matrix for  the
projected  2PCFs,  we  use  200  bootstrap  resamplings  of  the  galaxies  in
consideration. The covariance matrix is obtained using
\begin{equation}
  {\bf C}_{ij}\equiv \mbox{Cov}(w_{p,i},w_{p,j})=
  \frac{1}{N-1}\sum^N_{l=1}(w^l_{p,i}-\bar{w}_{p,i})(w^l_{p,j}-\bar{w}_{p,j}),
  \label{eq:cov}
\end{equation}
where $N=200$, and $w^l_{p,i}$ represents the value of the projected 2PCF of 
the $i$th bin in the $l$th resampling.  

The projected  2PCFs $w_\rmpp(r_\rmpp)$ for  all the galaxy  samples described
above   are   shown   in   Fig.~\ref{fig:wrp_SDSS},  with   different   panels
corresponding to galaxies in different  stellar mass bins, as indicated.  Open
circles with error  bars correspond to the `Volume1'  samples, while solid and
dashed  lines   correspond  to  the  `Volume2'   and  `Mass-limited'  samples,
respectively. Note  that the results  for the same  stellar mass bin  are very
similar regardless of the sample used,  except in the lowest stellar mass bin.
The  big difference between  the `Volume2'  from other  samples in  the lowest
stellar mass bin is probably produced  by the fact that the excess galaxies in
other  samples are  mainly  blue  galaxies which  are  relatively bright.   In
general, `Volume2'  sample is unbiased  with respect to galaxy  color, however
suffer more severely from the  survey volume effect.  Therefore, for our model
constraints in the main text, we  obtain the averaged projected 2PCFs over the
`Volume1',  `Volume2' and  `Mass-limited' samples.   The  resulting covariance
matrixes are updated using the  total 600 bootstrap resampling values as well.
Thus obtained values of  $w_\rmpp(r_\rmpp)$ are listed in Table~\ref{tab:wrp}.
The related covariance  matrixes are available from the  authors upon request.
As we have tested, our results are in good agreement with those obtained by Li
\etal (2006) from  the SDSS DR2 data  and those obtained by Guo  et al. (2011)
from  the SDSS  DR7 data,  which are  based on  stellar masses  estimated from
galaxy  spectra  and photometries,  respectively.   For  consistency with  our
stellar mass  functions, which  are all based  on the stellar  masses obtained
using  the  model  of Bell  \etal  (2003),  we  use  our own  measurements  of
$w_\rmpp(r_\rmpp)$  together with the  covariance matrixes  based on  the same
stellar masses.

\section{B. The stellar mass functions and conditional stellar mass functions
  measured from SDSS DR7}

 We have constructed group catalougues from the latest SDSS DR7
using the same method described in Yang et al. (2007)
\footnote{See also Yang et al. (2012, in preparation)}.
Here we provide the updated stellar mass functions and 
conditional stellar mass functions obtained from these
group catalougues. Readers are referred to Yang et al. (2009b) 
for the details of these measurements. 

 Listed in Table \ref{tab:gax_MF} are the stellar mass functions 
of galaxies. Galaxies are classified into three categories
according to their memberships in groups: 
all galaxies (ALL), central galaxies (CENTRAL) and satellites 
(SATELLITE). For galaxies in each category , results are provided
for all galaxies (all), red galaxies (red) and blue galaxies 
(blue), using the same color separation as in  
Yang et al. (2009b).  

 For the present paper, 
investigation in this study, only data for all galaxies in `ALL' 
groups are used. This stellar mass function is shown in
Fig. \ref{fig:SMF} as open circles with error bars. 
Fitting the observational data with the Schechter function, 
\begin{equation}\label{eq:phi_M}
\Phi(M_{\ast}) = \phi^{\star}\left ( {M_{\ast}\over M^{\star}}\right )^{(\alpha+1)}
  {\rm exp} \left[-  {M_{\ast}\over M^{\star}}\right ]\,,
\end{equation}
we obtain the best fit parameters, $\phi^{\star}=0.0083635$, 
$\alpha=-1.117$, and $\log M^{\star}=10.673$.  The best fit is 
shown as the solid line in Fig.\,\ref{fig:SMF}.

Listed in Table \ref{tab:CSMF_DR7} are the conditional stellar mass functions
of galaxies. Here halo masses are obtained assuming the mass function
of the WMAP7 cosmology. 

\begin{turnpage}
\begin{deluxetable*}{cccccccccccc}
  \tabletypesize{\scriptsize}
  \tablecaption{The galaxy stellar mass functions $\Phi(M_{\ast})$ }
  \tablewidth{0pt}
  \tablehead{  &\multicolumn{3}{c}{ALL} && \multicolumn{3}{c}{CENTRAL} && \multicolumn{3}{c}{SATELLITE}\\
  \cline{2-4} \cline{6-8} \cline{10-12}\\ (1) & (2) & (3) & (4) & & (5) & (6) &
  (7) && (8) & (9) & (10) \\
   $\log M_{\ast}$ & all & red & blue && all & red & blue && all & red & blue  }

\startdata
  8.2  &  3.7705 $\pm$ 1.5258  &  0.9436 $\pm$ 0.7870  &  2.8269 $\pm$ 1.2665  &&  3.0870 $\pm$ 1.6328  &  0.9436 $\pm$ 0.7870  &  2.1434 $\pm$ 1.3832  &&  0.6835 $\pm$ 0.9345  &  0.0000 $\pm$ 0.0000  &  0.6835 $\pm$ 0.9345 \\
  8.3  &  3.4598 $\pm$ 0.7363  &  1.2416 $\pm$ 0.4523  &  2.2182 $\pm$ 0.5867  &&  2.1801 $\pm$ 0.5884  &  0.6520 $\pm$ 0.3011  &  1.5281 $\pm$ 0.4684  &&  1.2796 $\pm$ 0.5566  &  0.5896 $\pm$ 0.3436  &  0.6900 $\pm$ 0.4418 \\
  8.4  &  4.1293 $\pm$ 0.5891  &  1.1804 $\pm$ 0.2965  &  2.9489 $\pm$ 0.4627  &&  2.7961 $\pm$ 0.4415  &  0.5128 $\pm$ 0.2023  &  2.2833 $\pm$ 0.3748  &&  1.3332 $\pm$ 0.4736  &  0.6676 $\pm$ 0.2879  &  0.6656 $\pm$ 0.2905 \\
  8.5  &  3.6421 $\pm$ 0.5547  &  0.9305 $\pm$ 0.2886  &  2.7116 $\pm$ 0.3771  &&  2.4913 $\pm$ 0.3387  &  0.4905 $\pm$ 0.1597  &  2.0008 $\pm$ 0.2727  &&  1.1508 $\pm$ 0.3368  &  0.4400 $\pm$ 0.1997  &  0.7108 $\pm$ 0.2176 \\
  8.6  &  3.3055 $\pm$ 0.4245  &  0.8003 $\pm$ 0.2345  &  2.5052 $\pm$ 0.2674  &&  2.2182 $\pm$ 0.2612  &  0.3709 $\pm$ 0.1511  &  1.8474 $\pm$ 0.2058  &&  1.0873 $\pm$ 0.2604  &  0.4294 $\pm$ 0.1365  &  0.6578 $\pm$ 0.1831 \\
  8.7  &  3.1321 $\pm$ 0.3100  &  0.8215 $\pm$ 0.1561  &  2.3106 $\pm$ 0.2224  &&  2.1598 $\pm$ 0.2294  &  0.3756 $\pm$ 0.1010  &  1.7842 $\pm$ 0.1819  &&  0.9723 $\pm$ 0.1686  &  0.4459 $\pm$ 0.1020  &  0.5264 $\pm$ 0.1109 \\
  8.8  &  3.0391 $\pm$ 0.2499  &  0.8716 $\pm$ 0.1181  &  2.1675 $\pm$ 0.1865  &&  1.8100 $\pm$ 0.1428  &  0.3005 $\pm$ 0.0669  &  1.5095 $\pm$ 0.1253  &&  1.2291 $\pm$ 0.1729  &  0.5711 $\pm$ 0.0936  &  0.6580 $\pm$ 0.1253 \\
  8.9  &  2.7949 $\pm$ 0.2433  &  0.8404 $\pm$ 0.1442  &  1.9545 $\pm$ 0.1538  &&  1.7266 $\pm$ 0.1265  &  0.2997 $\pm$ 0.0582  &  1.4269 $\pm$ 0.1027  &&  1.0683 $\pm$ 0.1745  &  0.5407 $\pm$ 0.1141  &  0.5276 $\pm$ 0.0929 \\
  9.0  &  3.1430 $\pm$ 0.1822  &  0.9815 $\pm$ 0.1210  &  2.1614 $\pm$ 0.1161  &&  1.9179 $\pm$ 0.0875  &  0.3476 $\pm$ 0.0524  &  1.5702 $\pm$ 0.0968  &&  1.2251 $\pm$ 0.1483  &  0.6339 $\pm$ 0.0941  &  0.5912 $\pm$ 0.0871 \\
  9.1  &  3.1047 $\pm$ 0.2357  &  1.0438 $\pm$ 0.1518  &  2.0609 $\pm$ 0.1199  &&  1.8162 $\pm$ 0.1129  &  0.3595 $\pm$ 0.0608  &  1.4568 $\pm$ 0.0765  &&  1.2884 $\pm$ 0.1579  &  0.6843 $\pm$ 0.1131  &  0.6042 $\pm$ 0.0723 \\
  9.2  &  2.9365 $\pm$ 0.1816  &  1.0557 $\pm$ 0.1398  &  1.8808 $\pm$ 0.0760  &&  1.6895 $\pm$ 0.1021  &  0.3411 $\pm$ 0.0622  &  1.3484 $\pm$ 0.0606  &&  1.2470 $\pm$ 0.1151  &  0.7146 $\pm$ 0.0969  &  0.5324 $\pm$ 0.0409 \\
  9.3  &  2.8092 $\pm$ 0.1786  &  1.0230 $\pm$ 0.1329  &  1.7861 $\pm$ 0.0766  &&  1.5992 $\pm$ 0.0730  &  0.3624 $\pm$ 0.0458  &  1.2368 $\pm$ 0.0476  &&  1.2100 $\pm$ 0.1358  &  0.6607 $\pm$ 0.1043  &  0.5493 $\pm$ 0.0518 \\
  9.4  &  2.8013 $\pm$ 0.0925  &  1.0764 $\pm$ 0.0703  &  1.7249 $\pm$ 0.0477  &&  1.6116 $\pm$ 0.0621  &  0.4012 $\pm$ 0.0381  &  1.2104 $\pm$ 0.0420  &&  1.1897 $\pm$ 0.0549  &  0.6753 $\pm$ 0.0479  &  0.5145 $\pm$ 0.0292 \\
  9.5  &  2.5093 $\pm$ 0.1140  &  1.0816 $\pm$ 0.0917  &  1.4277 $\pm$ 0.0418  &&  1.4360 $\pm$ 0.0522  &  0.4290 $\pm$ 0.0420  &  1.0070 $\pm$ 0.0354  &&  1.0733 $\pm$ 0.0804  &  0.6526 $\pm$ 0.0637  &  0.4208 $\pm$ 0.0290 \\
  9.6  &  2.3481 $\pm$ 0.1002  &  1.0112 $\pm$ 0.0787  &  1.3369 $\pm$ 0.0362  &&  1.3756 $\pm$ 0.0508  &  0.4339 $\pm$ 0.0302  &  0.9416 $\pm$ 0.0307  &&  0.9725 $\pm$ 0.0653  &  0.5772 $\pm$ 0.0598  &  0.3953 $\pm$ 0.0173 \\
  9.7  &  2.0970 $\pm$ 0.0640  &  1.0132 $\pm$ 0.0488  &  1.0837 $\pm$ 0.0286  &&  1.2562 $\pm$ 0.0420  &  0.4760 $\pm$ 0.0288  &  0.7802 $\pm$ 0.0231  &&  0.8408 $\pm$ 0.0368  &  0.5373 $\pm$ 0.0304  &  0.3035 $\pm$ 0.0153 \\
  9.8  &  1.9927 $\pm$ 0.0653  &  1.0453 $\pm$ 0.0526  &  0.9473 $\pm$ 0.0239  &&  1.2189 $\pm$ 0.0254  &  0.5060 $\pm$ 0.0210  &  0.7129 $\pm$ 0.0171  &&  0.7738 $\pm$ 0.0509  &  0.5393 $\pm$ 0.0407  &  0.2345 $\pm$ 0.0168 \\
  9.9  &  1.8551 $\pm$ 0.0555  &  1.0426 $\pm$ 0.0446  &  0.8125 $\pm$ 0.0210  &&  1.1423 $\pm$ 0.0240  &  0.5284 $\pm$ 0.0176  &  0.6139 $\pm$ 0.0143  &&  0.7128 $\pm$ 0.0398  &  0.5142 $\pm$ 0.0337  &  0.1986 $\pm$ 0.0119 \\
 10.0  &  1.7485 $\pm$ 0.0555  &  1.0329 $\pm$ 0.0448  &  0.7156 $\pm$ 0.0184  &&  1.1068 $\pm$ 0.0241  &  0.5713 $\pm$ 0.0197  &  0.5355 $\pm$ 0.0110  &&  0.6417 $\pm$ 0.0393  &  0.4616 $\pm$ 0.0320  &  0.1801 $\pm$ 0.0119 \\
 10.1  &  1.6715 $\pm$ 0.0430  &  1.0297 $\pm$ 0.0343  &  0.6418 $\pm$ 0.0156  &&  1.0844 $\pm$ 0.0161  &  0.5863 $\pm$ 0.0100  &  0.4981 $\pm$ 0.0121  &&  0.5871 $\pm$ 0.0326  &  0.4434 $\pm$ 0.0293  &  0.1436 $\pm$ 0.0066 \\
 10.2  &  1.6340 $\pm$ 0.0417  &  1.0560 $\pm$ 0.0331  &  0.5780 $\pm$ 0.0142  &&  1.0963 $\pm$ 0.0122  &  0.6465 $\pm$ 0.0088  &  0.4498 $\pm$ 0.0082  &&  0.5377 $\pm$ 0.0346  &  0.4095 $\pm$ 0.0289  &  0.1282 $\pm$ 0.0087 \\
 10.3  &  1.5273 $\pm$ 0.0419  &  1.0368 $\pm$ 0.0355  &  0.4905 $\pm$ 0.0113  &&  1.0326 $\pm$ 0.0104  &  0.6491 $\pm$ 0.0078  &  0.3835 $\pm$ 0.0069  &&  0.4947 $\pm$ 0.0356  &  0.3877 $\pm$ 0.0318  &  0.1071 $\pm$ 0.0064 \\
 10.4  &  1.3308 $\pm$ 0.0339  &  0.9331 $\pm$ 0.0266  &  0.3978 $\pm$ 0.0105  &&  0.9275 $\pm$ 0.0104  &  0.6129 $\pm$ 0.0081  &  0.3146 $\pm$ 0.0067  &&  0.4033 $\pm$ 0.0273  &  0.3202 $\pm$ 0.0232  &  0.0831 $\pm$ 0.0057 \\
 10.5  &  1.0870 $\pm$ 0.0292  &  0.7817 $\pm$ 0.0237  &  0.3052 $\pm$ 0.0084  &&  0.7882 $\pm$ 0.0095  &  0.5383 $\pm$ 0.0066  &  0.2499 $\pm$ 0.0051  &&  0.2988 $\pm$ 0.0225  &  0.2435 $\pm$ 0.0193  &  0.0553 $\pm$ 0.0045 \\
 10.6  &  0.8692 $\pm$ 0.0265  &  0.6337 $\pm$ 0.0210  &  0.2354 $\pm$ 0.0077  &&  0.6445 $\pm$ 0.0097  &  0.4523 $\pm$ 0.0069  &  0.1922 $\pm$ 0.0047  &&  0.2247 $\pm$ 0.0194  &  0.1814 $\pm$ 0.0164  &  0.0432 $\pm$ 0.0040 \\
 10.7  &  0.6629 $\pm$ 0.0208  &  0.4876 $\pm$ 0.0164  &  0.1753 $\pm$ 0.0061  &&  0.5122 $\pm$ 0.0089  &  0.3643 $\pm$ 0.0064  &  0.1479 $\pm$ 0.0039  &&  0.1507 $\pm$ 0.0138  &  0.1232 $\pm$ 0.0116  &  0.0274 $\pm$ 0.0029 \\
 10.8  &  0.4749 $\pm$ 0.0168  &  0.3555 $\pm$ 0.0133  &  0.1194 $\pm$ 0.0045  &&  0.3796 $\pm$ 0.0078  &  0.2776 $\pm$ 0.0056  &  0.1020 $\pm$ 0.0031  &&  0.0953 $\pm$ 0.0101  &  0.0779 $\pm$ 0.0088  &  0.0173 $\pm$ 0.0019 \\
 10.9  &  0.3130 $\pm$ 0.0133  &  0.2368 $\pm$ 0.0104  &  0.0762 $\pm$ 0.0038  &&  0.2598 $\pm$ 0.0083  &  0.1923 $\pm$ 0.0063  &  0.0675 $\pm$ 0.0029  &&  0.0532 $\pm$ 0.0057  &  0.0445 $\pm$ 0.0047  &  0.0087 $\pm$ 0.0013 \\
 11.0  &  0.1913 $\pm$ 0.0086  &  0.1491 $\pm$ 0.0066  &  0.0422 $\pm$ 0.0026  &&  0.1636 $\pm$ 0.0059  &  0.1260 $\pm$ 0.0044  &  0.0376 $\pm$ 0.0021  &&  0.0277 $\pm$ 0.0032  &  0.0231 $\pm$ 0.0027  &  0.0046 $\pm$ 0.0007 \\
 11.1  &  0.1055 $\pm$ 0.0056  &  0.0840 $\pm$ 0.0041  &  0.0215 $\pm$ 0.0019  &&  0.0943 $\pm$ 0.0044  &  0.0745 $\pm$ 0.0031  &  0.0198 $\pm$ 0.0015  &&  0.0112 $\pm$ 0.0015  &  0.0095 $\pm$ 0.0012  &  0.0016 $\pm$ 0.0004 \\
 11.2  &  0.0540 $\pm$ 0.0028  &  0.0447 $\pm$ 0.0021  &  0.0092 $\pm$ 0.0009  &&  0.0495 $\pm$ 0.0023  &  0.0408 $\pm$ 0.0017  &  0.0087 $\pm$ 0.0008  &&  0.0045 $\pm$ 0.0006  &  0.0039 $\pm$ 0.0006  &  0.0006 $\pm$ 0.0001 \\
 11.3  &  0.0245 $\pm$ 0.0015  &  0.0207 $\pm$ 0.0013  &  0.0039 $\pm$ 0.0004  &&  0.0231 $\pm$ 0.0014  &  0.0194 $\pm$ 0.0011  &  0.0037 $\pm$ 0.0004  &&  0.0015 $\pm$ 0.0002  &  0.0013 $\pm$ 0.0002  &  0.0001 $\pm$ 0.0001 \\
 11.4  &  0.0104 $\pm$ 0.0007  &  0.0086 $\pm$ 0.0006  &  0.0019 $\pm$ 0.0002  &&  0.0101 $\pm$ 0.0006  &  0.0083 $\pm$ 0.0005  &  0.0018 $\pm$ 0.0002  &&  0.0003 $\pm$ 0.0001  &  0.0003 $\pm$ 0.0001  &  0.0000 $\pm$ 0.0000 \\
 11.5  &  0.0042 $\pm$ 0.0003  &  0.0034 $\pm$ 0.0003  &  0.0008 $\pm$ 0.0001  &&  0.0041 $\pm$ 0.0003  &  0.0033 $\pm$ 0.0003  &  0.0008 $\pm$ 0.0001  &&  0.0001 $\pm$ 0.0000  &  0.0001 $\pm$ 0.0000  &  0.0000 $\pm$ 0.0000 \\
 11.6  &  0.0013 $\pm$ 0.0001  &  0.0010 $\pm$ 0.0001  &  0.0003 $\pm$ 0.0001  &&  0.0013 $\pm$ 0.0001  &  0.0010 $\pm$ 0.0001  &  0.0003 $\pm$ 0.0001  &&  0.0000 $\pm$ 0.0000  &  0.0000 $\pm$ 0.0000  &  0.0000 $\pm$ 0.0000 \\
 11.7  &  0.0003 $\pm$ 0.0001  &  0.0002 $\pm$ 0.0001  &  0.0001 $\pm$ 0.0000  &&  0.0003 $\pm$ 0.0001  &  0.0002 $\pm$ 0.0001  &  0.0001 $\pm$ 0.0000  &&  0.0000 $\pm$ 0.0000  &  0.0000 $\pm$ 0.0000  &  0.0000 $\pm$ 0.0000 \\
\enddata

\tablecomments{ Column (1): the median  of the logarithm of the galaxy stellar
  mass with  bin width  $\Delta \log M_{\ast}  = 0.05$.   Column (2 -  4): the
  stellar mass functions of all, red  and blue for 'ALL' group members. Column
  (5 - 7): the stellar mass functions of all, red and blue for 'CENTRAL' group
  members.  Column (8  - 10): the stellar mass functions of  all, red and blue
  for  'SATELLITE' group  members.   Note  that all  the  galaxy stellar  mass
  functions listed  in this table are  in units of  $10^{-2} h^3{\rm Mpc}^{-3}
  {\rm    d}   \log    M_{\ast}$,    where   $\log$    is    the   10    based
  logarithm. }\label{tab:gax_MF}
\end{deluxetable*}
\end{turnpage}

\begin{turnpage}
\begin{deluxetable*}{cccccccccc}
  \tabletypesize{\scriptsize} \tablecaption{The conditional stellar mass
    functions of galaxies $\Phi(M_{\ast}|M_h)$ } \tablewidth{0pt} 

\tablehead{ (1) & (2) & (3) & (4) & (5) & (6) & (7) & (8) & (9) & (10)
  \\ $\log M_{\ast}$ & [12.1, 12.4] & [12.4, 12.7] & [12.7, 13.0] & [13.0,
    13.3] & [13.3, 13.6] & [13.6, 13.9] & [13.9, 14.2] & [14.2, 14.5] & $>14.5$ }

\startdata
  8.9  &   0.857 $\pm$   0.238  &   1.534 $\pm$   0.558  &   3.944 $\pm$   1.745  &   7.232 $\pm$   3.599  &    0.000 $\pm$   0.000 &   0.000 $\pm$   0.000  &   0.000 $\pm$   0.000  &   0.000 $\pm$   0.000  &   0.000 $\pm$   0.000 \\
  9.0  &   0.867 $\pm$   0.191  &   1.793 $\pm$   0.364  &   2.569 $\pm$   0.723  &   7.962 $\pm$   1.834  &  10.542 $\pm$   4.496  &  19.005 $\pm$   6.481  &  38.365 $\pm$  10.604  & 120.947 $\pm$  91.664  &  0.000 $\pm$   0.000 \\
  9.1  &   0.860 $\pm$   0.138  &   1.117 $\pm$   0.237  &   2.984 $\pm$   0.662  &   7.154 $\pm$   1.573  &  12.582 $\pm$   4.204  &  22.436 $\pm$   4.295  &  44.760 $\pm$  17.909  & 105.092 $\pm$  53.583  & 232.801 $\pm$ 187.113 \\
  9.2  &   0.740 $\pm$   0.099  &   1.353 $\pm$   0.209  &   2.221 $\pm$   0.437  &   6.045 $\pm$   0.845  &  11.677 $\pm$   3.663  &  20.303 $\pm$   4.000  &  35.106 $\pm$  10.618  & 120.722 $\pm$  59.000  & 273.560 $\pm$ 191.005 \\
  9.3  &   0.626 $\pm$   0.080  &   1.278 $\pm$   0.180  &   2.412 $\pm$   0.321  &   4.447 $\pm$   0.903  &   7.831 $\pm$   1.935  &  19.684 $\pm$   3.393  &  37.028 $\pm$   9.060  & 114.475 $\pm$  54.627  & 240.362 $\pm$ 124.617 \\
  9.4  &   0.572 $\pm$   0.071  &   1.252 $\pm$   0.136  &   3.118 $\pm$   0.316  &   4.153 $\pm$   0.540  &   8.288 $\pm$   1.085  &  20.259 $\pm$   2.992  &  39.099 $\pm$   8.065  & 100.398 $\pm$  30.349  & 268.824 $\pm$ 134.078 \\
  9.5  &   0.476 $\pm$   0.057  &   1.064 $\pm$   0.110  &   2.034 $\pm$   0.222  &   4.965 $\pm$   0.653  &   9.177 $\pm$   1.263  &  21.391 $\pm$   2.823  &  37.373 $\pm$   4.539  &  87.398 $\pm$  26.248  & 251.692 $\pm$  57.480 \\
  9.6  &   0.522 $\pm$   0.054  &   0.991 $\pm$   0.093  &   2.512 $\pm$   0.218  &   4.064 $\pm$   0.455  &   8.755 $\pm$   1.012  &  16.330 $\pm$   1.757  &  33.938 $\pm$   5.091  &  94.456 $\pm$  22.537  & 188.324 $\pm$  47.538 \\
  9.7  &   0.414 $\pm$   0.037  &   0.873 $\pm$   0.078  &   1.691 $\pm$   0.160  &   3.553 $\pm$   0.382  &   6.781 $\pm$   1.065  &  15.900 $\pm$   1.777  &  41.002 $\pm$   4.031  &  85.229 $\pm$  17.790  & 206.746 $\pm$  46.094 \\
  9.8  &   0.419 $\pm$   0.046  &   0.817 $\pm$   0.088  &   1.700 $\pm$   0.138  &   3.288 $\pm$   0.283  &   7.694 $\pm$   0.865  &  14.748 $\pm$   1.399  &  36.148 $\pm$   3.253  &  78.295 $\pm$  13.852  & 170.928 $\pm$  37.206 \\
  9.9  &   0.401 $\pm$   0.028  &   0.802 $\pm$   0.055  &   1.733 $\pm$   0.133  &   2.959 $\pm$   0.243  &   7.274 $\pm$   1.067  &  16.538 $\pm$   1.313  &  38.849 $\pm$   5.484  &  78.331 $\pm$  10.478  & 163.676 $\pm$  24.944 \\
 10.0  &   0.339 $\pm$   0.052  &   0.766 $\pm$   0.057  &   1.598 $\pm$   0.104  &   2.983 $\pm$   0.197  &   6.379 $\pm$   0.630  &  15.380 $\pm$   1.293  &  30.516 $\pm$   2.681  &  71.385 $\pm$   8.285  & 154.757 $\pm$  21.411 \\
 10.1  &   0.283 $\pm$   0.055  &   0.673 $\pm$   0.037  &   1.388 $\pm$   0.073  &   3.167 $\pm$   0.173  &   6.209 $\pm$   0.453  &  13.795 $\pm$   0.807  &  30.357 $\pm$   2.111  &  67.792 $\pm$   8.082  & 133.214 $\pm$  26.301 \\
 10.2  &   0.176 $\pm$   0.040  &   0.610 $\pm$   0.039  &   1.391 $\pm$   0.079  &   2.850 $\pm$   0.133  &   5.491 $\pm$   0.486  &  13.024 $\pm$   0.811  &  26.062 $\pm$   2.053  &  60.936 $\pm$   5.756  & 117.930 $\pm$  16.417 \\
 10.3  &   0.088 $\pm$   0.034  &   0.547 $\pm$   0.098  &   1.300 $\pm$   0.097  &   2.473 $\pm$   0.147  &   5.500 $\pm$   0.242  &  11.631 $\pm$   0.710  &  23.506 $\pm$   1.265  &  48.402 $\pm$   4.657  & 109.039 $\pm$  10.113 \\
 10.4  &   0.010 $\pm$   0.013  &   0.357 $\pm$   0.061  &   1.077 $\pm$   0.071  &   2.332 $\pm$   0.130  &   4.577 $\pm$   0.218  &  10.016 $\pm$   0.700  &  20.390 $\pm$   1.332  &  45.767 $\pm$   4.143  & 103.090 $\pm$   8.596 \\
 10.5  &   0.002 $\pm$   0.002  &   0.113 $\pm$   0.039  &   0.828 $\pm$   0.120  &   1.918 $\pm$   0.094  &   3.793 $\pm$   0.199  &   8.246 $\pm$   0.323  &  16.756 $\pm$   1.288  &  35.299 $\pm$   3.344  &  76.226 $\pm$   6.337 \\
 10.6  &   0.000 $\pm$   0.001  &   0.014 $\pm$   0.017  &   0.418 $\pm$   0.078  &   1.546 $\pm$   0.158  &   3.206 $\pm$   0.217  &   6.514 $\pm$   0.337  &  12.571 $\pm$   0.726  &  24.642 $\pm$   1.687  &  60.269 $\pm$   5.612 \\
 10.7  &   0.001 $\pm$   0.003  &   0.001 $\pm$   0.002  &   0.094 $\pm$   0.048  &   0.843 $\pm$   0.154  &   2.318 $\pm$   0.197  &   4.879 $\pm$   0.407  &   9.591 $\pm$   0.418  &  19.094 $\pm$   1.294  &  42.580 $\pm$   3.151 \\
 10.8  &   0.002 $\pm$   0.004  &   0.000 $\pm$   0.000  &   0.005 $\pm$   0.009  &   0.295 $\pm$   0.073  &   1.433 $\pm$   0.201  &   3.261 $\pm$   0.261  &   7.104 $\pm$   0.555  &  13.486 $\pm$   0.764  &  27.614 $\pm$   1.719 \\
 10.9  &   0.013 $\pm$   0.016  &   0.002 $\pm$   0.003  &   0.001 $\pm$   0.001  &   0.036 $\pm$   0.025  &   0.583 $\pm$   0.104  &   1.891 $\pm$   0.255  &   4.694 $\pm$   0.363  &   9.536 $\pm$   0.396  &  19.100 $\pm$   1.067 \\
 11.0  &   0.032 $\pm$   0.036  &   0.004 $\pm$   0.004  &   0.000 $\pm$   0.000  &   0.003 $\pm$   0.004  &   0.117 $\pm$   0.043  &   0.991 $\pm$   0.189  &   2.748 $\pm$   0.289  &   6.312 $\pm$   0.364  &  13.341 $\pm$   0.736 \\
 11.1  &   0.109 $\pm$   0.106  &   0.014 $\pm$   0.017  &   0.001 $\pm$   0.003  &   0.000 $\pm$   0.000  &   0.005 $\pm$   0.005  &   0.233 $\pm$   0.062  &   1.201 $\pm$   0.143  &   3.402 $\pm$   0.489  &   9.205 $\pm$   0.568 \\
 11.2  &   0.341 $\pm$   0.247  &   0.052 $\pm$   0.047  &   0.009 $\pm$   0.010  &   0.001 $\pm$   0.005  &   0.001 $\pm$   0.002  &   0.021 $\pm$   0.015  &   0.296 $\pm$   0.071  &   1.387 $\pm$   0.188  &   5.054 $\pm$   0.459 \\
 11.3  &   0.740 $\pm$   0.421  &   0.143 $\pm$   0.105  &   0.029 $\pm$   0.025  &   0.007 $\pm$   0.008  &   0.000 $\pm$   0.000  &   0.003 $\pm$   0.003  &   0.034 $\pm$   0.018  &   0.277 $\pm$   0.061  &   2.132 $\pm$   0.290 \\
 11.4  &   1.817 $\pm$   0.184  &   0.471 $\pm$   0.161  &   0.108 $\pm$   0.071  &   0.028 $\pm$   0.020  &   0.005 $\pm$   0.008  &   0.001 $\pm$   0.002  &   0.000 $\pm$   0.000  &   0.026 $\pm$   0.022  &   0.704 $\pm$   0.168 \\
 11.5  &   3.559 $\pm$   0.827  &   1.043 $\pm$   0.291  &   0.342 $\pm$   0.143  &   0.121 $\pm$   0.055  &   0.017 $\pm$   0.019  &   0.000 $\pm$   0.000  &   0.004 $\pm$   0.006  &   0.017 $\pm$   0.018  &   0.210 $\pm$   0.089 \\
 11.6  &   2.973 $\pm$   0.558  &   1.889 $\pm$   0.352  &   0.869 $\pm$   0.154  &   0.351 $\pm$   0.140  &   0.123 $\pm$   0.048  &   0.050 $\pm$   0.025  &   0.000 $\pm$   0.000  &   0.000 $\pm$   0.000  &   0.000 $\pm$   0.000 \\
 11.7  &   0.511 $\pm$   0.489  &   3.753 $\pm$   0.869  &   1.629 $\pm$   0.224  &   0.940 $\pm$   0.162  &   0.410 $\pm$   0.135  &   0.141 $\pm$   0.051  &   0.021 $\pm$   0.019  &   0.005 $\pm$   0.023  &   0.000 $\pm$   0.000 \\
 11.8  &   0.052 $\pm$   0.060  &   2.449 $\pm$   0.269  &   2.606 $\pm$   0.293  &   1.687 $\pm$   0.139  &   0.961 $\pm$   0.166  &   0.482 $\pm$   0.123  &   0.162 $\pm$   0.064  &   0.013 $\pm$   0.023  &   0.000 $\pm$   0.000 \\
 11.9  &   0.009 $\pm$   0.010  &   0.294 $\pm$   0.287  &   3.360 $\pm$   0.650  &   2.320 $\pm$   0.133  &   1.854 $\pm$   0.116  &   1.112 $\pm$   0.171  &   0.551 $\pm$   0.106  &   0.183 $\pm$   0.058  &   0.009 $\pm$   0.035 \\
 12.0  &   0.003 $\pm$   0.003  &   0.030 $\pm$   0.034  &   1.158 $\pm$   0.200  &   2.866 $\pm$   0.491  &   2.449 $\pm$   0.059  &   1.997 $\pm$   0.137  &   1.336 $\pm$   0.140  &   0.597 $\pm$   0.130  &   0.080 $\pm$   0.067 \\
 12.1  &   0.000 $\pm$   0.001  &   0.007 $\pm$   0.008  &   0.082 $\pm$   0.089  &   1.769 $\pm$   0.303  &   2.381 $\pm$   0.219  &   2.533 $\pm$   0.069  &   2.298 $\pm$   0.207  &   1.613 $\pm$   0.120  &   0.785 $\pm$   0.150 \\
 12.2  &   0.000 $\pm$   0.000  &   0.002 $\pm$   0.003  &   0.012 $\pm$   0.013  &   0.227 $\pm$   0.110  &   1.732 $\pm$   0.321  &   2.107 $\pm$   0.085  &   2.310 $\pm$   0.090  &   2.360 $\pm$   0.141  &   1.481 $\pm$   0.197 \\
 12.3  &   0.000 $\pm$   0.000  &   0.000 $\pm$   0.001  &   0.006 $\pm$   0.007  &   0.021 $\pm$   0.023  &   0.432 $\pm$   0.107  &   1.410 $\pm$   0.266  &   1.791 $\pm$   0.114  &   2.135 $\pm$   0.141  &   2.194 $\pm$   0.226 \\
 12.4  &   0.000 $\pm$   0.000  &   0.000 $\pm$   0.000  &   0.001 $\pm$   0.001  &   0.004 $\pm$   0.005  &   0.034 $\pm$   0.026  &   0.527 $\pm$   0.083  &   1.189 $\pm$   0.176  &   1.659 $\pm$   0.132  &   2.157 $\pm$   0.223 \\
 12.5  &   0.000 $\pm$   0.000  &   0.000 $\pm$   0.000  &   0.001 $\pm$   0.001  &   0.002 $\pm$   0.003  &   0.009 $\pm$   0.010  &   0.069 $\pm$   0.036  &   0.445 $\pm$   0.132  &   1.023 $\pm$   0.140  &   1.674 $\pm$   0.217 \\
 12.6  &   0.000 $\pm$   0.000  &   0.000 $\pm$   0.000  &   0.000 $\pm$   0.000  &   0.001 $\pm$   0.002  &   0.003 $\pm$   0.004  &   0.011 $\pm$   0.007  &   0.091 $\pm$   0.020  &   0.418 $\pm$   0.077  &   1.007 $\pm$   0.174 \\
 12.7  &   0.000 $\pm$   0.000  &   0.000 $\pm$   0.000  &   0.000 $\pm$   0.000  &   0.000 $\pm$   0.000  &   0.001 $\pm$   0.001  &   0.001 $\pm$   0.002  &   0.007 $\pm$   0.007  &   0.046 $\pm$   0.023  &   0.445 $\pm$   0.124 \\
 12.8  &   0.000 $\pm$   0.000  &   0.000 $\pm$   0.000  &   0.000 $\pm$   0.000  &   0.000 $\pm$   0.000  &   0.000 $\pm$   0.000  &   0.001 $\pm$   0.002  &   0.003 $\pm$   0.005  &   0.005 $\pm$   0.011  &   0.128 $\pm$   0.081 \\
 12.9  &   0.000 $\pm$   0.000  &   0.000 $\pm$   0.000  &   0.000 $\pm$   0.000  &   0.000 $\pm$   0.000  &   0.000 $\pm$   0.000  &   0.000 $\pm$   0.000  &   0.000 $\pm$   0.000  &   0.000 $\pm$   0.000  &   0.014 $\pm$   0.029 \\
 13.0  &   0.000 $\pm$   0.000  &   0.000 $\pm$   0.000  &   0.000 $\pm$   0.000  &   0.000 $\pm$   0.000  &   0.000 $\pm$   0.000  &   0.000 $\pm$   0.000  &   0.000 $\pm$   0.000  &   0.000 $\pm$   0.000  &   0.028 $\pm$   0.043 \\

\enddata

\tablecomments{ Column (1): the median of the logarithm of the galaxy stellar
  mass with bin width $\Delta \log M_{\ast} = 0.05$.  Column (2 - 10): the
  conditional stellar mass functions in halos of different mass ranges as
  indicated. The average halo masses in these bins are $\log <M_h> = 12.26,
  12.55, 12.85, 13.15, 13.44, 13.74, 14.04, 14.33, 14.72$, respectively. Note
  that all the conditioanl galaxy stellar mass functions listed in this table
  are in units of $ {\rm d} \log M_{\ast}$, where $\log$ is the 10 based
  logarithm. Here results are listed for satellite (upper part) and central
  (lower part) galaxies separately, which can be distinguished with zero
  measurements. Note here the stellar masses for central galaxies should be
  converted using $\log M_{\ast, c} = \log M_{\ast} -1.0$.  }\label{tab:CSMF_DR7}
\end{deluxetable*}
\end{turnpage}


\clearpage

\end{document}